\documentclass[pdflatex,sn-mathphys-num]{sn-jnl}
\usepackage{amsmath,amssymb,amsfonts}
\usepackage{setspace}
\usepackage{hyperref}
\usepackage{booktabs}
\usepackage{pdflscape}
\usepackage{soul}
 
\usepackage{color}
\usepackage{graphicx}
\usepackage{textcomp}
\usepackage[utf8]{inputenc}
\usepackage[british]{babel}

\usepackage{subcaption}
\usepackage{siunitx}
\usepackage{multirow}
\usepackage{amsthm}
\usepackage{mathrsfs}
\usepackage[title]{appendix}
\usepackage{xcolor}
\usepackage{manyfoot}
\usepackage{algorithm}
\usepackage{algorithmicx}
\usepackage{algpseudocode}
\usepackage{listings}

\DeclareSIUnit\vitesse{\meter\per\second}

\newcommand{\beginsupplement}{%
      \setcounter{section}{0}
        \renewcommand{\thesection}{S\arabic{section}}%
        \setcounter{table}{0}
        \renewcommand{\thetable}{S\arabic{table}}%
        \setcounter{figure}{0}
        \renewcommand{\thefigure}{S\arabic{figure}}%
        \setcounter{page}{1}
        \onecolumn
     }

\definecolor{dgreen}{rgb}{0,0.5,0}
\definecolor{palered}{rgb}{1,0.5,0.5}
\definecolor{purple}{rgb}{1,0,1}
\definecolor{dpurple}{rgb}{0.75,0,0.75}

\setstcolor{dgreen}

\newcommand{\vire}[1]{{}} 
\newcommand{\modif}[1]{\textcolor{black}{#1}}  

\begin{document}

\title{Geometric Confinement Reveals Scale-Free Velocity Correlations in Epithelial Cell Monolayer}
\author[1]{Guillaume Duprez$^{*}$}
\author[2]{M\'elina Durande$^{*}$}
\author[2]{Fran\c{c}ois Graner}
\author[1]{H\'el\`ene Delano\"e-Ayari$^{\dagger}$}

\affil[1]{Universit\'e Claude Bernard - Lyon 1, CNRS, ILM, UMR5306, F-69100 Villeurbanne, France}
\affil[2]{Universit\'e Paris Cit\'e, CNRS, MSC, UMR7057, F-75006 Paris, France}

\date{\today}
\maketitle
\begingroup
\renewcommand\thefootnote{*}
\footnotetext{Authors contributed equally to this work.}
\renewcommand\thefootnote{$\dagger$}
\footnotetext{Corresponding author: helene.ayari@univ-lyon1.fr}
\endgroup

\medskip
\medskip

\medskip
\medskip

\begin{center}
{\bf Abstract}\\ 
\end{center}

{\small
\noindent 
\vire{Collective cell flows are a hallmark of tissue dynamics in development, wound healing, and various diseases. Here, we investigate how the size, shape, topology and rigidity of patterned substrate influence the organization of flows and mechanical fields in an epithelial MDCK cell monolayer at several time and space scales. Using micropatterned substrates with and without free front (a strip and a closed racetrack), 
we show that  confinement and obstacles induce spatial heterogeneities in velocity and force fields, leading to the emergence of domains, waves, and long-range correlations. We evidence that spatial velocity correlations are scale-free, following a power law whose exponent is sensitive to cytoskeletal perturbations. This challenges the notion of a single intrinsic correlation length. We also show that in absence of free front, spontaneous collective motions are stronger on soft than on hard substrate. Our findings provide new insights into the rheology of epithelial tissues and the interplay between mechanics and collective migration.}%
\modif{Collective cell flows are a hallmark of tissue dynamics in development, wound healing, and various diseases. Here, we perform experiments on epithelial MDCK cell monolayers, over tens of hours without jamming, on millimeter-scale micropatterned substrates with or without free front (a strip or a closed racetrack).
During maturation in time, domains and long-range correlations of the velocity field appear. Enstrophy  increases (along with kinetic energy) during 5 hours, then passes through a maximum and decreases.
Spatial velocity correlations are scale-free, following a power law, which challenges the notion of a single intrinsic correlation length. 
It suggests that  the monolayer behaves as a critical-like system where information is transmitted across its entire size, a feature consistent with models of active solids capable of long-range stress propagation. 
The spatial correlation exponent significantly evolves with time, probably reflecting the monolayer maturation.
The size, shape, topology and rigidity of patterned substrate influence the organization of flows and mechanical fields.
Spontaneous collective motions are stronger on soft than on hard substrate.
The presence of a free front accumulates vimentin on a much larger length scale than a fixed boundary ($65\pm 4$~$\mu$m vs $3.2 \pm 0.7$~$\mu$m), possibly revealing an underlying polarizing cue on the cell velocity field.}
}

\medskip
\medskip

\medskip
\medskip
\medskip
\medskip

\medskip
\medskip

\section{Introduction}

Collective cellular movement plays a crucial role in various biological processes, including tissue development, wound healing, and cancer metastasis~\cite{friedl2009a,rorth2009}. It involves coordinated interactions between cells, enabling them to move as a cohesive group. This behavior is influenced by a combination of biochemical signals, mechanical forces, and environmental cues, creating a complex interplay that governs their collective dynamics~\cite{trepat2009,banerjee2019,porta2019}.

\subsection{State of the art}

Experiments on confluent epithelial monolayers since two decades~\cite{poujade2007b} have probed {\it in vitro} migration by culturing cells on solid substrates patterned to create various geometries delineated by non-adhesive regions, for instance within a strip~\cite{vedula2012}, within an annulus~\cite{jain2020,lovecchio2024}, within a circle~\cite{doxzen2013,saraswathibhatla2020}, or around it~\cite{kim2013}. If one of the non-adhesive boundaries is mobile, removing it creates a front for the cell layer and a free space in which migration becomes suddenly possible, as in the so-called ``wound healing assay" where cells migrate to invade or expand from rectangular~\cite{poujade2007b,das2015} or circular~\cite{cochet-Escartin2014,rausch2013} regions. Such biophysical experiments established several results that deepen our understanding of collective migration.

Cells near an unfilled space exert pulling forces towards it regardless of their migration direction~\cite{kim2013}, and the presence of a free front modulates both cell self-propulsion and alignment strength, favoring flowing states that propagate into the monolayer bulk~\cite{chepizhko2018}. Fronts may undergo fingering instabilities where lateral heterogeneities penetrate the epithelium over distances comparable to finger lengths~\cite{alert2019,reffay11}. Cells at finger tips extend protrusions with distinct morphologies characterized by large, highly active lamellipodia, their number and direction tuned by local front curvature whether spontaneous or experimentally imposed~\cite{rausch2013}.

Further behind the front, cells contribute traction forces on the substrate with some actively driving motion and others acting as mechanical resistors~\cite{trepat2009,deng2021}. The balance of traction forces and substrate friction is captured by the epithelization coefficient, distinguishing bulk- versus boundary-dominated behaviors~\cite{cochet-Escartin2014}. Competition between traction and cell-cell interaction forces can lead to active wetting transitions, where tissues either spread or retract based on their size~\cite{perez-gonzalez2019}. Cell-cell interactions also stimulate motility and regulate transitions between mesenchymal and epithelial behaviors, facilitating the onset of collective migration~\cite{mishra2019,christiansen2006,vedula2012,Friedl2012}.

Velocity and density variations within cell cohorts contribute to continuum mechanical descriptions, where velocity typically decreases with increasing cell density both in strips~\cite{tlili2018} and circular corrals~\cite{doxzen2013}. Boundary conditions allowing sufficient space lead to large-scale solid-body rotational movements (``swirls" or ``vortices") with uniform angular velocity over tens of cell lengths, as observed across various geometries including strips, circles, and annuli~\cite{hiraiwa2022,vedula2012,doxzen2013,jain2020}.

From a theoretical perspective, collective migration of epithelial monolayers is well-suited to continuum mechanics descriptions via coarse-graining from the individual cell scale to the cohort scale~\cite{khalilgharibi2016}. The cell sheet exhibits properties akin to a glass~\cite{angelini2011} or a liquid crystal~\cite{armengol-collado2024}. Internally, cell-cell interaction forces are represented by a stress tensor, either extensile or contractile in nature~\cite{balasubramaniam2021}, while traction and friction forces between cells and the substrate form external forces per unit area. These external forces balance the divergence of the internal stress, multiplied by tissue thickness, linking internal and external mechanical cues~\cite{tlili2015}.

\subsection{Current approach}

Here, we take advantage of MDCK II epithelial cells being a well established model of confluent cell monolayer migration, and we build on the existing experimental protocols (recalled in details in Section~\ref{sec:mat_meth}). 
We want  to study and disentangle the effects on collective cell movement of pattern size, shape, topology and rigidity. 
More precisely, we want to focus on the correlation between cell movements, not only to understand how far cells coordinate within the bulk, but also to detect how boundaries (whether free or confining) propagate their influence towards the bulk. These multi-scale questions require large systems, long durations and strong statistics.

Two of our\vire{ preceding} papers have introduced a method for that purpose,  applied to cells migrating within a cell-adhesive strip which was either bare~\cite{tlili2018} or with the inclusion of a circular obstacle  (i.e. a non-adhesive circle)~\cite{tlili2020}. 
The duration of \vire{preceding}\modif{anterior} experiments reported in the literature had been limited to a few hours by cell jamming, due to the increase in cell density. 
We have added the mytomicin drug before beginning the experiment (see Section~\ref{sec:culture}) to decrease the cell division rate by at least a factor of five. This, combined with the cell density decrease due to the monolayer spreading,  has prevented the cell density to increase beyond the jamming point. We have imaged the monolayer for more than a day, during which time the cells  have migrated over several millimeters. With an interframe time of 5 min and a cell size of order \vire{or}\modif{of} 10~$\mu$m, our collective migration movies have covered more than two decades both in time and in space.
This has enabled us to obtain  with unprecedented signal-to-noise ratio several  results, including the observation of  twelve successive velocity and density waves~\cite{tlili2018}, and the determination of the monolayer visco-elastic time~\cite{tlili2020}.

In the present paper, we analyze images  from sub-cellular scale to hundred-cells scale while comparing two different pattern geometries, both having a large size and a non-trivial topology. One is the already used~\cite{tlili2020} strip \modif{made favorable to cell adhesion thanks to fibronectin} with the inclusion  of a \modif{non-adherent}  circular obstacle  (Section~\ref{sec:strip}), within which cells migrate with a free front after removal of a mobile boundary (which, for technical reasons, requires a hard substrate); we perform 16 \vire{experiments}\modif{measurements}. 
The other is a racetrack~\cite{giuglaris2024}  without free front (Section~\ref{sec:racetrack}), with circular obstacles too, with variable dimensions \modif{(see Supp. Movies 1-2) or chiral $V$-shaped  design obstacles (see Supp. Movie 3)}; we performed \vire{47 measurements, all on a hard substrate}\modif{66 measurements, on 48 hard substrates} (both for convenience and to legitimate comparisons with the strip), see Table~S1 and Fig.~\ref{fig:distrib_width}. 
In addition, we  realize \vire{one racetrack experiment}\modif{racetracks} on soft substrate to measure cell traction forces through the deformation they induce in the substrate \modif{and perform 4 measurements (2 with $W=1000$~$\mu$m and 2 with $W=600$~$\mu$m)}.
Finally, we focus on spatial and temporal correlations
 (Section~\ref{sec:correlations}).
 
\modif{Overall, the three main novel features of the present study are: (i) analysis of velocity correlations on larger time- and length-scales than in the literature; (ii) comparison between open geometry with free front, and closed periodic geometry, performed in similar experimental conditions; (iii) correlation with vimentin at boundaries.}

\begin{figure}[t!]
    \centering
    \begin{subfigure}{\textwidth}
        \includegraphics[width=\linewidth]{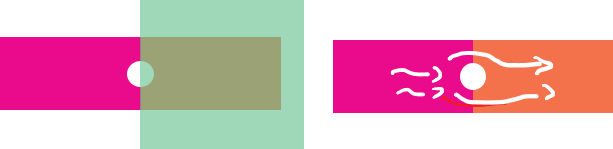}
         \caption{}        
    \end{subfigure}
    
    \begin{subfigure}{0.9\textwidth}
        \includegraphics[width=\linewidth]{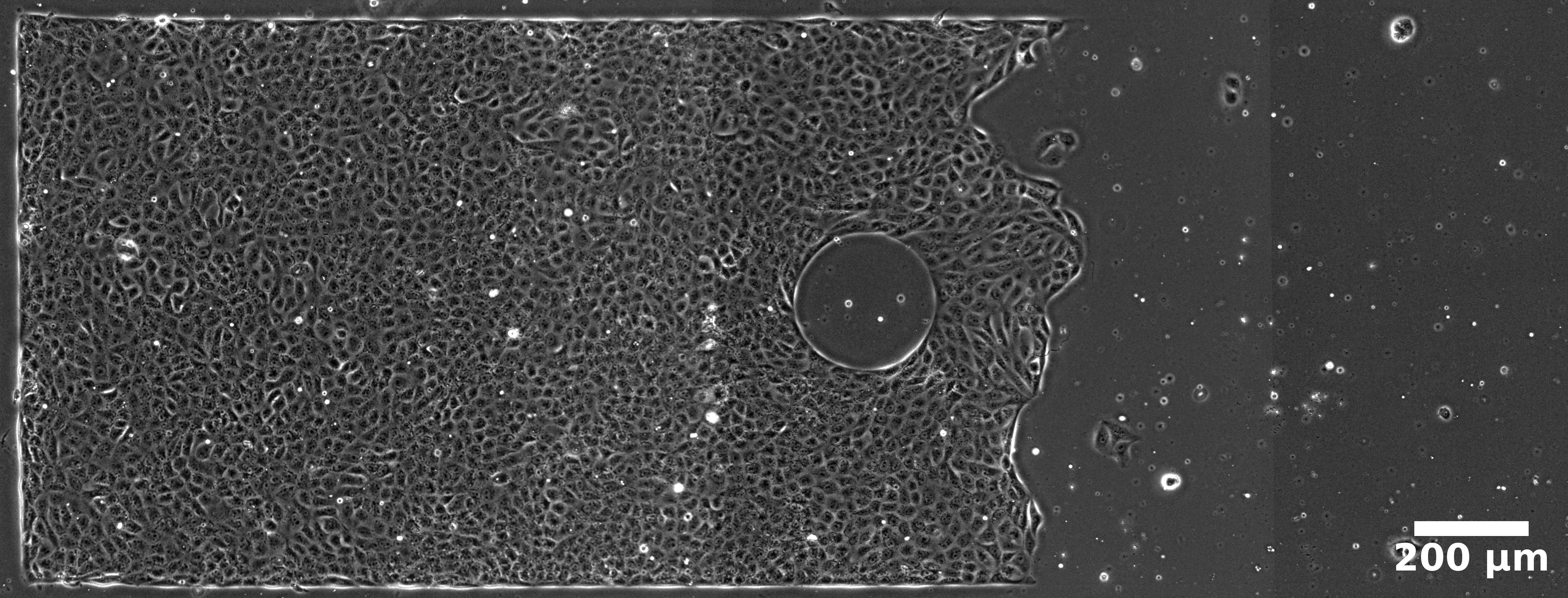}
        \caption{}
        \label{fig:phase_bande_avt}
    \end{subfigure}    
        \begin{subfigure}{0.9\textwidth}
        \includegraphics[width=\linewidth]{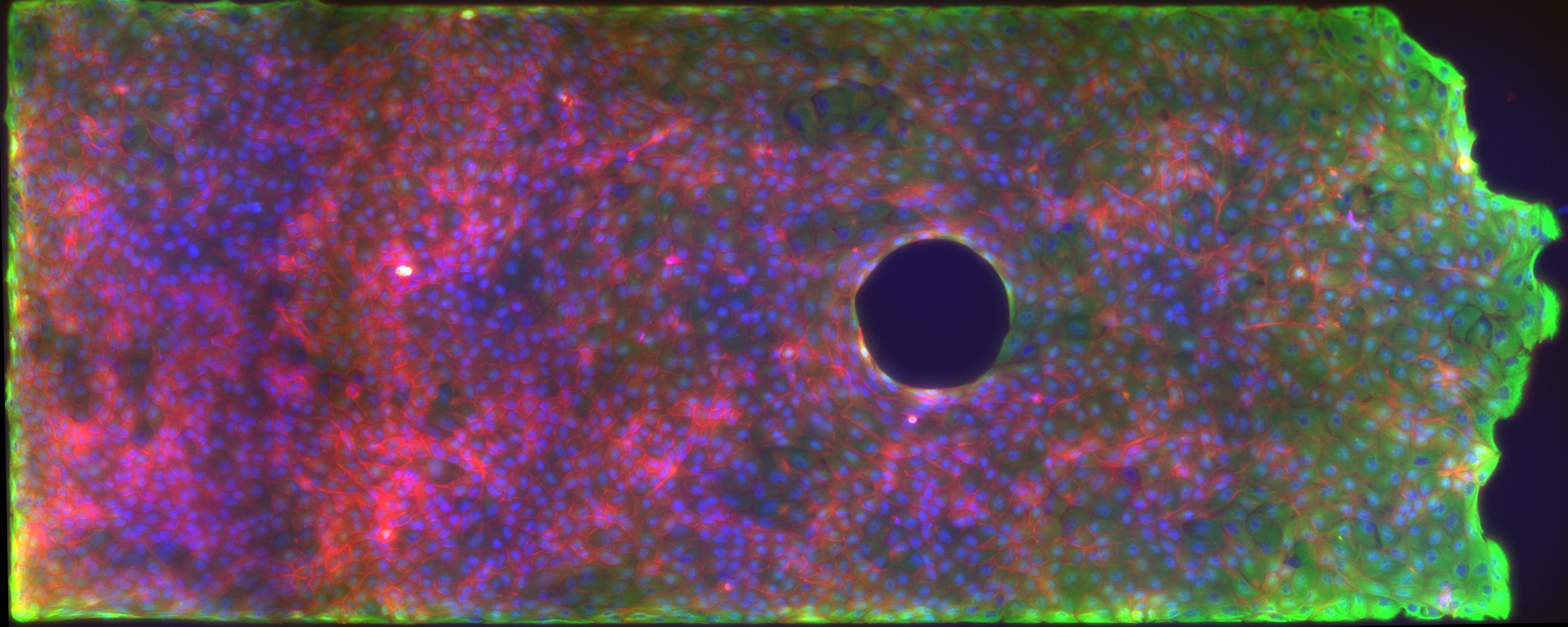}
        \caption{}
        \label{fig:phase_bande_after}
    \end{subfigure}
\caption{{\bf Strip setup.}   
{\bf (a)} Schematics, with adhesive substrate regions in orange when bare and in pink when covered with cells; left: strip at $t<0$, i.e., before removal of the mobile non-adhesive boundary; right: at $t>0$, after removal, white arrows mark directional flow.
{\bf (b)} Beginning of the experiment, live imaging, phase contrast. For legibility, \vire{only part of the strip length is}\modif{the strip bare parts on the picture right are not} shown.
{\bf (c)} 16 h later, fixed image, triple staining: nuclei (blue), actin (red), vimentin (green). 
}
\label{Fig_setup_strip}
\end{figure}

\section{Strip, with  free front}
\label{sec:strip}

The strip geometry (Fig.~\ref{Fig_setup_strip}) is as previously described~\cite{,tlili2020}. 
Briefly, a rectangular strip, of length $L \approx 3000$~$\mu$m and width $W=1000\pm10$~$\mu$m is made adhesive over all its length except that, in its center, a circular region of radius $130 \pm 10$~$\mu$m is made non-adhesive (the ``obstacle"). A mobile boundary is placed in the middle of the strip. 
MDCK cells are then seeded  in the accessible adhesive zone (left part of the strip) and grow until confluence. At $t=0$, the mobile boundary is removed. This manipulation  involves some delay before the movie can be recorded, and requires a hard substrate to preserve the cell monolayer and substrate integrities (here the PDMS substrate rigidity modulus is 1 to 2~MPa, much too large for cells to deform it).  By active migration, cells at the free front spread towards the newly accessible adhesive zone (right part of the strip) and the collective movement begins. Since cells conserve their volume, their height steadily decreases with time until it plateaus.

\begin{figure}
    \centering
    \begin{subfigure}{0.66\textwidth}
        \centering
        \includegraphics[width=\linewidth]{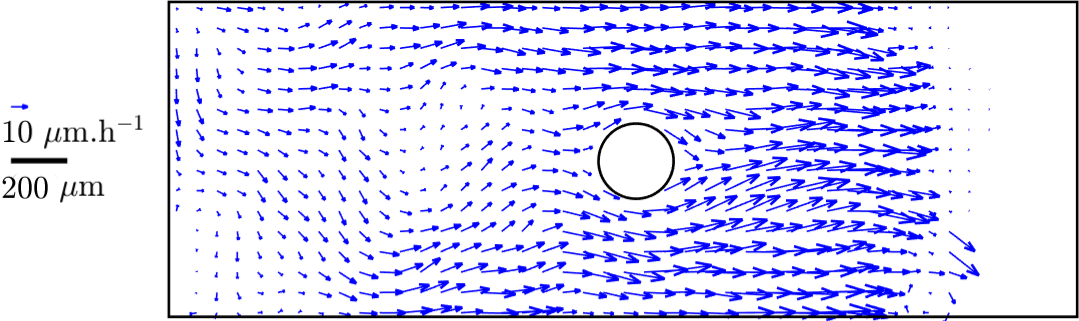}
        \caption{}
        \label{fig:vel_vector_not_averaged}
    \end{subfigure}
    \begin{subfigure}{0.44\textwidth}
        \centering
        \includegraphics[width=\linewidth]{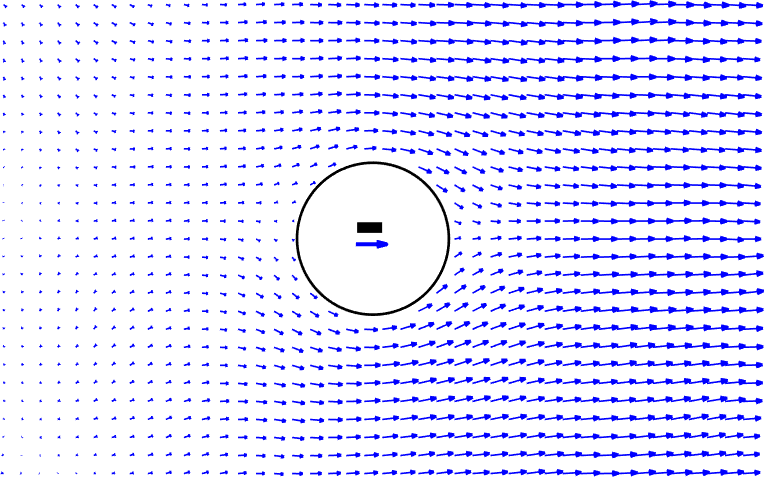}
        \caption{}
        \label{fig:vel_vector_ensemble_averaged}
    \end{subfigure}
    \hfill 
      \begin{subfigure}{0.44\textwidth}
        \centering
        \includegraphics[width=0.7\linewidth]{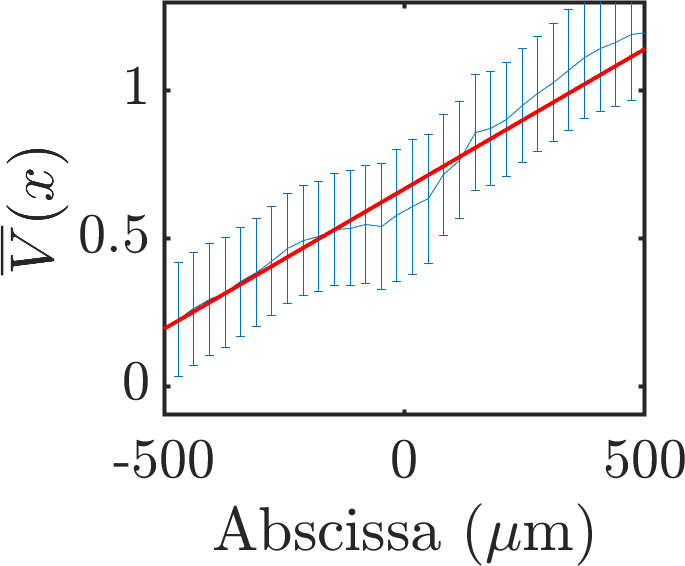}
        \caption{}
        \label{fig:mes_gradient}
    \end{subfigure}
        \begin{subfigure}{0.44\textwidth}
        \centering
        \includegraphics[width=\linewidth]{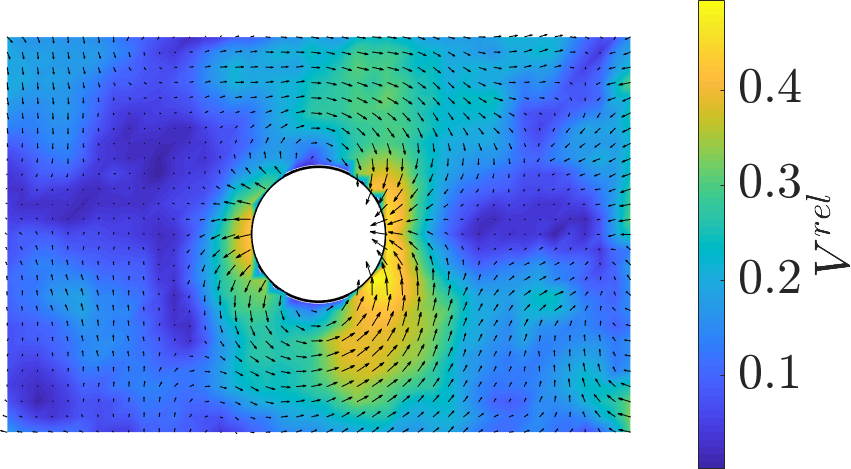}
        \caption{}
        \label{fig:map_speed_stripe_ref_fluid}
    \end{subfigure} 
     \hfill 
       \begin{subfigure}{0.44\textwidth}
        \centering
        \includegraphics[width=\linewidth]{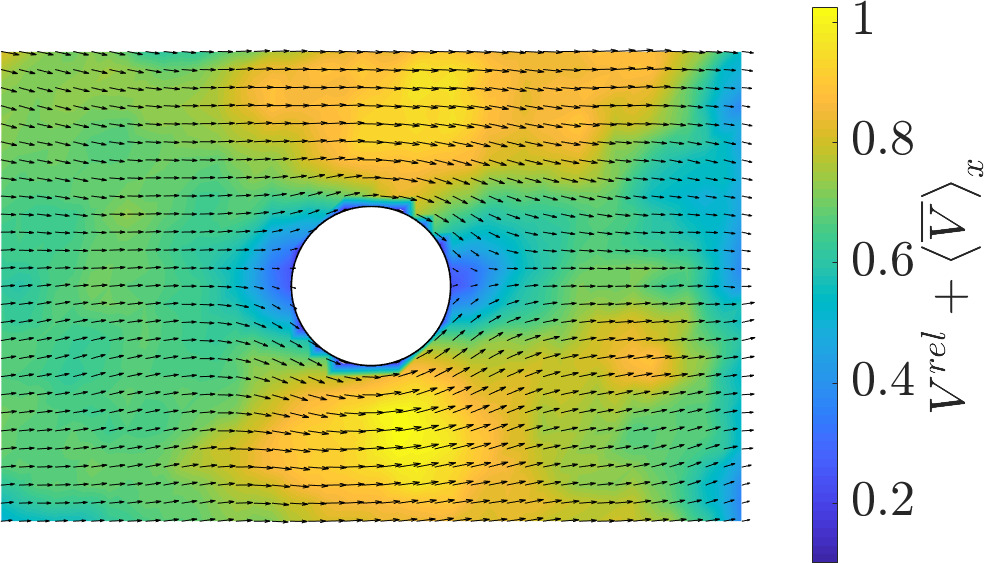}
        \caption{}
        \label{fig:map_speed_stripe_ref_obst}
    \end{subfigure}
    \caption{{\bf Velocity field in a strip with a circular obstacle.}    Axis $x$ is parallel to the strip long axis (here horizontal), oriented towards the right.
 {\bf (a)} \modif{Snapshot of velocity field. {\bf (b)} Average v}\vire{V}elocity field $\vec{v}(x,y)$\vire{ near the obstacle}.  \modif{The ``region of interest" (ROI) is the rectangle centered around the obstacle, vertical width $W$, horizontal length 1.5~$W$.}
    For each frame, the velocity field is adimensioned by its median value within this frame. It is then averaged over time $t$ (over 8~h) and over 16 different experiments.
    Black scale bar: 50~$\mu$m. 
    \vire{Pink scale arrow: velocity 2, in units adimensioned by the median.}\modif{Blue scale arrow for adimensioned velocity: twice the median velocity.}
    \vire{(b)}\modif{\bf (c)} 
    Average  of \vire{(a)}\modif{(b)} across the strip width of the velocity $x$-component,  
    \modif{$\overline{V}(x)$. Red straight line: linear fit.}
    \vire{(c)}\modif{\bf (d)} 
    Relative velocity  $\vec{V}^{rel}(x,y)= \vec{v}(x,y) - \overline{V}(x)\vec{x}$, i.e. in the virtual reference frame of the monolayer: same velocity field as in \vire{(a)}\modif{(b)} after substraction of the speed gradient \vire{(b)}\modif{(c)}.
    \vire{(d)}\modif{\bf (e)} 
    Gradient-less velocity back in the substrate reference frame, i.e. same velocity field as in \vire{(c)}\modif{(d)} after addition of the spatial average of velocity.}
    \label{fig: velocity_strip}
\end{figure}

\subsection{Velocity field is spatially graded}

The presence of a free boundary  induces a directional flow with a gradient in both velocity and density (number of cells per unit surface). 
We measure cell velocities using the Kanade-Lucas-Tomasi algorithm~\cite{lucas1981, bouguet}  which tracks on successive images landmarks of high local intensity gradient (Section~\ref{sec:KLT}).
The resulting velocity field \modif{Fig.~\ref{fig: velocity_strip}a)} displays a high inter-experiments variability, mostly due to variations in initial cell density. We first normalize each velocity field by dividing it by its median velocity measured in the \modif{region of interest (}ROI\modif{, i.e.} field of view centered around the obstacle, Fig.~\ref{fig: velocity_strip}\vire{a}\modif{b}). This adimensioned velocity field  is noted $\vec{v}(x, y, t)$, where $x$ is the axis parallel to the strip \vire{long angle}\modif{length $L$} and $y$ across the strip width $W$. It has a reduced variability, and if we  average it  both over time (here, over 8 hours) and across different experiments we retrieve a  smooth  field \modif{(Fig.~\ref{fig: velocity_strip}b)}.

Whether averaged over time, over space or over experiment repeats, the flow has a non-zero average parallel to the strip. 
\vire{Averaging the $x$-component of $\vec{v}(x, y, t)$ over $y$ defines a one-dimensional velocity field $\overline{V}(x,t)= \int_{y=0}^{y=W} v_x(x, y, t)dy / W$. 
We  measure the average of $\overline{V}(x,t)$  over time and experiments, which as expected~[21] exhibits a well-defined spatial gradient characteristic of the migration with a free front. Thanks to the multiple averaging, it displays a good signal-to-noise ratio ($\sim$10)  despite a high  standard deviation ($\sim$0.2 in adimensioned units), 
see Fig.~2b.}\modif{To improve the signal-to-noise ratio, we perform a multiple averaging of the $x$-component of $\vec{v}(x, y, t)$, as follows. 
We first average it over $y$. This defines a one-dimensional velocity field $\overline{V}(x,t)= \int_{y=0}^{y=W} v_x(x, y, t)dy / W$. We  then perform the average of $\overline{V}(x,t)$  over time, then ensemble average it over experiments. As expected~\cite{tlili2018}, it exhibits a well-defined spatial gradient characteristic of the migration with a free front. 
The final signal-to-noise ratio reaches $\sim$10  despite a high  standard deviation ($\sim$0.2 in adimensioned units), see Fig.~\ref{fig: velocity_strip}c.}

\subsection{Obstacle influence range is anisotropic}

The obstacle locally perturbs the flow direction, and this pertubation too has a non-zero average whether  
over time, over space or over experiment repeats. In the past, the presence of an obstacle has helped \modif{us} probe the cell monolayer visco-elastic nature~\cite{,tlili2020} and has been shown to be a discriminant benchmark between numerical models of active collective motions~\cite{beatrici2023}. Here we take advantage of it to measure the deviation of cell velocity  around this non-zero average, either locally around the obstacle, or globally across the whole strip.

To evidence how far the obstacle influences the flow, we  subtract the spatial gradient to obtain the relative velocity $\vec{V}^{rel}(x,y,t)= \vec{v}(x,y,t) - \overline{V}(x,t)\vec{x}$ and obtain the velocity field in the (virtual) reference frame of the cells within which the obstacle would be moving.
Upstream and downstream, there are arrest points which are very localized ($\sim$1/4 of the obstacle diameter) and on either lateral sides an acceleration as wide as the strip itself (up to the top and bottom boundaries of Fig.~\ref{fig: velocity_strip}\vire{c,d}\modif{d,e}). 
This representation facilitates the comparison with passive incompressible materials, where the upstream and downstream obstacle influence extend\modif{s} to several obstacle diameters~\cite{Cheddadi2011}, and with active collective movements of particles, where the lateral obstacle influence is much narrower~\cite{beatrici2023}.

\begin{figure}
    \centering
    \begin{subfigure}{0.78\textwidth}
        \centering
        \includegraphics[width=\linewidth]{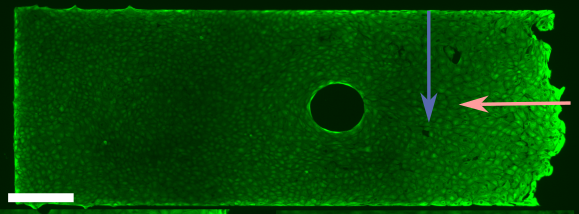}
        \caption{}
        \label{fig:vim_stripes}
    \end{subfigure}
     \begin{subfigure}{0.34\textwidth}
        \centering
        \includegraphics[width=\linewidth]{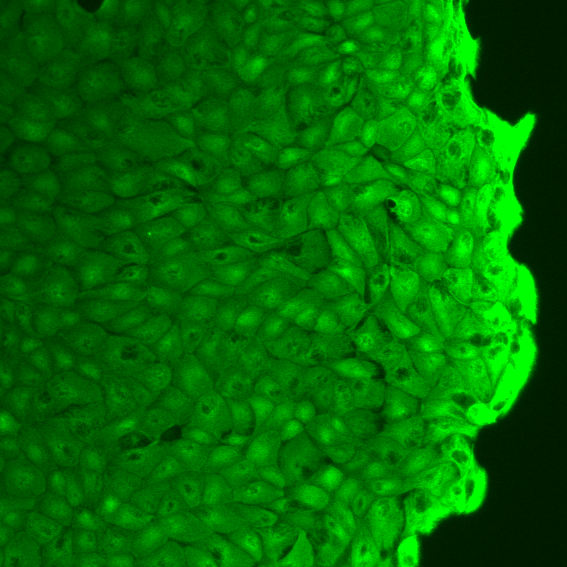}
        \caption{}
        \label{fig:vim_front}
    \end{subfigure}
        \begin{subfigure}{0.34\textwidth}
        \centering
        \includegraphics[width=\linewidth]{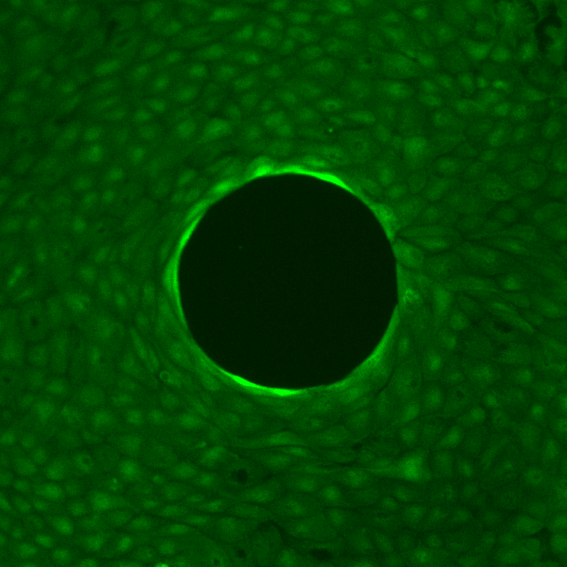}
        \caption{}
        \label{fig:vim_obst}
    \end{subfigure}
        \begin{subfigure}{0.34\textwidth}
        \centering
        \includegraphics[width=\linewidth]{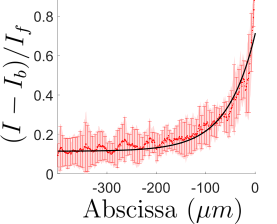}
        \caption{}
        \label{fig:vim_curve_side}
    \end{subfigure} 
            \begin{subfigure}{0.34\textwidth}
        \centering
        \includegraphics[width=\linewidth]{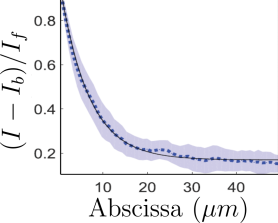}
        \caption{}
        \label{fig:vim_curve_front}
    \end{subfigure} 
    \caption{{\bf Vimentin in a strip}. 
    	{\bf (a)} Global view of vimentin in a strip, showing a brighter staining at boundaries. 
        {\bf (b)} Zoom on the front. 
        {\bf (c)} Zoom on the obstacle. 
         {\bf (d)} Fluorescence intensity vs distance from the free boundary, measured along the \modif{horizontal} red arrow in (a).
        {\bf (e)} Fluorescence intensity vs distance from the confining boundary, measured along the \modif{vertical} blue arrow in (a). 
  }
    \label{fig : vimentin}
\end{figure}

\begin{figure}[t!]
(a) \hfill ~\\
\includegraphics[width=0.811\textwidth]{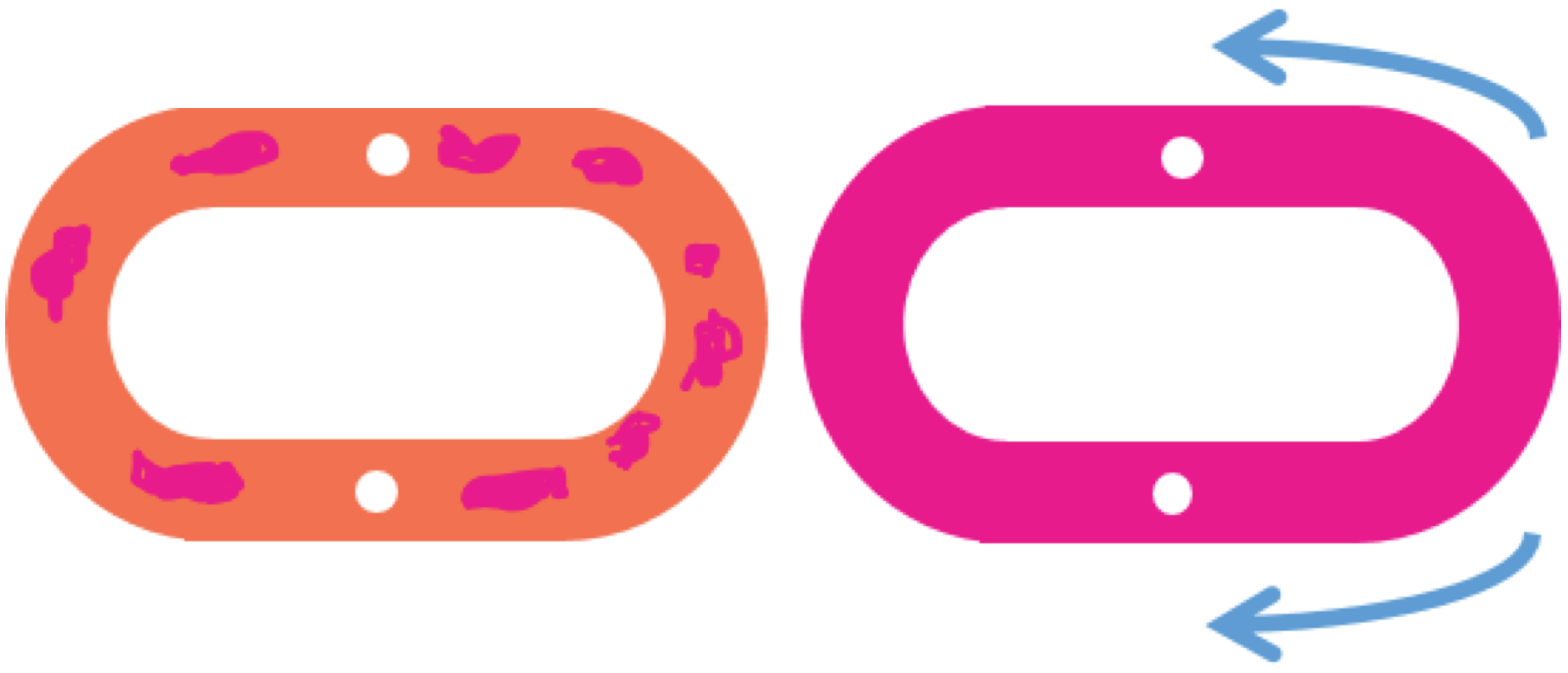}\\
(b) \hfill (c) \hfill ~\\
\includegraphics[height=0.338\textwidth]{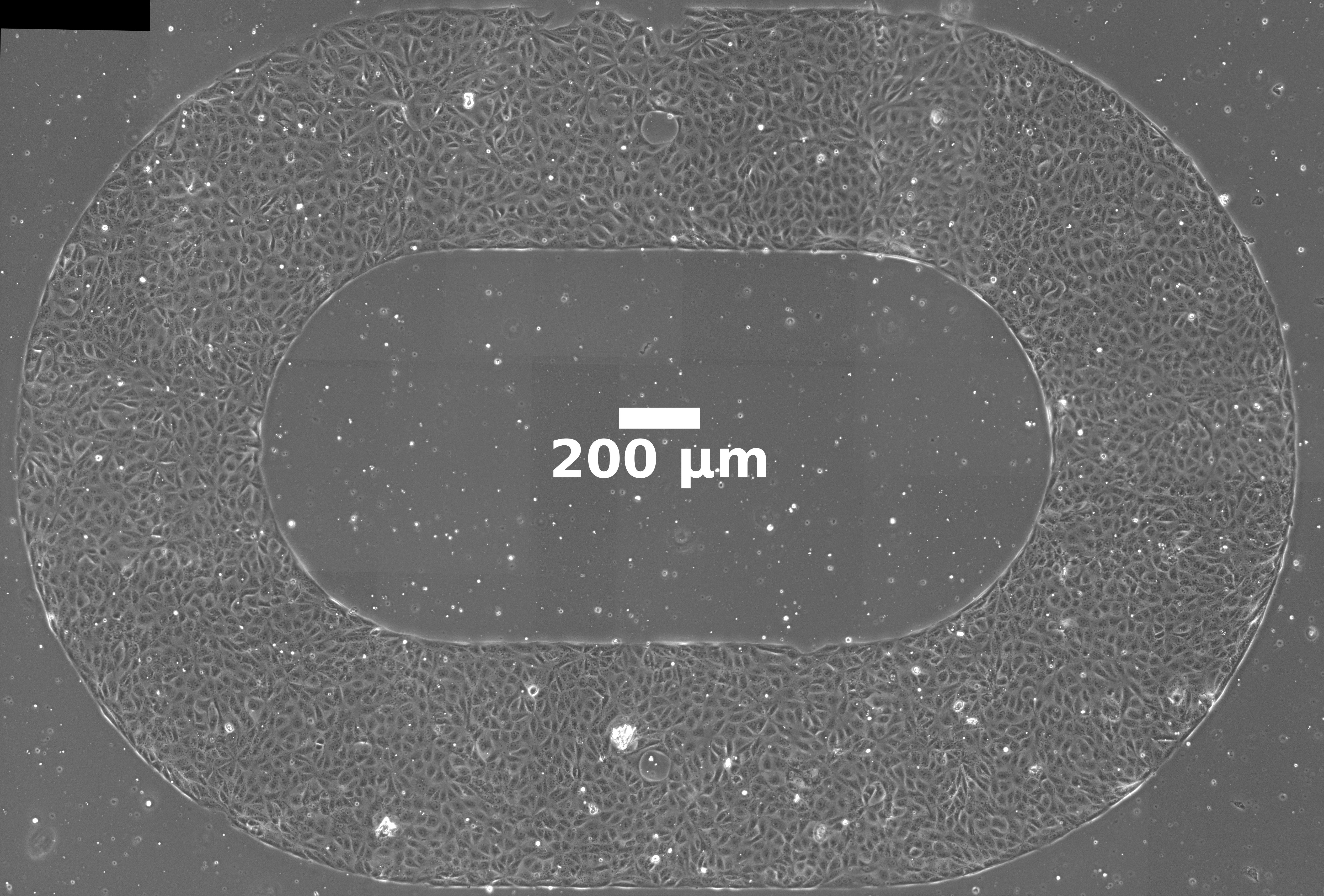}
\includegraphics[height=0.338\textwidth]{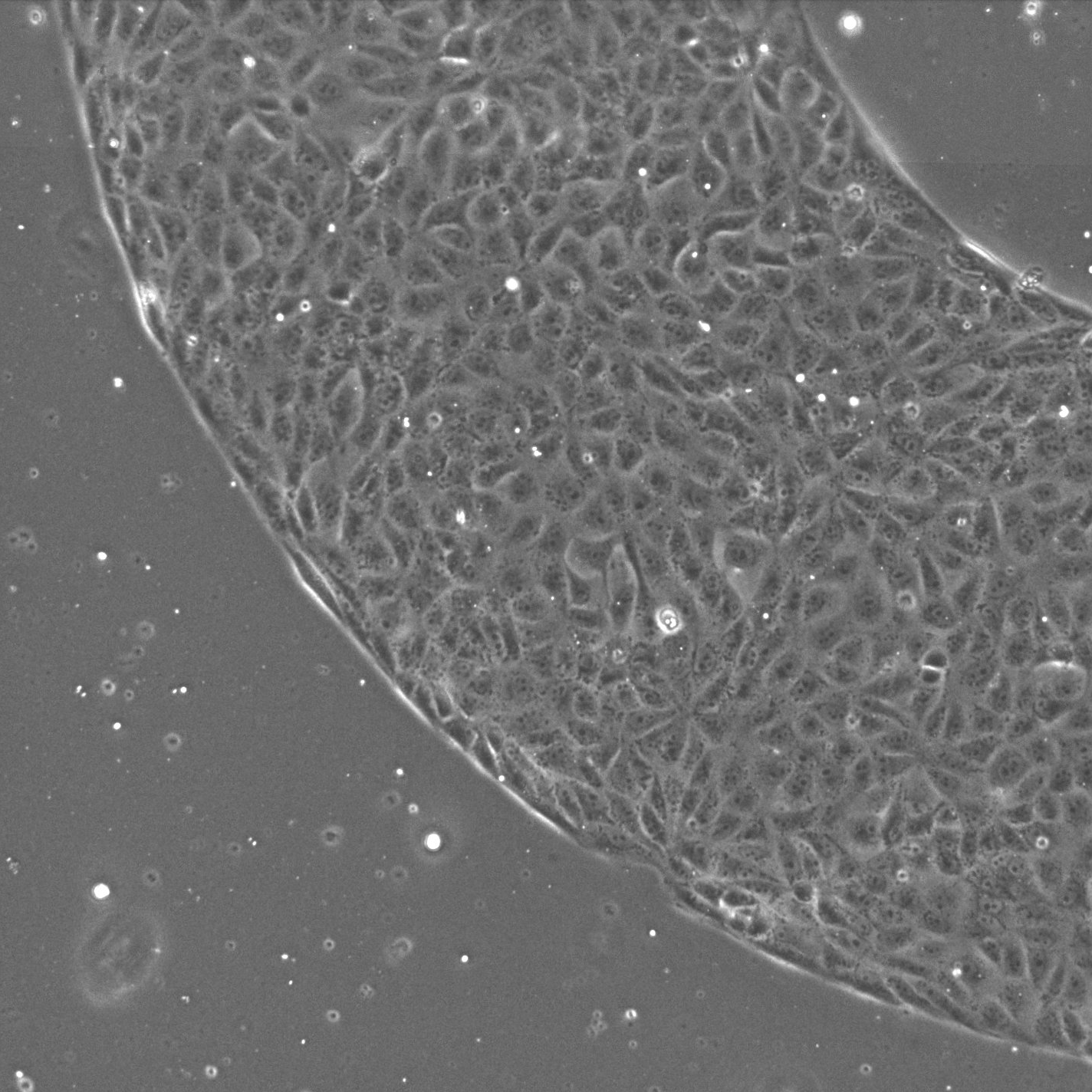}
\hfill ~ \\
(d) \hfill (e) \hfill ~\\
\includegraphics[height=0.348\textwidth]{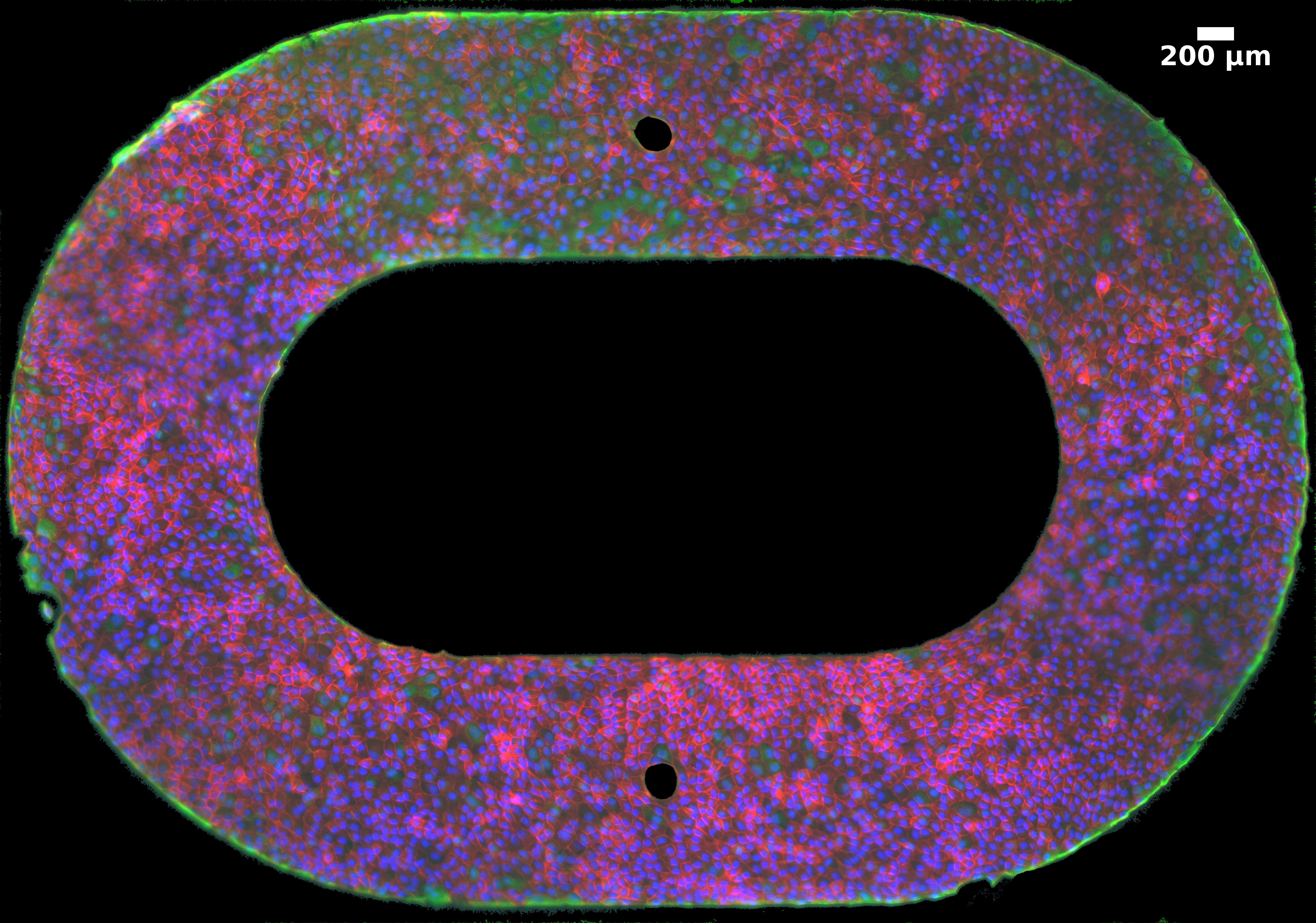}
\includegraphics[height=0.348\textwidth]{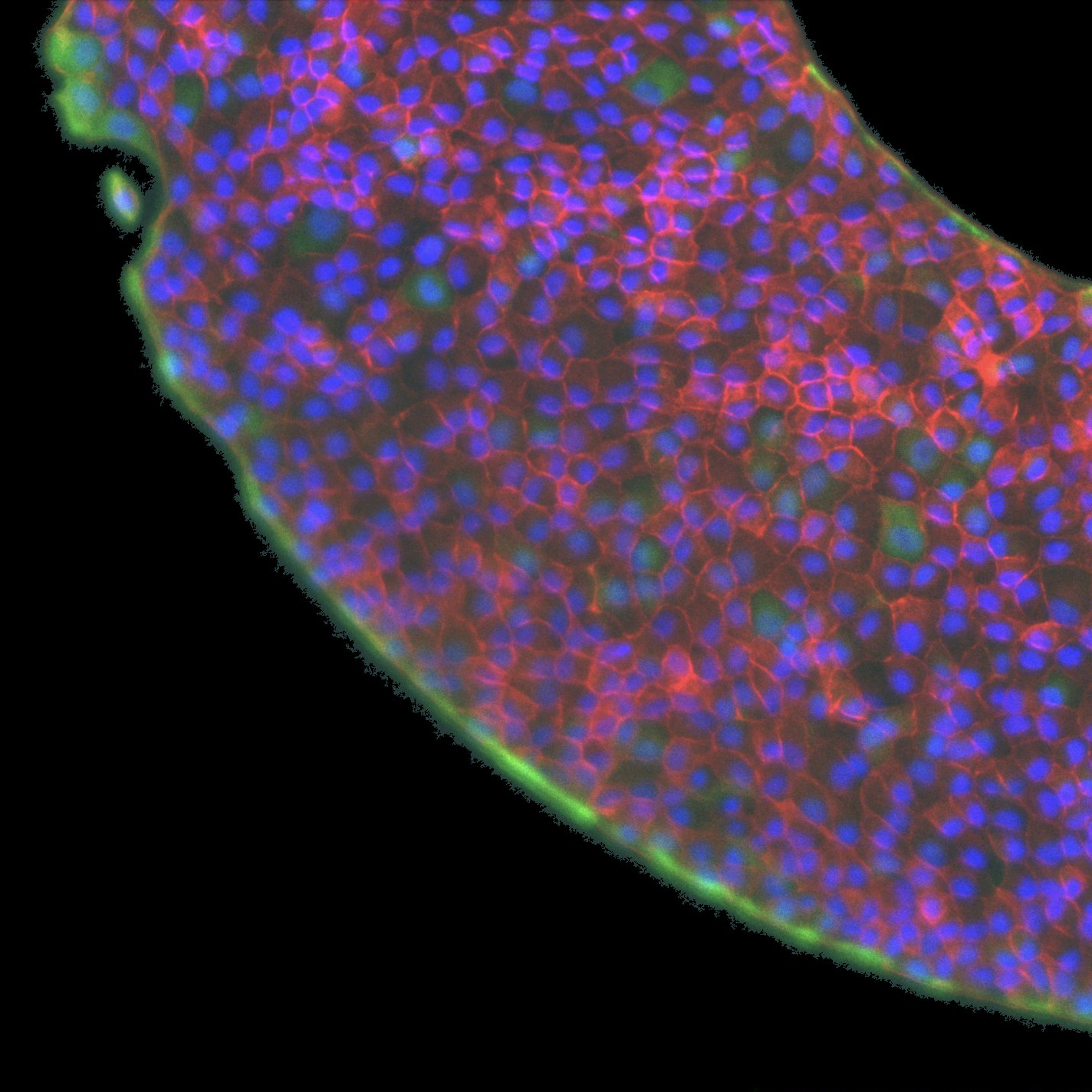}
\hfill ~ \\
\caption{{\bf Racetrack setup.} 
{\bf (a)} Schematics, with adhesive substrate regions in orange when bare and in pink when covered with cells; left:  at  $t<0$, i.e., before seeded cells are confluent, and at $t>0$, cells are confluent, blue arrows represent alternate flow directions.
{\bf (b)} Beginning of the experiment, live imaging, phase contrast. 
\modif{{\bf (c)} Zoom of the bottom left part of (b).}
\vire{(c)}\modif{{\bf (d,e)} Same as (b,c),}
16 h later, fixed image, triple staining: nuclei (blue), actin (red), vimentin (green).
In (b\vire{,c)}\modif{-e)}, substrate is hard.
}
\label{Fig_setup_hippo}
\end{figure}

\subsection{\modif{Effect of the free front}}
{\bf \large \vire{2.3 The free front acts as an external field \\ }}

In free front experiments, mechanical forces mediate intercellular signaling, providing and propagating guidance cues for collective cell migration~\cite{hino2020,boocockTheoryMechanochemicalPatterning2021}. 
 Cells can be highly polarized in the direction of the free front, even far from the front itself. Hence the \modif{free} front \modif{boundary condition} plays the role of an external polarization \vire{field}\modif{cue}. 
We do not observe any actin cable near the boundaries (Fig.~\ref{Fig_setup_strip}c). 
In order to test how cells might detect their distance from this external cue, we stained them for vimentin, an intermediate filament. 
Vimentin plays a pro\vire{e}minent role in force generation~\cite{wu2022,nunesvicente2022} and is also a marker of the epithelial-to-mesenchymal transition, which is important in cancer and wound healing.
Vimentin is overexpressed when cells begin to migrate, in particular near the free front~\cite{balcioglu2020}. 

Here we observe that vimentin marks in fact all boundaries, whether confining or free (Fig.~\ref{fig : vimentin}a-c). Its expression level is spatially graded at the vicinity of the boundaries  (Fig.~\ref{fig : vimentin}b,c).
The intensity curve vs distance to the boundary can be fitted by a decreasing exponential (Fig.~\ref{fig : vimentin}d,e). Strikingly, the characteristic length near the free boundary, $65\pm 4$~$\mu$m,  is an order of magnitude greater than that near the confining boundary, $3.2 \pm 0.7$~$\mu$m.

As opposed to keratin, another intermediate filament which resists high externally applied stress or strain~\cite{latorre2018}, here vimentin marks the spontaneous stress or strain internally generated by the migrating cells. 
Interestingly, in presence of Simvastatin, a drug that causes the vimentin network to collapse near the nucleus (without affecting microtubules or actin, and without depolymerizing vimentin)~\cite{trogden2018}, cells become able to escape the patterning (data not shown).

\vire{In conclusion, through vimentin, cells convert the presence of the free front into a biochemical signal which significantly impacts their migratory behavior.}
\modif{In conclusion, in addition to defining the boundary conditions, the presence of cell front seems to affect the cell behaviour itself. The vimentin gradient possibly reveals an underlying polarizing cue on the cell velocity field. As vimentin is part of several biochemical signaling cascades, it would be interesting to obtain a spatial mapping of the RNA expression.}

\begin{figure}[t!]
\begin{center}
(a) \hfill  \hfill   \hfill ~ ~  (b) \hfill   \hfill  ~ ~ ~ (c) \hfill   \hfill  \hfill  \hfill ~ \\
    \includegraphics[height=0.25\textwidth]{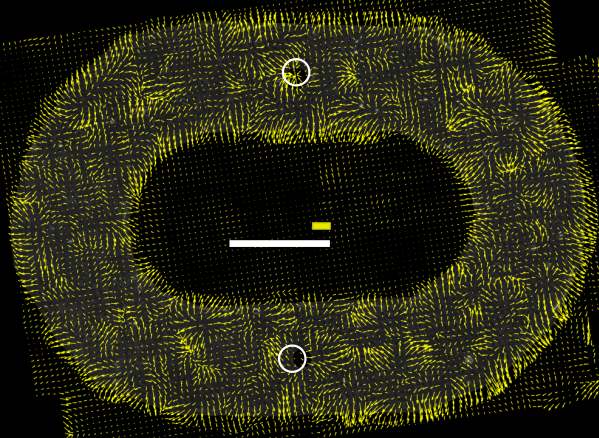}
    \hfill
    \includegraphics[height=0.25\textwidth]{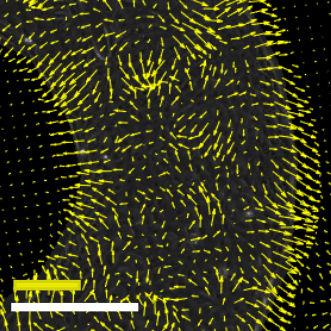}
    \hfill
    \includegraphics[height=0.25\textwidth]{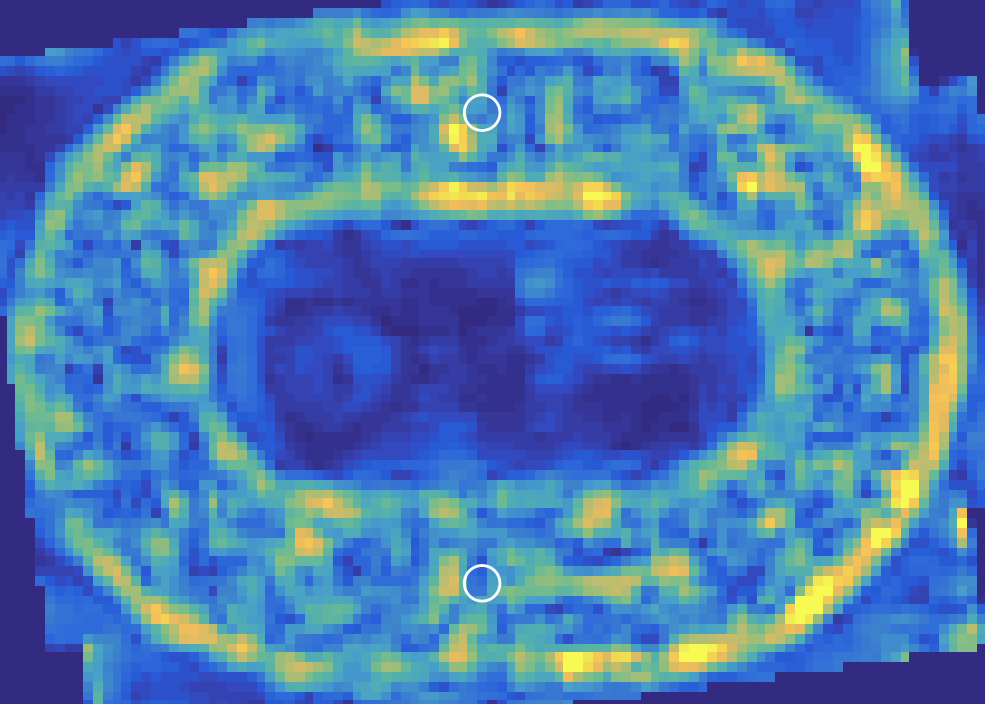}
    \includegraphics[height=0.26\textwidth]{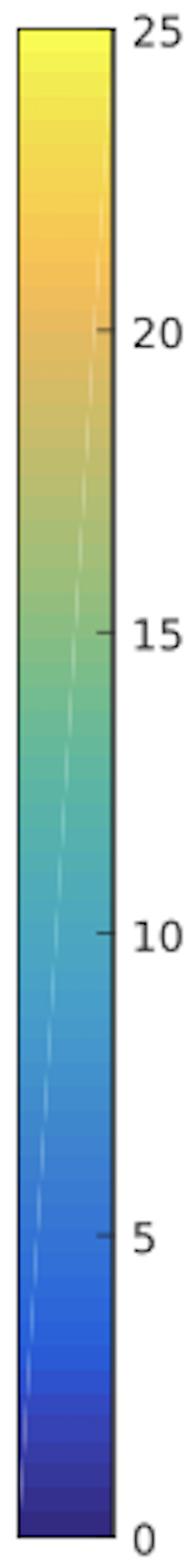}\\
~ ~ ~ (d) \hfill  \hfill  ~ ~ ~  ~   (e) \hfill  \hfill ~ ~  ~  ~  ~   (f)  \hfill  \hfill  \hfill  \hfill  ~ \\
~ \hfill 
    \includegraphics[height=0.25\textwidth]{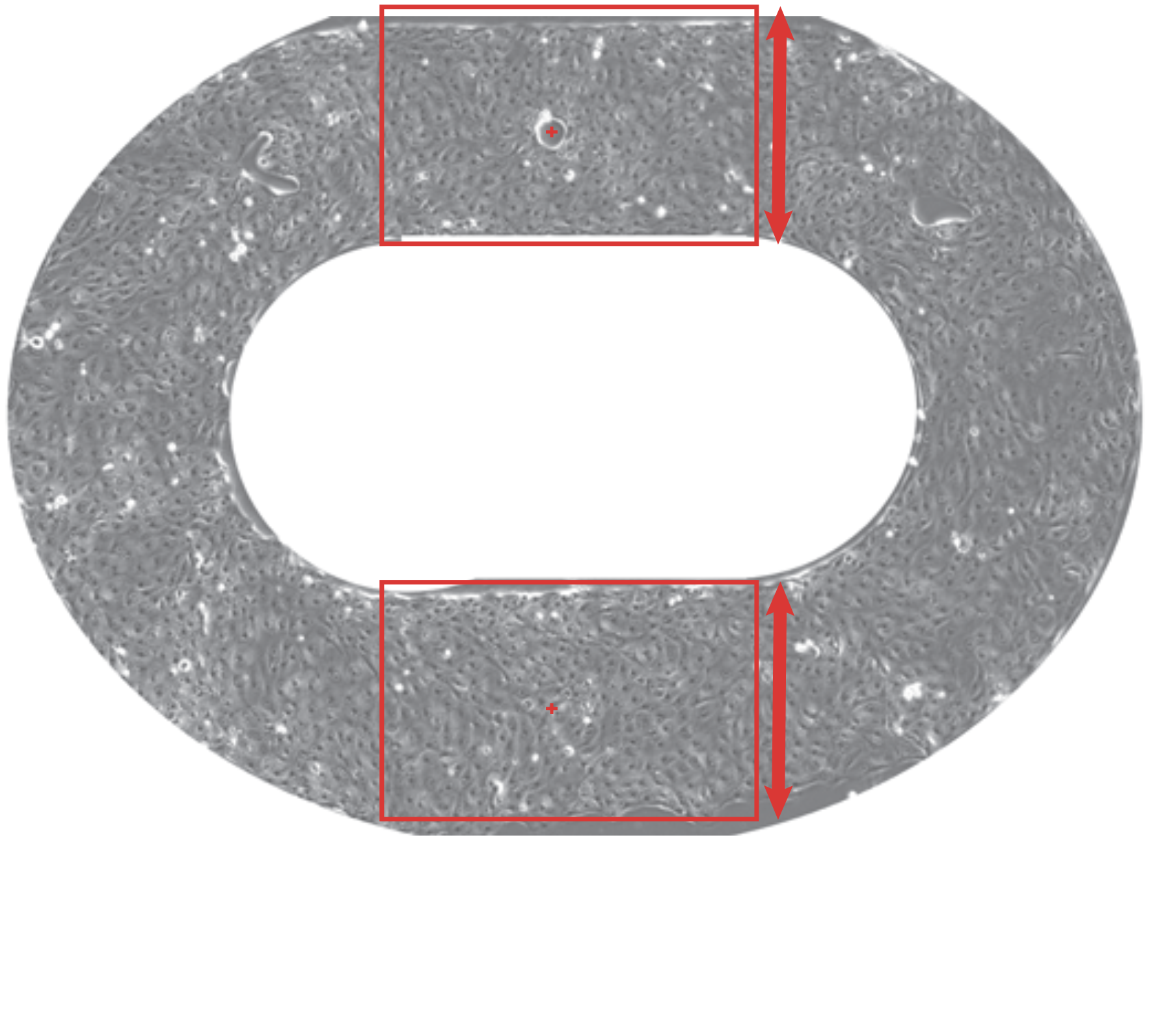}
    \includegraphics[height=0.29\textwidth]{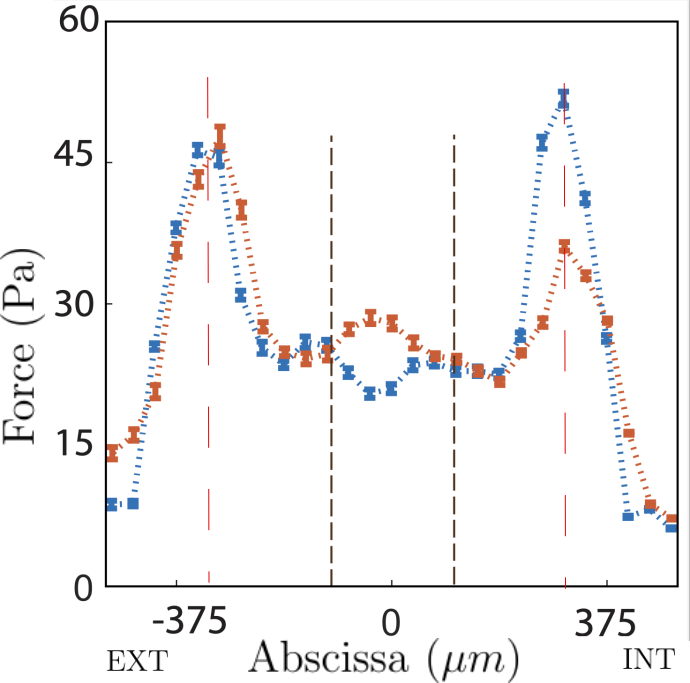}
  \includegraphics[height=0.29\textwidth]{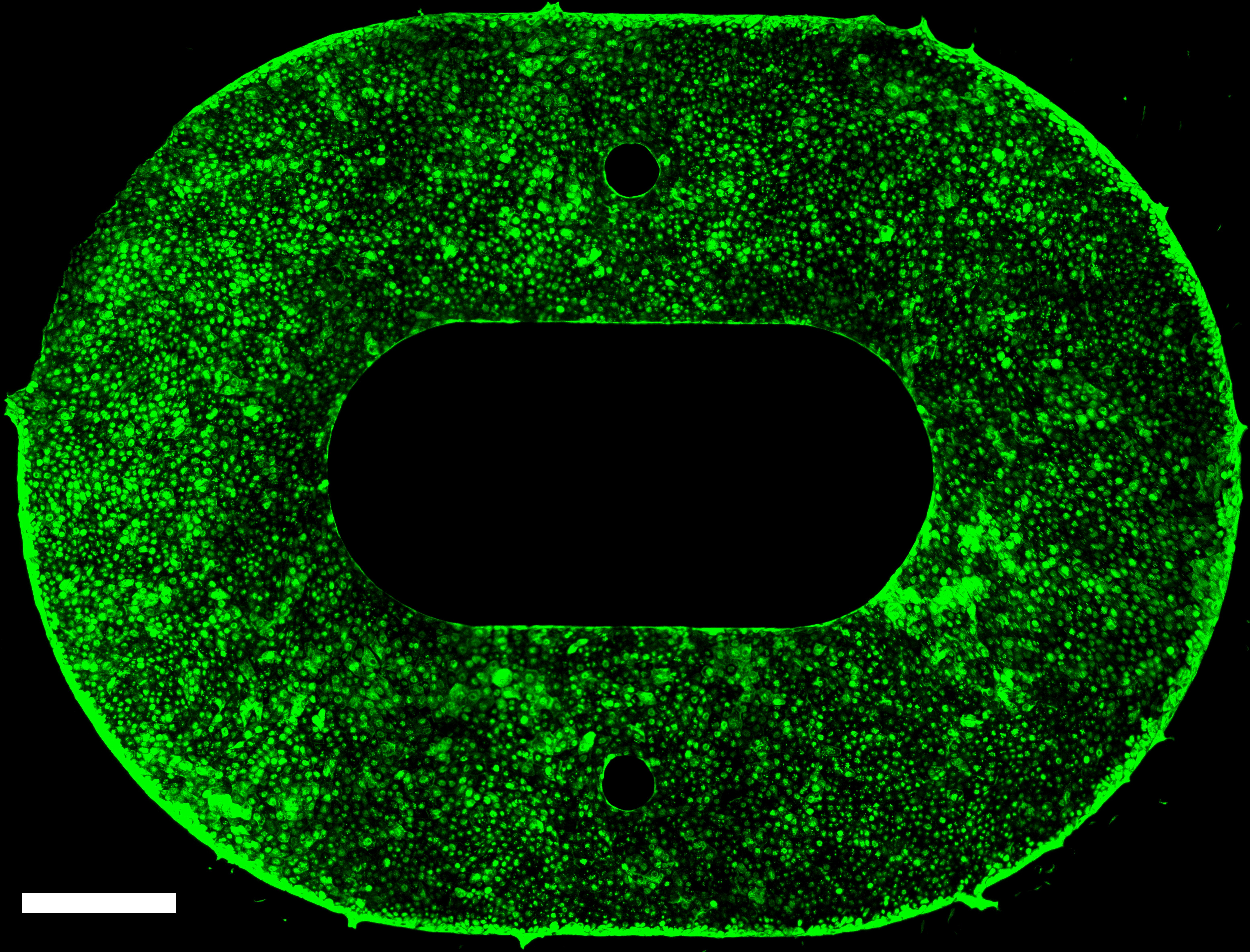}
 \end{center}
\caption{{\bf Traction forces o\vire{f}\modif{n} soft substrate.}
{\bf (a)} Forces per unit area (here exerted by the substrate on the cells) in the soft racetrack, with force vectors averaged over 200-pixel boxes. White scale bar: 750~$\mu$m, yellow scale bar: 30~Pa. 
White circles indicate obstacle locations.
{\bf (b)} Zoom on the right part of (a). White scale bar: 325~$\mu$m; yellow scale bar: 30~Pa. 
{\bf (c)} Force magnitude, time averaged over 150 frames ($\sim$13~h), color coded from 0 to 60~Pa. 
{\bf (d)} Regions of interest (rectangles \modif{of vertical width $W$, horizontal length 1.5~$W$, centered on an obstacle}), directions of measurements (arrows) and obstacles (dots). 
{\bf (e)} Force profile across the width. Force   in (a) is  measured in a red rectangle in (d), averaged in space parallel to the racetrack midline, then plotted  vs the distance to obstacle center along the red arrow in (d), with ``EXT" outside the racetrack and ``INT" inside. Bars: standard deviation of time average.
Vertical dashes: obstacle \modif{(black dashes) and track (red dashes)}  boundaries at the beginning of the experiment. Blue curve corresponds to the top obstacle in (a), red curve to the bottom obstacle which is engulfed by the cells at long times.
{\bf (f)} Vimentin staining on a hard racetrack.
}
\label{fig: wholeracetrack}
\label{fig: force_profile}
     \label{fig:vim_hippo}
\end{figure}

\section{Racetrack, without front}
\label{sec:racetrack}

To discriminate more clearly the cell monolayer's intrinsic bulk  properties from the external impact of the free front, we removed the latter by turning to a closed geometry: the racetrack  (Fig.~\ref{Fig_setup_hippo}). An adhesive periodic circuit is made of two half circles, linked with two straight portions in which a circular obstacle is inserted, so that we can compare the effect of different boundary curvatures: concave, flat or convex. To enable a multiscale study comparable to that of strips, we keep a large size in each of the \vire{47 patterns: the overall length $L$ is chosen in the range 4800 to 6200~$\mu$m, and each portion has the same width $W$ which is chosen in three sizes: either $\sim$600, 900 or 1000~$\mu$m (see size distributions in Fig.~S1).}\modif{48 patterns. Each portion of the racetrack has a uniform width $W$ which is chosen as: either $W\sim600$~$\mu$m, in which case the midline perimeter length $L$ is 6700~$\mu$m; or $W\sim 900-1000$~$\mu$m, in which case $L=8000$~$\mu$m; see substrate distributions in Table~S1 and actual measurements in Fig.~\ref{fig:distrib_width}. }


\begin{figure}[t!]
    \centering
    \begin{subfigure}{0.45\textwidth}
        \centering
        \includegraphics[width=\linewidth]{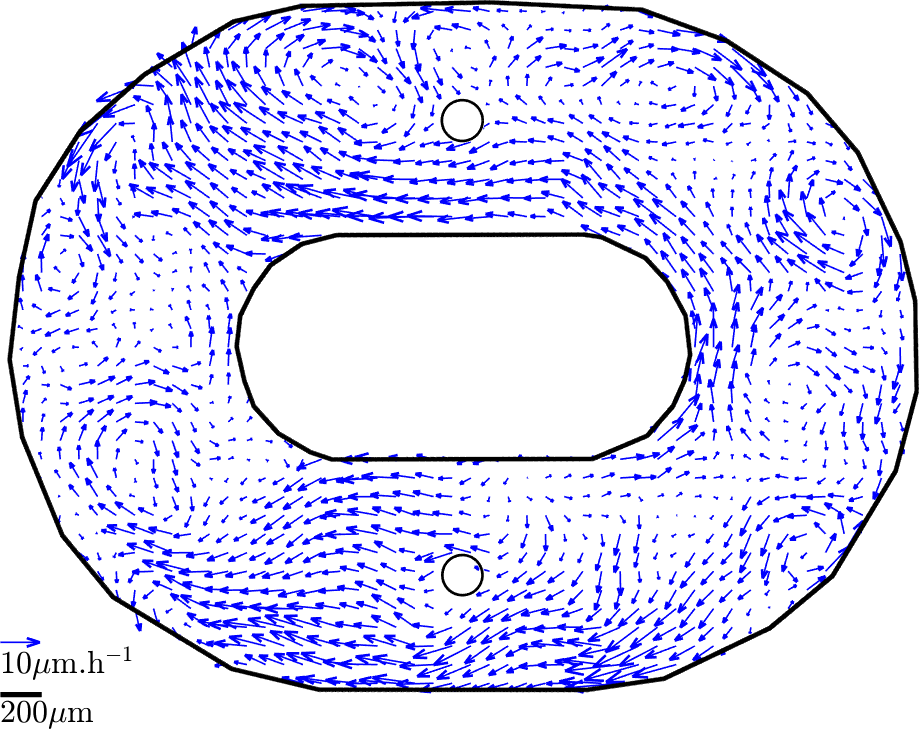}
        \caption{}
        \label{fig:vel_full_hippo}
    \end{subfigure}
    \hfill
     \begin{subfigure}{0.51\textwidth}
        \centering
        \includegraphics[width=\linewidth]{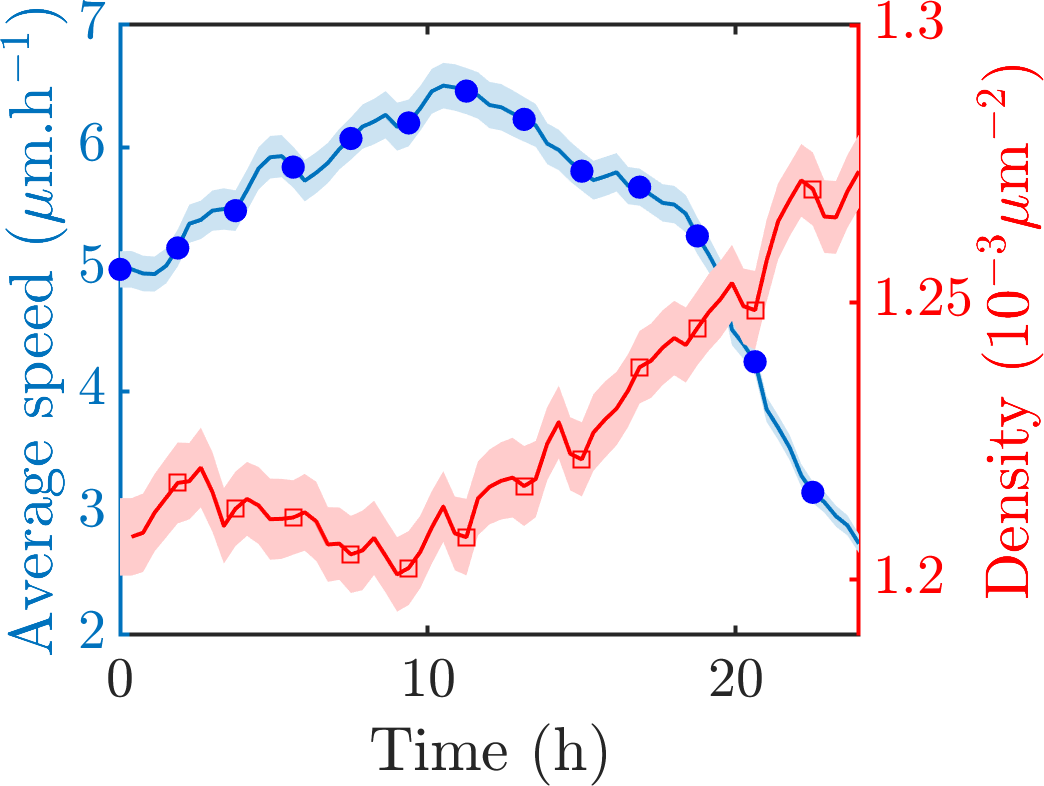}
        \caption{}
        \label{fig:density}
    \end{subfigure}
    \begin{subfigure}[t]{0.45\textwidth}
        \centering
        \includegraphics[width=\linewidth]{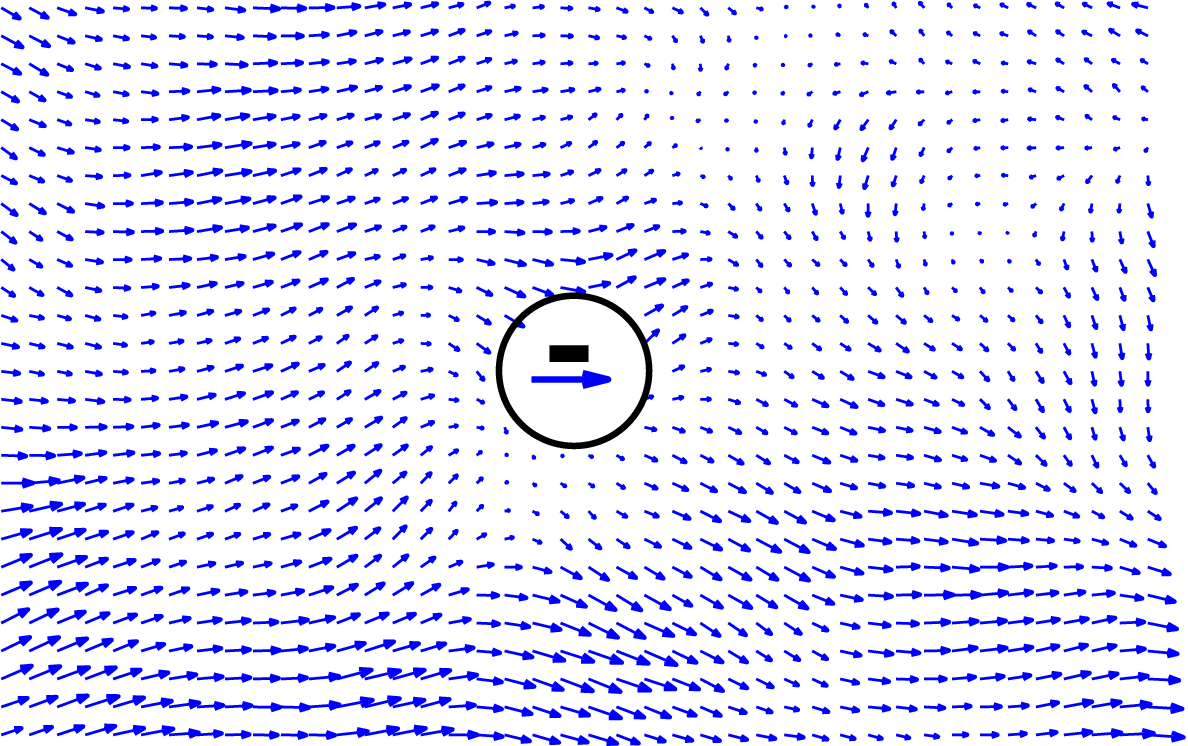}
        \caption{}
        \label{fig:hippo_obst_quiver}
    \end{subfigure}
    \hfill
    \begin{subfigure}[t]{0.51\textwidth}
        \centering
        \includegraphics[width=\linewidth]{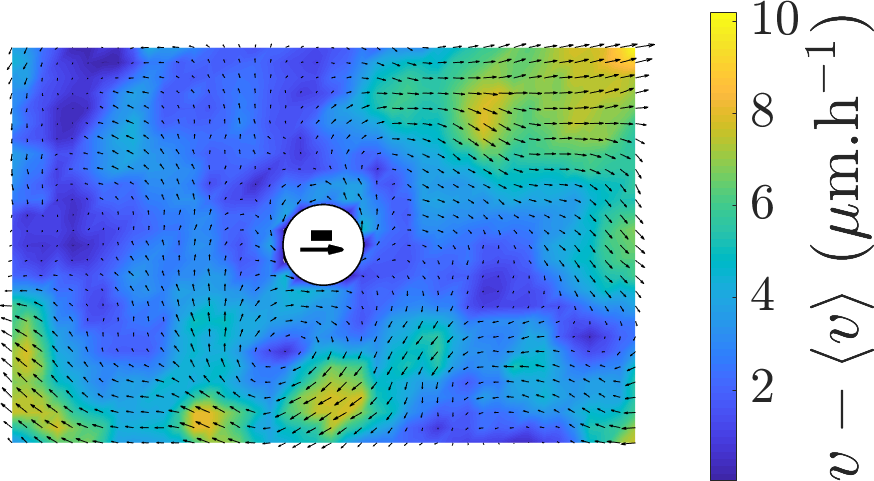}
        \caption{}
        \label{fig:hippo_obst_ref_fluid}
    \end{subfigure}
    \caption{\label{fig:presentation} {\bf Velocity in a hard racetrack.}
    {\bf (a)} Velocity field averaged on 100 frames 
    \modif{$\sim$8~h}
    (much larger than fluctuation correlation time, see Fig.~\ref{fig:time_corr} below).
    {\bf (b)} 
   \vire{Average density and speed  vs  time}\modif{Time evolution of a  typical experiment. Speed (closed blue circles, left axis) and density (open red squares, right axis); averages ($\pm$~s.e.m.) in space over $\sim$1500 boxes.}
    {\bf (c)} Velocity field around an obstacle, zoom from (a). 
    {\bf (d)} Velocity field around the obstacle, same as (c) but in the virtual referential of the cell monolayer.
    \modif{Scale bars in (c,d): 50~$\mu$m.} Scale arrows in (c,d): 30~$\mu$m~h$^{-1} = 0.5$~$\mu$m~min$^{-1}$.
    }
\end{figure}

\subsection{Forces}
In the absence of significant inertia, mechano-chemical coupling relies on the equilibrium of mechanical forces. 
Fig.~\ref{fig: wholeracetrack}a-c presents the \modif{traction} force distribution measured on the racetrack with a PDMS substrate soft enough  (rigidity modulus: 3 kPa) that\vire{ cell} traction forces can deform it, and with $W=600$~$\mu$m.
As already  observed by Kim \textit{et al.}~\cite{kim2013},
\vire{significant forces are exerted at the confining  boundaries, perpendicularly to them and  directed inward into the tissues.}
\modif{significant forces are exerted on the cells by the confining  boundaries, perpendicularly to them and  directed outward (i.e. cells pull inward into the tissue).}
Fig.~\ref{fig: force_profile}d,e quantitatively evidences the force peak at boundaries.
 It is consistent with the vimentin staining peak at the  same boundaries   (Fig.~\ref{fig:vim_hippo}f) 
and with the  role of vimentin in force generation~\cite{wu2022,nunesvicente2022}.

\subsection{Spontaneous velocity field}

Cells are seeded at low density, and once they reach confluence spontaneous migration movements are observed  (Fig.~\ref{fig:presentation}a).
As expected~\cite{green2020}, we do not detect any simple correlation  between forces and velocities (data not shown). In two studies using annuli or racetracks up to millimeter-sized diameter, with MDCK cells~\cite{jain2020} and with the more coordinated HBEC cells~\cite{giuglaris2024}, the periodic boundary conditions and  the non-trivial topology based on  the hole in the center of the racetrack  enabled a coherent cell circulation to set in.  

Here cells slowly reorganize, and as time progresses, the spatially averaged velocity modulus increases during the first few hours, reaches a plateau at longer timescales, and eventually decreases (Fig.~\ref{fig:presentation}b), inversely correlated with spatially averaged density. 
A similar plateau has been observed in unconfined HBEC cells and interpreted as a jamming transition~\cite{garcia2015}. 
Here, the transition to slower movement is delayed thanks to mitomycin treatment.
We do not detect any significant effect of confining boundary curvature, whether concave, flat or convex. 


Unlike in the strip,  in the racetrack  the velocity field fluctuates around a zero  time average and a zero space average (and we do not adimension the velocity field). 
Locally, due to symmetry breaking, self-organized collective cell movements appear.
Within a straight portion of the racetrack, directed migration  can be observed
over short time and length scales, and the inserted obstacle locally probes this migration (Fig.~\ref{fig:presentation}c). 
The obstacle creates a hole, i.e. locally a  non-trivial topology, and a circulation can set in (Fig.~\ref{fig:presentation}d).

\begin{figure}[t!]
	\centering
		\includegraphics[width=0.7\textwidth]{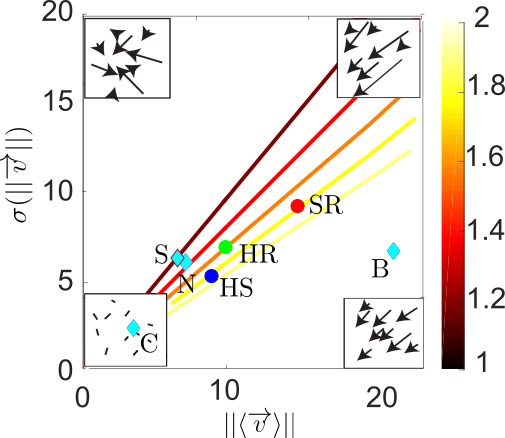}
    \caption{{\bf \modif{Velocity coordination.}} 
    \modif{Phase diagram of observed velocity fields. 
    Horizontal axis: modulus of average velocity vector ($\mu$m.h$^{-1}$). 
    Vertical axis: standard deviation of the velocity modulus ($\mu$m.h$^{-1}$). 
    Four extreme archetypes are schematized in the corners. 
    To guide the eye, the coordination parameter (ratio of vertical to horizontal axis) is indicated by color-coded straight lines. 
    Measurements without drug (circles): hard strips (HS), hard racetracks (HR), soft racetracks (SR).  
    Measurements on racetracks are performed on a rectangle of vertical width $W$, horizontal length 1.5~$W$, centered around a $V$-shaped obstacle. 
    Measurements with drug on hard strips (lozenges): Simvasatin (S), Nocodazole (N), CK666 (C), Blebbistatin (B).
    Averages are performed over time (100~min, i.e. 21 frames); then each average and standard deviation is calculated over several measurements performed on the same day in the same conditions.
    HS, $N=16$ ($W=1000$~$\mu$m); HR, $N=14$, with chiral $V$-shaped obstacle design (8 with $W=1000$~$\mu$m and 6 with $W=600$~$\mu$m); SR, $N=4$  (2 with $W=1000$~$\mu$m and 2 with $W=600$~$\mu$m); drugs, $N=4$.
    }
    }
\label{fig:phase_diag_order}
\end{figure}

\subsection{\modif{Velocity coordination in different conditions}}
{\bf \large \vire{3.3 Cells migrate on soft racetrack faster than on hard one \\ }}

\vire{It is difficult to compare absolute velocities on strips vs racetracks because density values and initial conditions differ too much.}

\vire{Comparing migrations on soft and hard racetracks, we observe they are qualitatively similar but significantly differ quantitatively (Student test $p< 0.01$): on soft substrate, there is a nearly 50\% increase  both in the average of velocity modulus (17.7$\pm$1.1 vs 12.3$\pm$1.6), and in the modulus of the velocity average (13.1$\pm$4.0 vs 8.8$\pm$2.0). The standard deviation of the velocity modulus is also 25\% higher  on soft substrate (8.5$\pm$0.6 vs 6.7$\pm$0.9).}
\modif{Migration on soft racetracks is qualitatively similar but significantly higher  (Student test $p< 0.01$) than on hard substrate: there is a 50\% increase  in the modulus of the velocity average and 30\% increase in the  standard deviation of the velocity modulus, see Fig.~\ref{fig:phase_diag_order}.}

This means that the individual cell migration velocity is higher on soft substrate, as opposed to what is usually observed with a free front (in fact, a velocity higher on soft substrate than on hard one had been reported once~\cite{balcioglu2020}, but is was at 60 kPa, more than a decade above our rigidity modulus).
It also means that the velocity fluctuates more; and also that  the inter-cell coordination is higher, maybe thanks to cell-cell interactions mediated by the substrate elastic deformation~\cite{bischofs2004}. 

%
%
%
%
%

\modif{Fig.~\ref{fig:phase_diag_order} shows velocity and coordination are similar on strips vs racetracks. It is difficult to interpret these data because density values and initial conditions differ too much.}

\modif{Fig.~\ref{fig:phase_diag_order} also shows a reduction in average velocity on strip when adding three drugs (see Supp. Movies). 
CK666, a selective inhibitor of the Arp2/3 complex, nucleates branched actin filament networks and thus significantly impacts formation and remodeling of branched actin arrays such as those in the lamellipodia~\cite{latorre2018,chen2019}. The two other drugs  alter  the monolayer mechanical resistance~\cite{chang2008,trogden2018}: Nocodazole specifically binds to $\beta-$tubulin and reversibly inhibits microtubule polymerization causing rapid microtubule depolymerization;
Simvastatin inhibits HMG-CoA reductase and thereby blocks the prenylation of small GTPases that regulate actin filament organization and intermediate filament dynamics.}

\modif{Conversely, adding Blebbistatin, which affects myosin activity and cellular activity, increases the average velocity and its coordination. This is surprising and we speculate that the increase in coordination might be due to a decrease in cell activity fluctuations. }

\begin{figure}[t!]
	   \centering
    \begin{subfigure}[c]{0.44\textwidth}
        \centering
        \includegraphics[width=\linewidth]{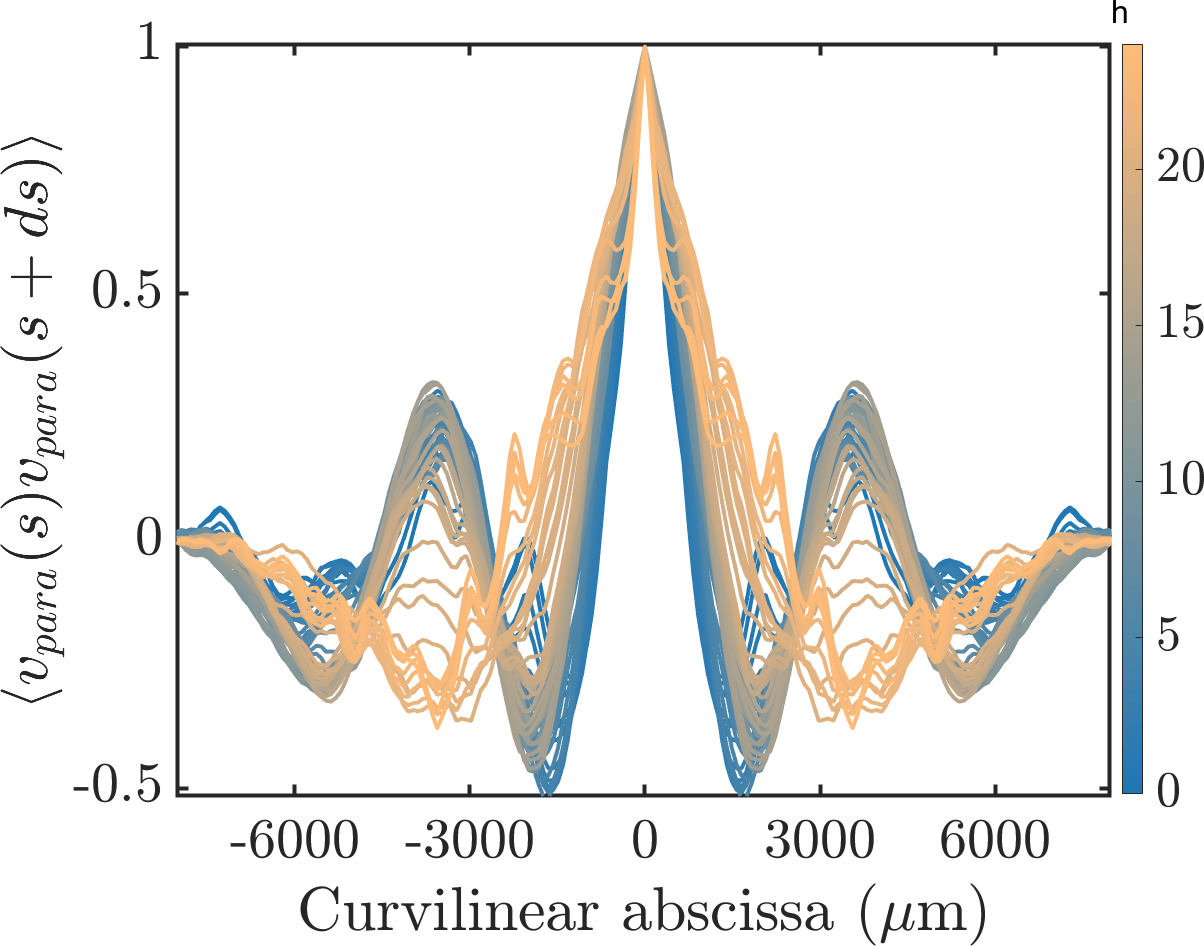}
        \caption{}
        \label{fig:wave_corr_para}
    \end{subfigure}
    \hfill
     \begin{subfigure}[c]{0.54\textwidth}
        \centering
        \includegraphics[width=\linewidth]{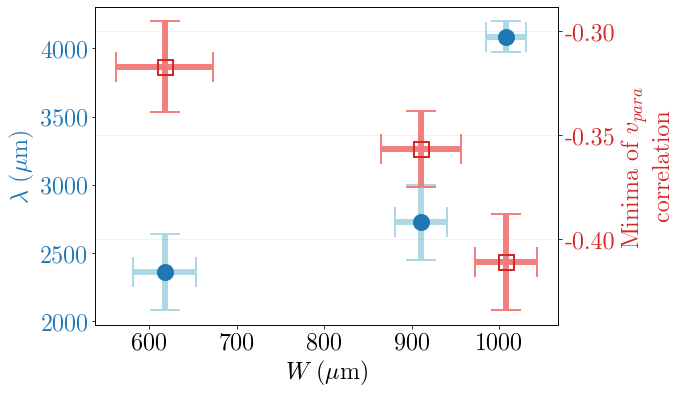}
        \caption{}
        \label{fig:wave_mini_corr}
    \end{subfigure}

	\caption{{\bf Spatial velocity correlations in hard racetrack.} 
	{\bf (a)} Autocorrelation function of a given experiment, along the racetrack midline (curvilinear abscissa $s$), of the  velocity component locally parallel to this midline, $v_{para}(s)$. Color code: time (h). Here $W=1000$~$\mu$m, midline perimeter $L=8000$~$\mu$m, symmetrized $s \to -s$. 
	{\bf (b)} Statistics of typical domain length $\lambda$ (\vire{red}\modif{blue closed} circles, left scale),  and value of the first peak minimum (\vire{blue}\modif{red open} squares, right scale), vs racetrack width $W$. Bars \vire{=}\modif{are} standard errors of the mean (s.e.m.).
 	}
	\label{fig:space_corr}
\end{figure}

\subsection{Velocity domains and waves}

Jain \textit{et al.} have performed experiments 
of MDCK migration in annuli, mainly of 200~$\mu$m in  outer diameter (at most 1000~$\mu$m) and 20~$\mu$m in width (at most  200~$\mu$m)~\cite{jain2020}. They aimed to limit cell migration along the ring to one-dimensional track, in order to minimizes lateral intercellular interactions. They have observed a   cell flow coordinated  in the same direction over the entire tracks. 
In Giuglaris' PhD, similar observations have been made with more coordinated HBEC cells on racetracks up to 1800~$\mu$m in length, $W=150$~$\mu$m in width~\cite{giuglaris2024}.

Conversely, to observe the coexistence of several correlated domains, two studies have used large enough confined patterns~\cite{petrolli2019,peyret2019}.
They have both evidenced velocity waves, a feature which is generic in active cell migration~\cite{yabunaka2017,blanch-mercader2017,tlili2020,boocockTheoryMechanochemicalPatterning2021,boocock2023}.
In one-dimensional strips up to length $L=2000$~$\mu$m and width $W=40$~$\mu$m,
Petrolli \textit{et al.} observed on MDCK cells  that $v_x$ showed a multimodal standing wave with a wavelength and pulsation which both increase with $L$ up to  $L>500$~$\mu$m, then plateau~\cite{petrolli2019}; this led them to the hypothesis  that there exist\modif{s} an unique tissue-intrinsic pattern length, $\sim$380~$\mu$m, and time, $\sim$270~min. 
On two-dimensional  rectangles up to $L=3500$~$\mu$m and  $W=1000$~$\mu$m, 
Peyret \textit{et al.}, focusing on human keratinocytes (HaCaT) and enterocytes (Caco2), observed that $v_y$ presented  standing wave patterns with a pulsation and amplitude that were dependent on  $W$~\cite{peyret2019}. They also observed that $v_x$ showed a mix of propagating waves and multimodal standing waves, with a characteristic wavelength dependent on  $W$.

We too observe domains of velocities oriented in the same direction, which alternate along the racetrack midline, schematized on Fig.~\ref{Fig_setup_hippo}a.
This is quantified by the autocorrelation function of  the velocity component parallel to the   midline.
The position  $\lambda$ of its first minimum is a robust, model-free characterisation of the typical domain length scale, especially since the minimum is pronounced.
In a rare example, we observed \vire{in}\modif{a} time evolution of $\lambda$  (Fig.~\ref{fig:space_corr}a).
The velocity component perpendicular to the  midline displays a similar behaviour (Fig.~\ref{fig:wave_corr_perp}).
\vire{Performing the statistics only on experiments stable in time, when increasing $W$ from $\sim$900~$\mu$m to
 $\sim$1500~$\mu$m  (see distribution of $W$ in Fig.~S1),
 we observe the}\modif{We perform the statistics  on 66 measurements stable in time (Fig.~\ref{fig:distrib_width}). When increasing $W$ from $\sim$600~$\mu$m to
 900$\sim$1000~$\mu$m, the}
coherence (marked by minus the minimum value of the auto-correlation) gets more and more pronounced from $\sim$0.3 to  $\sim$0.4 (Fig.~\ref{fig:space_corr}b).
This increase in coherence is reminiscent of the increase of wave amplitude with $W$ in Peyret  \textit{et al.}'s study~\cite{peyret2019}.
 We also observe  that the domain size (marked by $\lambda$) increases from $\sim$1500~$\mu$m to
 $\sim$3000~$\mu$m, suggesting  that there is no unique tissue-intrinsic patterning length, and that the smallest confining length (here the width $W$) plays a determinant role.
The characteristic time of domains is so large ($>10-20$~h) that we find \modif{it} difficult to  quantify it, except for the rare cases like in Fig.~\ref{fig:space_corr}a.
It is much larger than the period found on thinner systems by Peyret  \textit{et al.}, which is consistent with their finding that the period increases with $W$~\cite{peyret2019}.

\modif{Symmetry breakings are limited in time and space: our experiments have large enough time-scales and length-scales to be overall symmetric in average. Obstacles do not have significant long-range effect on symmetry breaking, even if they are deliberately created with a chiral $V$-shaped obstacle design (see Supp. Movie 3).}

\subsection{\modif{Swirls, kinetic energy and enstrophy}}

\modif{Lin \textit{et al.} measured the distribution of kinetic energy and enstrophy during 10~h while a monolayer was jamming~\cite{lin2021}. They showed that both kinetic energy and enstrophy decreased with time and  correlated.}

\begin{figure}[t!]
    \centering
    \begin{subfigure}[t]{0.32\textwidth}
        \centering
        \includegraphics[width=\linewidth]{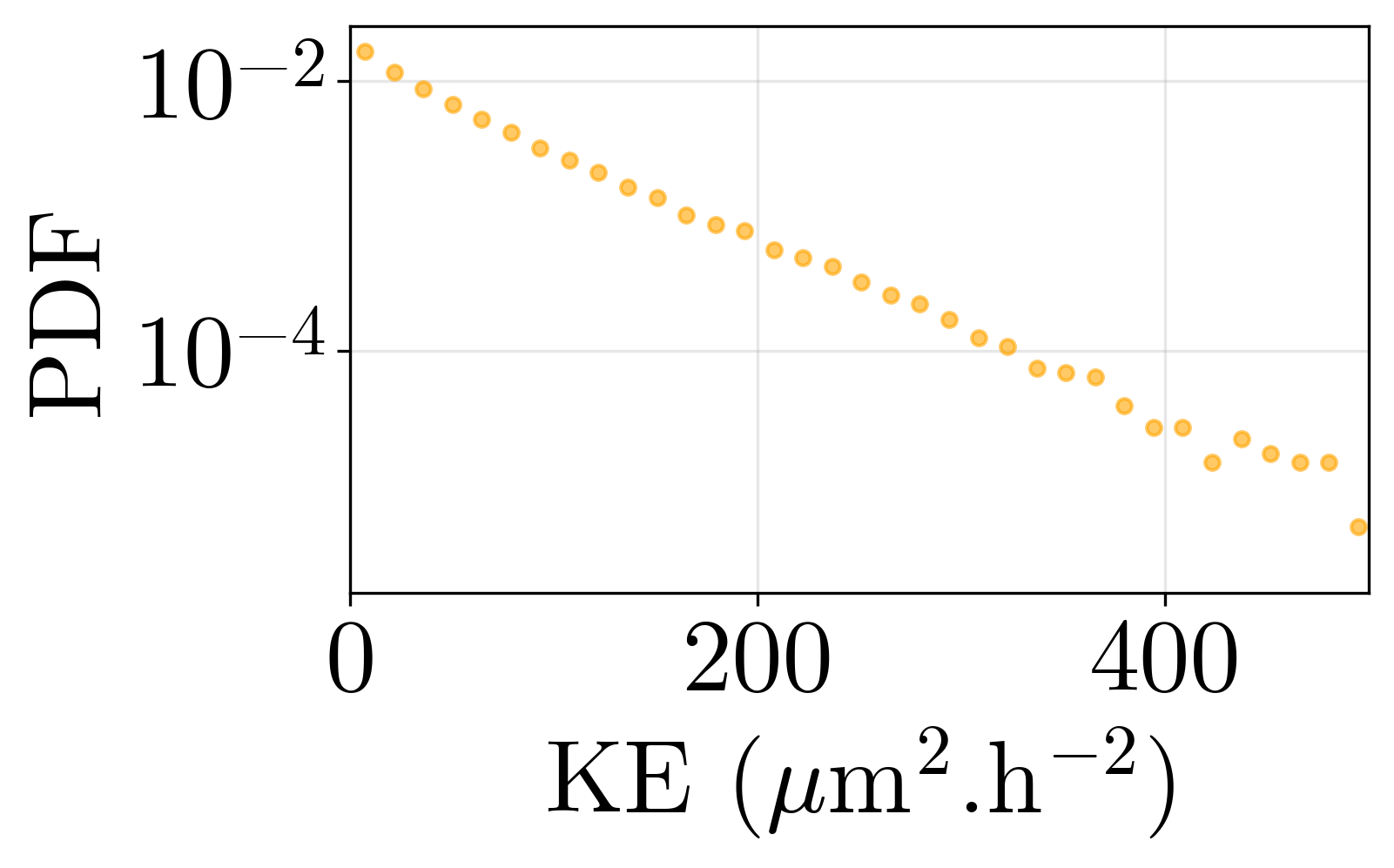}
        \caption{}
        \label{fig:pdf_nrj_comb}
    \end{subfigure}
        \hfill
    \begin{subfigure}[t]{0.32\textwidth}
        \centering
        \includegraphics[width=\linewidth]{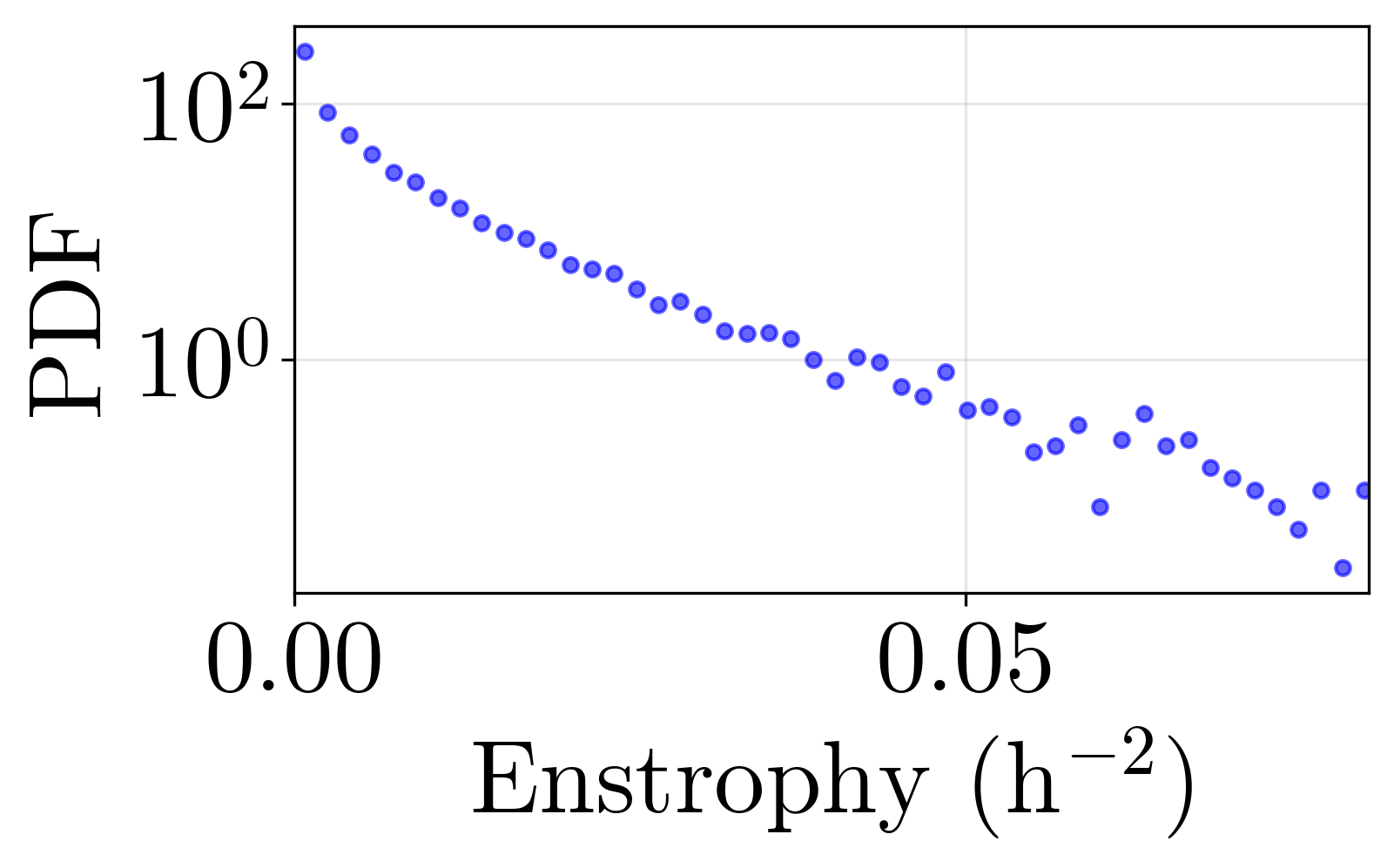}
        \caption{}
        \label{fig:pdf_ens_comb}
    \end{subfigure}
        \hfill
    \begin{subfigure}[t]{0.34\textwidth}
        \centering
        \includegraphics[width=\linewidth]{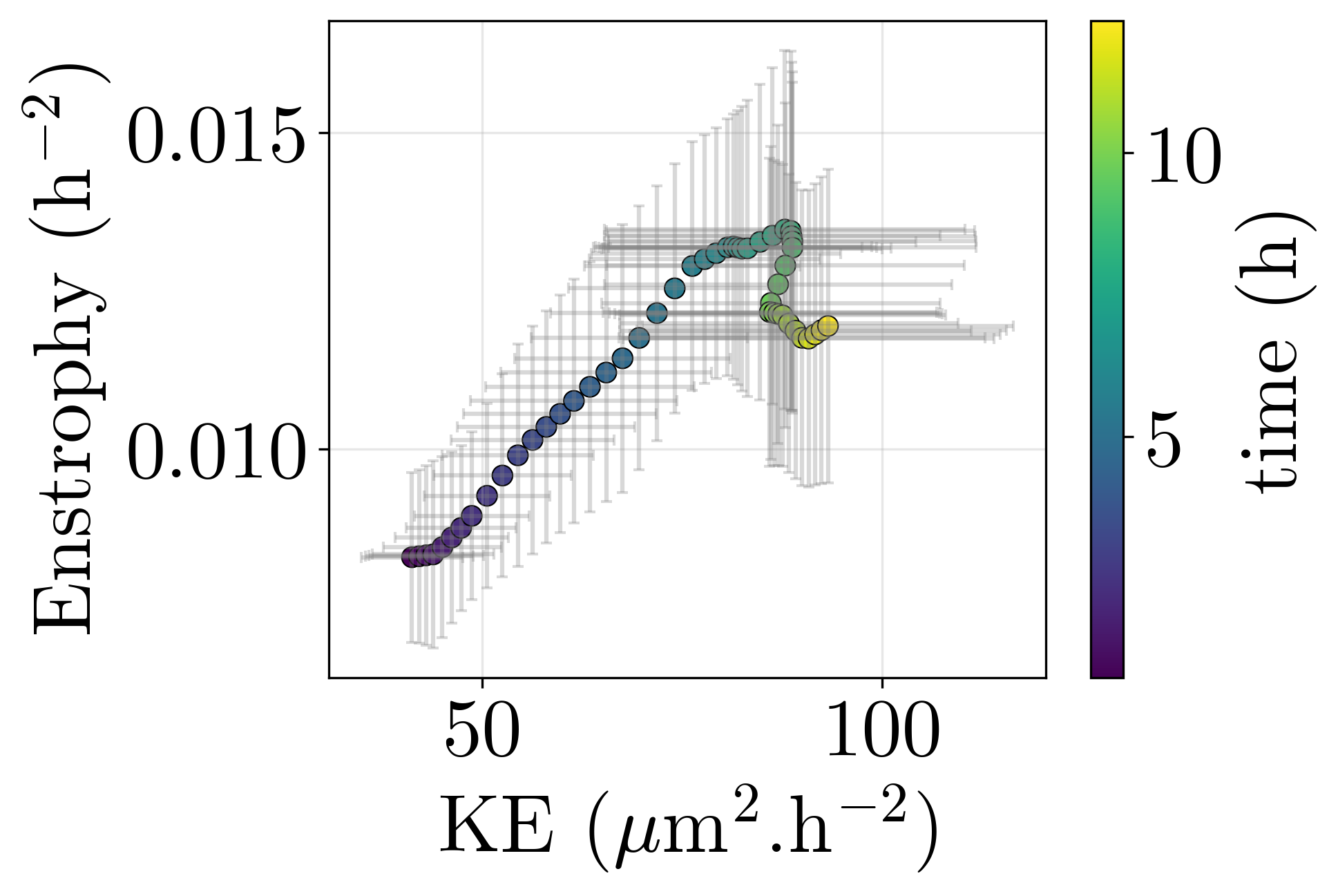}
        \caption{}
        \label{fig:ens_vs_nrj_comb}
    \end{subfigure}
    \caption{\label{fig:energy_enstrophy} 
    \modif{
    {\bf Kinetic energy and enstrophy.}
    Statistics averaged  in space over 64 pixels, in time over 50 frames, and over 8 measurements performed on racetracks with circular obstacles and different sizes (Fig.~\ref{fig:energy_enstrophy_supp}a-i).
     {\bf (a)} PDF of kinetic energy.
    {\bf (b)} PDF of enstrophy.
    {\bf (c)} Enstrophy versus kinetic energy for successive times (color-coded). }}
 
\end{figure}

\modif{Since we also observe swirls in the racetrack geometry (see Fig.~\ref{fig:presentation}a), we perform similar measurements over 10~h, in our case with a larger track size and without jamming. 
We performe 8 measurements on racetracks with different widths and circular obstacle sizes. They display the same behavior, suggesting that our results do not depend on set-up dimensions (Fig.~\ref{fig:energy_enstrophy_supp}a-i). We thus pool the results from all 8 measurements (Fig.~\ref{fig:energy_enstrophy}).
We observe a similar behaviour with the presence of chiral $V$-shaped obstacle design (Fig.~\ref{fig:energy_enstrophy_supp}j-l).}

\modif{Figs.~\ref{fig:energy_enstrophy},~\ref{fig:energy_enstrophy_supp} use the same representation as that of Lin \textit{et al.}~\cite{lin2021}. 
Contrary to their findings, in our case the kinetic energy and enstrophy follow a Boltzmann distribution (Fig.~\ref{fig:energy_enstrophy}a,b).
During the first 5 hours, while long-distance correlation grows, both kinetic energy and  enstrophy increase; during the next 5 hours, kinetic energy still increases while enstrophy passes through a maximum and decreases (Fig.~\ref{fig:energy_enstrophy}c). }

\modif{In the future, it would be interesting to measure the Okubo-Weiss field and compare it with its measurements by  Giomi~\cite{giomi2015}, and to measure the radial profile of velocity in the swirls, possibly pointing to an interpretation in terms of phonon mode superposition.
}

\section{Correlations}
\label{sec:correlations}

Taking advantage of the  large length scales and  long timescales  of our experiments, we now investigate
 the velocity spatial and temporal correlation functions~\cite{Cavagna2018,vicsek2012}.

\subsection{Spatial correlations}

For spatial correlations, the analysis is performed only on the hard racetrack, where it is much easier than on the strip thanks to its large size and to the absence of a steady spatial gradient, and where we \vire{have realiszed}\modif{realize} experiments with drugs.
We use  the spatial correlation function  defined as follows: 
\begin{equation}
    \label{eq:spatial_correlation}
    C_s(r_i,t) = 
    \frac{\left\langle  [\vec{V}(\vec{X},t)-\vec{V_0}] \cdot [\vec{V}(\vec{X}+\vec{r},t)-\vec{V_0}]\right\rangle_{\vec{X}}}{\left\langle [\vec{V}(\vec{X},t)-\vec{V_0}]^2 \right\rangle_{\vec{X}}}        
    \end{equation}
Here $r_i$ is the distance between two points, $ \vert \vec{r} \vert$ is in the interval $ \left[ r_i,r_{i+1}\right]$,
$\vec{V}(\vec{X},t)$ is the velocity field at the position $\vec{X}$ and time $t$. Finally, $\vec{V_0}$ is the average of the velocity field, which  should be chosen with care in active systems without stationnarity~\cite{Cavagna2018}. We do not choose  $\vec{V_0}$ to be the spatial  average  in a given window size $\mathrm{d}A$ around each point, because this would yield measurements depending on $\mathrm{d}A$ without any plateau (\modif{Figs.~\ref{fig: dA_effect_classic}, \ref{fig:effect_of_dA}}, Section~\ref{sec:measure_correlations}). We do choose $\vec{V_0}$ to be the time  average at position $\vec{X}$\vire{ and find the correlation function is well fitted using a power law $  C_s(r) \sim r^{\alpha}$ with an exponent $\alpha<0$ (Fig.~7c)}.

\begin{figure}[t!]
    \centering
    \begin{subfigure}{0.49\textwidth}
        \centering
        \includegraphics[width=\linewidth]{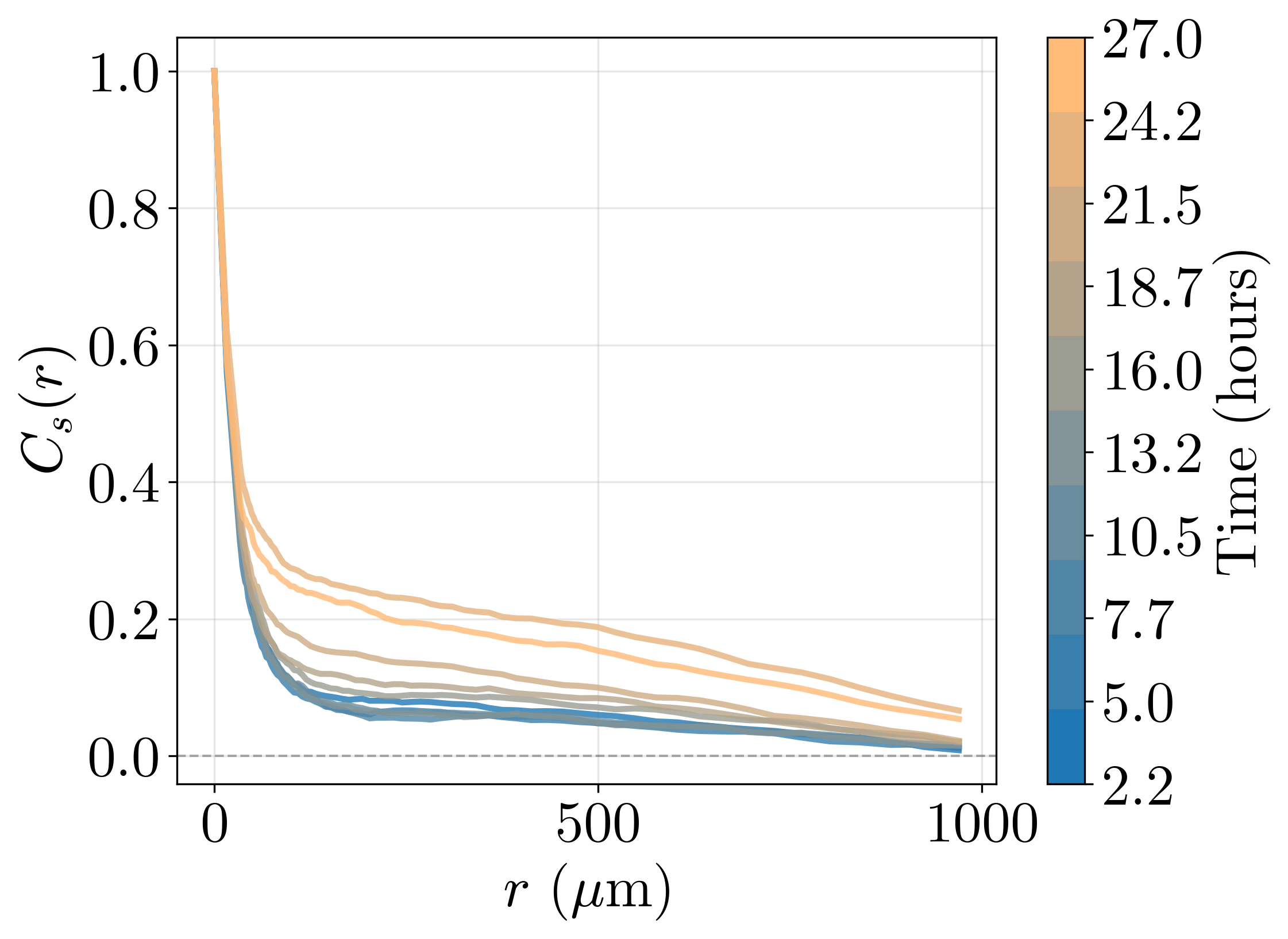}
        \caption{}
        \label{fig:corr_spat_lin}
    \end{subfigure}
    \hfill
     \begin{subfigure}{0.49\textwidth}
        \centering
        \includegraphics[width=\linewidth]{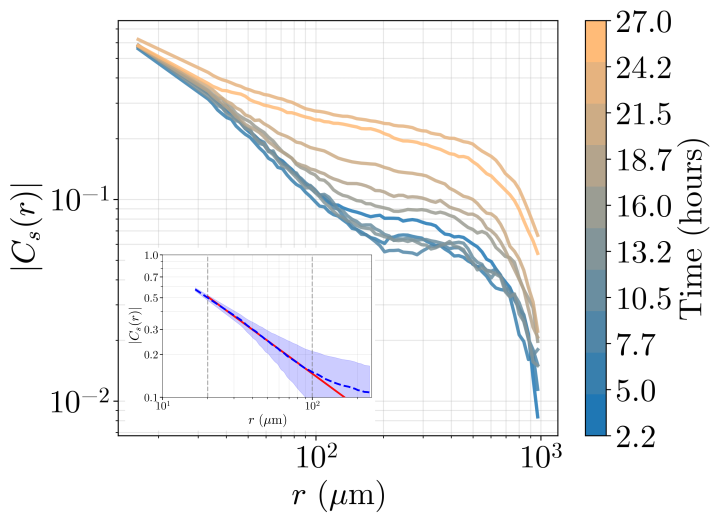}
        \caption{}
        \label{fig:corr_spat_log}
    \end{subfigure}
    \begin{subfigure}[t]{0.49\textwidth}
        \centering
        \includegraphics[width=\linewidth]{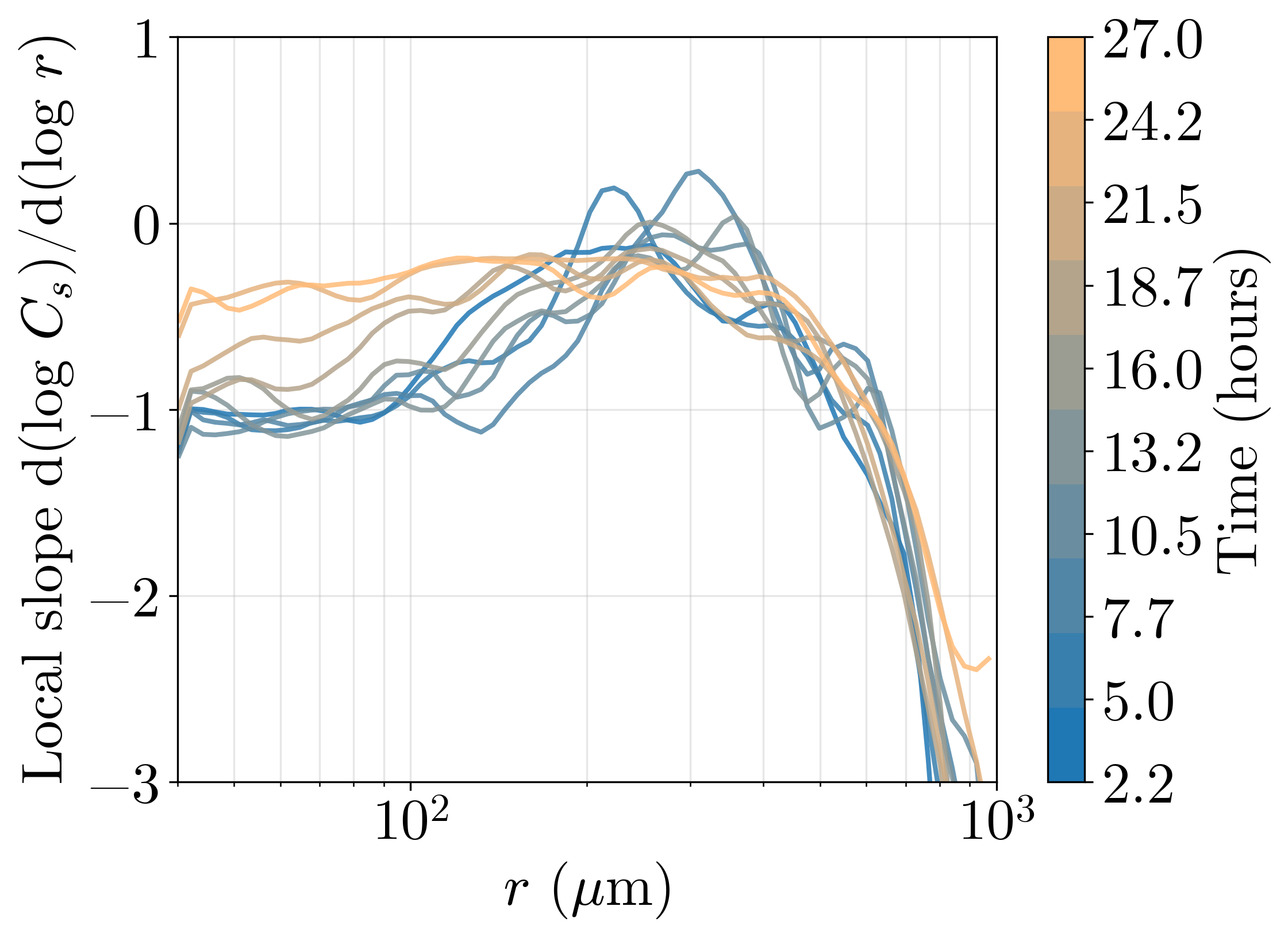}
        \caption{}
        \label{fig:corr_spat_slopes}
    \end{subfigure}
    \hfill
    \begin{subfigure}[t]{0.49\textwidth}
        \centering
        \includegraphics[width=\linewidth]{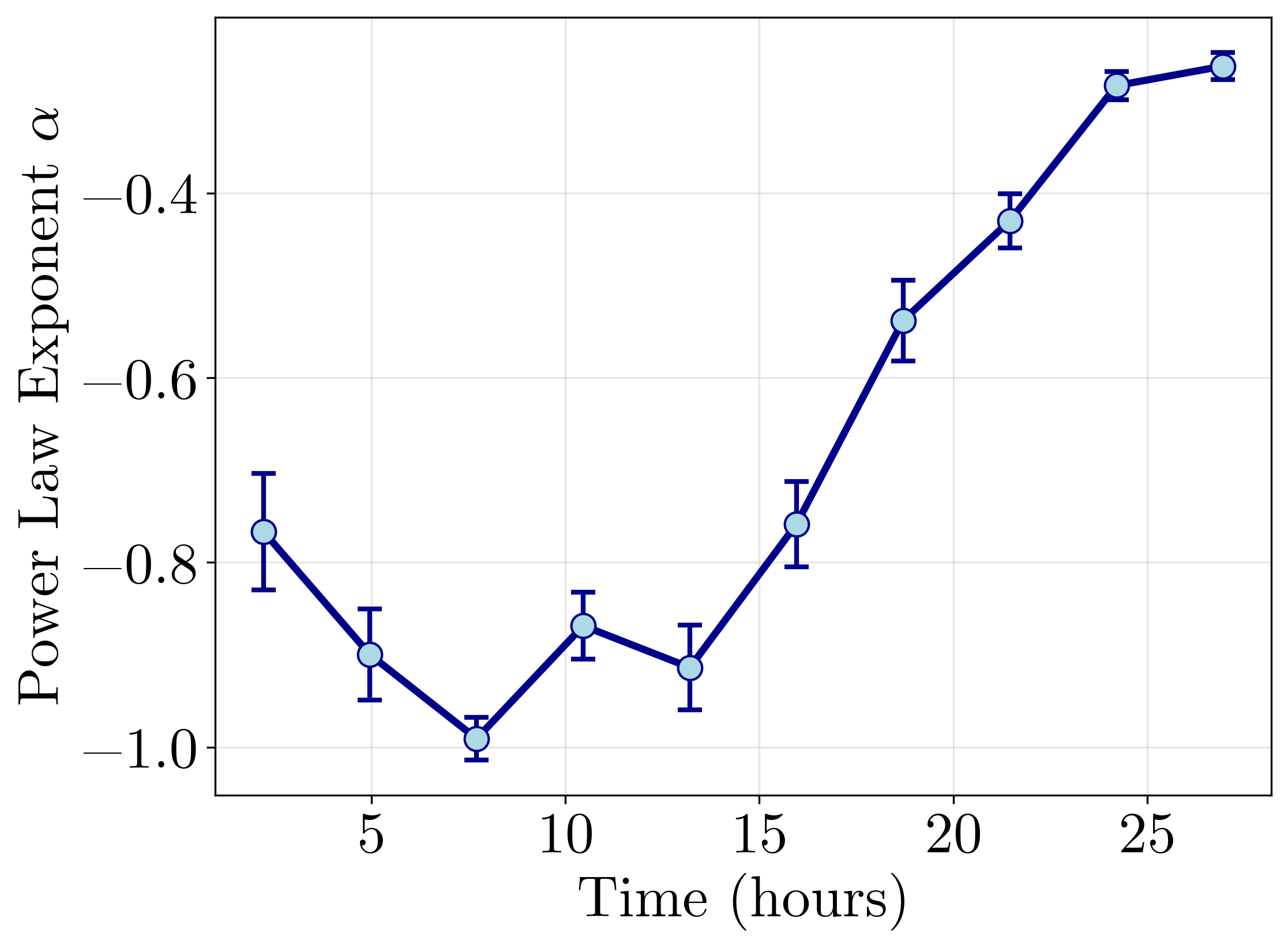}
        \caption{}
        \label{fig:corr_spat_alpha}
    \end{subfigure}
    \caption{\label{fig:temporal_evo} 
    \modif{
    {\bf Temporal evolution of the velocity correlation function.}  Racetrack width $W=600$~$\mu$m.
    {\bf (a)} Spatial correlation function of the velocity versus distance $r$, for successive times color-coded from 2.2 to 27~h.
    {\bf (b)} Same as (a) in log-log scale. Inset: spatial correlation function of the velocity averaged over all times (thick blue dashes) with its std (blue hatches); thick solid red line: linear fit to the log-log values, between $r=20$ and 100~$\mu$m (thin vertical blue dashes).
    {\bf (c)} Local exponent defined as the slope of (b), i.e. logarithmic derivative of (a).
    {\bf (d)} Time evolution of exponent $\alpha$, defined as the slope in (c),
    average ($\pm$~s.e.m.) between $r=30$ and 200~$\mu$m.
    }}
\end{figure}

\modif{Under these conditions, we find that the correlation function  is a power law with a local exponent  constant at small  length scale (below 200~$\mu$m), which drops at larger length scales
(Fig.~\ref{fig:temporal_evo}). The exponent $\alpha$ defined as the average exponent between 30 and 200~$\mu$m varies significantly with time, indicating that the monolayer matures. It would be interesting in the future to understand this time-dependence of $\alpha$, maybe due to the maturation of inter-cellular contacts. The experiment size and the addition of drugs affecting the intra-cellular mechanics have much less effect that this time-dependent maturation (Figs.~\ref{fig:temporal_evo_supp}, \ref{fig:drugs_slopes}). Whatever the drug, the measurements performed after drug addition (see Supp. Movies) are not significantly different from the measurements performed just before drug addition, which themselves differ significantly from measurements performed at earlier times.
}

\vire{Each experiment is performed before and after the injection of a drug. The correlation function retains the same shape after the injection, and can still be fitted by a power law. Exponents $\alpha$ are measured separately before and after the injection. The experiment is repeated 3 to 4 times for each drug (Fig.~7d). 
Before injection, the exponent is around $-0.5$. The control (i.e. injection in itself, without any drug) shows a slight increase in $\alpha$. CK666, a drug which only prevents actin polymerization in the lamellipodia, and which does not perturb the overall cell architecture, induces a  slight decrease in $\alpha$.  Strikingly, adding Nocodazole or  Simvastatin, which affect the cytoskeleton and hence, induces  a strong decrease in $\alpha$, which ranges between $-0.8$ and $-1$.}

\begin{figure}[t!]
    \centering
    \begin{subfigure}{0.49\textwidth}
        \centering
        \includegraphics[width=\linewidth]{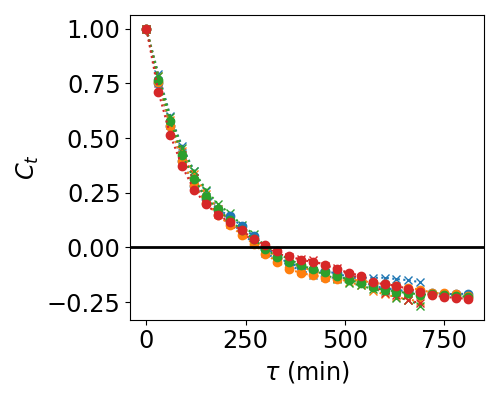}
        \caption{}
        \label{fig:corr_temp_lin}
    \end{subfigure}
    \hfill
     \begin{subfigure}{0.49\textwidth}
        \centering
        \includegraphics[width=\linewidth]{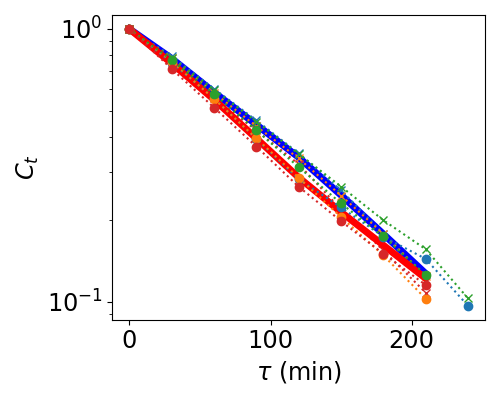}
        \caption{}
        \label{fig:corr_temp_log}
    \end{subfigure}
	\caption{{\bf Temporal velocity correlation.}   \modif{{\bf (a)}} Correlation of the velocity field fluctuations around the temporal average for 4 strip \vire{experiments}\modif{measurements} (crosses) and 4 hard racetrack \vire{experiments}\modif{measurements} (circles).  \vire{Inset:  s}\modif{{\bf (b)} S}ame data in semi-logarithmic scale\vire{, with an exponential fit}.
    Solid lines: blue,  average over the 4 strips; red, same for the 4 racetracks. }
\label{fig:time_corr}
\end{figure}

\subsection{Temporal correlations}

As we record cell velocity fields for several hours in  strips and racetracks, we study time correlation on both (for simplicity, we have not analyzed the soft racetrack). 
To define the velocity field on a regular spatial grid, 
we average it over a spatial window of size $w_s = 32.6$~$\mu$m,
and then over a small time window of size $w_t = 6$ frames (30~min).
The temporal correlation function is then computed for different time lags $\tau$ (Fig.~\ref{fig:time_corr}\modif{a}).
In a semi-logarithmic scale, the correlation function is then fitted by an exponential decay  (\vire{inset of }Fig.~\ref{fig:time_corr}\modif{b}).
Strikingly, we observe that for all  experiments, whether strip or racetrack, the decay time is the same, $\sim$100~min.

\section{Conclusion and perspectives}

We \vire{have made experiments on 16}\modif{perform 16 measurements using} open strips with an obstacle, on hard substrates. We compare them with \vire{47 experiments on}\modif{66 measurements with 48} closed racetracks of different sizes, also with obstacles, on hard substrates too\vire{ (}\modif{,} and \vire{1}\modif{4 using} racetrack\modif{s} on soft substrates\vire{)}. This enables us to investigate and disentangle how the size, shape, topology and rigidity of patterned substrate affect the collective migration of epithelial monolayers at several space and time scales. \modif{In the future, it might be interesting to perform open strip traction force experiments on a soft substrate with different approaches, including lithography or magnetically actuated barriers.}

\modif{The boundary conditions strongly affect the vimentin, which decays from the free boundary with a characteristic length $65\pm 4$~$\mu$m,  an order of magnitude greater than that near the confining boundary, $3.2 \pm 0.7$~$\mu$m.}
In the strip, the free front acts as a powerful external cue, imposing a global polarization and velocity gradient that dominates the tissue's intrinsic dynamics. 
In contrast, the periodic racetrack geometry in absence of any net global drift  isolates  complex intrinsic behaviors, which are stronger on soft than on hard substrate.
They are characterized by scale-free, power law fluctuating velocity domains,
rather than by a single correlation length. 
This result  suggests the monolayer behaves as a critical-like system where information is transmitted across its entire size, a feature consistent with models of active solids capable of long-range stress propagation. 
\vire{The power-law exponent  characterizes the tissue's mechanical state and is sensitive to cytoskeletal perturbations.}


Understanding the observed power-law correlations and predicting the exponent value  from sub-cellular properties\vire{is a}\modif{, as well as explaining the non-monotonous time evolution of enstrophy, are} theoretical challenge\modif{s}.
Integrating our force measurements with the velocity correlation analysis, as well as with future  high-resolution imaging of molecular effectors such as vimentin or actin, might be determinant toward building a complete mechanical model connecting the tissue-scale rheology to its sub-cellular origins.

\section*{Acknowledgments}

We warmly thank Sham Tlili for fruitful discussions and for her invaluable help to implement the experimental protocol. For the purpose of Open Access, a CC-BY 4.0 public copyright licence \url{<https://creativecommons. org/licenses/by/4.0>} has been applied by the authors to the present document and will be applied to all subsequent versions up to the Author Accepted Manuscript arising from this submission. 

\section*{Conflicts of interest}
The authors have no conflicts to disclose.


\section*{Author Contribution Statement}

Conceptualization: H. Delano\"e-Ayari, F. Graner; Methodology: G. Duprez, M. Durande, F. Graner, H. Delano\"e-Ayari; Formal analysis and investigation: G. Duprez, M. Durande, F. Graner, H. Delano\"e-Ayari; Writing - original draft preparation: F. Graner, H. Delano\"e-Ayari ; Resources: H. Delano\"e-Ayari, F. Graner; Supervision: H. Delano\"e-Ayari, F. Graner.


\begin{thebibliography}{62}
\ifx \bisbn   \undefined \def \bisbn  #1{ISBN #1}\fi
\ifx \binits  \undefined \def \binits#1{#1}\fi
\ifx \bauthor  \undefined \def \bauthor#1{#1}\fi
\ifx \batitle  \undefined \def \batitle#1{#1}\fi
\ifx \bjtitle  \undefined \def \bjtitle#1{#1}\fi
\ifx \bvolume  \undefined \def \bvolume#1{\textbf{#1}}\fi
\ifx \byear  \undefined \def \byear#1{#1}\fi
\ifx \bissue  \undefined \def \bissue#1{#1}\fi
\ifx \bfpage  \undefined \def \bfpage#1{#1}\fi
\ifx \blpage  \undefined \def \blpage #1{#1}\fi
\ifx \burl  \undefined \def \burl#1{\textsf{#1}}\fi
\ifx \doiurl  \undefined \def \doiurl#1{\url{https://doi.org/#1}}\fi
\ifx \betal  \undefined \def \betal{\textit{et al.}}\fi
\ifx \binstitute  \undefined \def \binstitute#1{#1}\fi
\ifx \binstitutionaled  \undefined \def \binstitutionaled#1{#1}\fi
\ifx \bctitle  \undefined \def \bctitle#1{#1}\fi
\ifx \beditor  \undefined \def \beditor#1{#1}\fi
\ifx \bpublisher  \undefined \def \bpublisher#1{#1}\fi
\ifx \bbtitle  \undefined \def \bbtitle#1{#1}\fi
\ifx \bedition  \undefined \def \bedition#1{#1}\fi
\ifx \bseriesno  \undefined \def \bseriesno#1{#1}\fi
\ifx \blocation  \undefined \def \blocation#1{#1}\fi
\ifx \bsertitle  \undefined \def \bsertitle#1{#1}\fi
\ifx \bsnm \undefined \def \bsnm#1{#1}\fi
\ifx \bsuffix \undefined \def \bsuffix#1{#1}\fi
\ifx \bparticle \undefined \def \bparticle#1{#1}\fi
\ifx \barticle \undefined \def \barticle#1{#1}\fi
\bibcommenthead
\ifx \bconfdate \undefined \def \bconfdate #1{#1}\fi
\ifx \botherref \undefined \def \botherref #1{#1}\fi
\ifx \url \undefined \def \url#1{\textsf{#1}}\fi
\ifx \bchapter \undefined \def \bchapter#1{#1}\fi
\ifx \bbook \undefined \def \bbook#1{#1}\fi
\ifx \bcomment \undefined \def \bcomment#1{#1}\fi
\ifx \oauthor \undefined \def \oauthor#1{#1}\fi
\ifx \citeauthoryear \undefined \def \citeauthoryear#1{#1}\fi
\ifx \endbibitem  \undefined \def \endbibitem {}\fi
\ifx \bconflocation  \undefined \def \bconflocation#1{#1}\fi
\ifx \arxivurl  \undefined \def \arxivurl#1{\textsf{#1}}\fi
\csname PreBibitemsHook\endcsname

\bibitem[\protect\citeauthoryear{Friedl and Gilmour}{2009}]{friedl2009a}
\begin{barticle}
\bauthor{\bsnm{Friedl}, \binits{P.}},
\bauthor{\bsnm{Gilmour}, \binits{D.}}:
\batitle{Collective cell migration in morphogenesis, regeneration and cancer}.
\bjtitle{Nature Reviews Molecular Cell Biology}
\bvolume{10}(\bissue{7}),
\bfpage{445}--\blpage{457}
(\byear{2009})
\doiurl{10.1038/nrm2720}
\end{barticle}
\endbibitem

\bibitem[\protect\citeauthoryear{R{\o}rth}{2009}]{rorth2009}
\begin{barticle}
\bauthor{\bsnm{R{\o}rth}, \binits{P.}}:
\batitle{{Collective Cell Migration}}.
\bjtitle{Annual Review of Cell and Developmental Biology}
\bvolume{25}(\bissue{Volume 25, 2009}),
\bfpage{407}--\blpage{429}
(\byear{2009})
\doiurl{10.1146/annurev.cellbio.042308.113231}
\end{barticle}
\endbibitem

\bibitem[\protect\citeauthoryear{Trepat et~al.}{2009}]{trepat2009}
\begin{barticle}
\bauthor{\bsnm{Trepat}, \binits{X.}},
\bauthor{\bsnm{Wasserman}, \binits{M.R.}},
\bauthor{\bsnm{Angelini}, \binits{T.E.}},
\bauthor{\bsnm{Millet}, \binits{E.}},
\bauthor{\bsnm{Weitz}, \binits{D.A.}},
\bauthor{\bsnm{Butler}, \binits{J.P.}},
\bauthor{\bsnm{Fredberg}, \binits{J.J.}}:
\batitle{Physical forces during collective cell migration}.
\bjtitle{Nature Physics}
\bvolume{5}(\bissue{6}),
\bfpage{426}--\blpage{430}
(\byear{2009})
\doiurl{10.1038/nphys1269}
\end{barticle}
\endbibitem

\bibitem[\protect\citeauthoryear{Banerjee and Marchetti}{2019}]{banerjee2019}
\begin{bchapter}
\bauthor{\bsnm{Banerjee}, \binits{S.}},
\bauthor{\bsnm{Marchetti}, \binits{M.C.}}:
\bctitle{Continuum models of collective cell migration}.
In: \beditor{\bsnm{La~Porta}, \binits{C.A.M.}},
\beditor{\bsnm{Zapperi}, \binits{S.}} (eds.)
\bbtitle{Cell Migrations: Causes and Functions},
pp. \bfpage{45}--\blpage{66}.
\bpublisher{Springer},
\blocation{Cham}
(\byear{2019}).
\doiurl{10.1007/978-3-030-17593-1\_4}
\end{bchapter}
\endbibitem

\bibitem[\protect\citeauthoryear{La~Porta and Zapperi}{2019}]{porta2019}
\begin{bchapter}
\bauthor{\bsnm{La~Porta}, \binits{C.A.M.}},
\bauthor{\bsnm{Zapperi}, \binits{S.}}:
\bctitle{Statistical features of collective cell migration}.
In: \beditor{\bsnm{La~Porta}, \binits{C.A.M.}},
\beditor{\bsnm{Zapperi}, \binits{S.}} (eds.)
\bbtitle{Cell Migrations: Causes and Functions},
pp. \bfpage{67}--\blpage{78}.
\bpublisher{Springer},
\blocation{Cham}
(\byear{2019}).
\doiurl{10.1007/978-3-030-17593-1\_5}
\end{bchapter}
\endbibitem

\bibitem[\protect\citeauthoryear{Poujade et~al.}{2007}]{poujade2007b}
\begin{botherref}
\oauthor{\bsnm{Poujade}, \binits{M.}},
\oauthor{\bsnm{{Grasland-Mongrain}}, \binits{E.}},
\oauthor{\bsnm{Hertzog}, \binits{A.}},
\oauthor{\bsnm{Jouanneau}, \binits{J.}},
\oauthor{\bsnm{Chavrier}, \binits{P.}},
\oauthor{\bsnm{Ladoux}, \binits{B.}},
\oauthor{\bsnm{Buguin}, \binits{A.}},
\oauthor{\bsnm{Silberzan}, \binits{P.}}:
Collective migration of an epithelial monolayer in response to a model wound.
Proceedings of the National Academy of Sciences of the United States of
  America,
15988--93
(2007)
\end{botherref}
\endbibitem

\bibitem[\protect\citeauthoryear{Vedula et~al.}{2012}]{vedula2012}
\begin{barticle}
\bauthor{\bsnm{Vedula}, \binits{S.R.K.}},
\bauthor{\bsnm{Leong}, \binits{M.C.}},
\bauthor{\bsnm{Lai}, \binits{T.L.}},
\bauthor{\bsnm{Hersen}, \binits{P.}},
\bauthor{\bsnm{Kabla}, \binits{A.J.}},
\bauthor{\bsnm{Lim}, \binits{C.T.}},
\bauthor{\bsnm{Ladoux}, \binits{B.}}:
\batitle{Emerging modes of collective cell migration induced by geometrical
  constraints.}
\bjtitle{Proceedings of the National Academy of Sciences of the United States
  of America}
\bvolume{109}(\bissue{32}),
\bfpage{12974}--\blpage{9}
(\byear{2012})
\doiurl{10.1073/pnas.1119313109}
\end{barticle}
\endbibitem

\bibitem[\protect\citeauthoryear{Jain et~al.}{2020}]{jain2020}
\begin{barticle}
\bauthor{\bsnm{Jain}, \binits{S.}},
\bauthor{\bsnm{Cachoux}, \binits{V.M.L.}},
\bauthor{\bsnm{Narayana}, \binits{G.H.N.S.}},
\bauthor{\bsnm{{de Beco}}, \binits{S.}},
\bauthor{\bsnm{D'Alessandro}, \binits{J.}},
\bauthor{\bsnm{Cellerin}, \binits{V.}},
\bauthor{\bsnm{Chen}, \binits{T.}},
\bauthor{\bsnm{Heuz{\'e}}, \binits{M.L.}},
\bauthor{\bsnm{Marcq}, \binits{P.}},
\bauthor{\bsnm{M{\`e}ge}, \binits{R.M.}},
\bauthor{\bsnm{Kabla}, \binits{A.J.}},
\bauthor{\bsnm{Lim}, \binits{C.T.}},
\bauthor{\bsnm{Ladoux}, \binits{B.}}:
\batitle{The role of single-cell mechanical behaviour and polarity in driving
  collective cell migration}.
\bjtitle{Nature Physics}
\bvolume{16}(\bissue{7}),
\bfpage{802}--\blpage{809}
(\byear{2020})
\doiurl{10.1038/s41567-020-0875-z}
\end{barticle}
\endbibitem

\bibitem[\protect\citeauthoryear{Lo~Vecchio et~al.}{2024}]{lovecchio2024}
\begin{barticle}
\bauthor{\bsnm{Lo~Vecchio}, \binits{S.}},
\bauthor{\bsnm{Pertz}, \binits{O.}},
\bauthor{\bsnm{Szopos}, \binits{M.}},
\bauthor{\bsnm{Navoret}, \binits{L.}},
\bauthor{\bsnm{Riveline}, \binits{D.}}:
\batitle{Spontaneous rotations in epithelia as an interplay between cell
  polarity and boundaries}.
\bjtitle{Nature Physics}
\bvolume{20}(\bissue{2}),
\bfpage{322}--\blpage{331}
(\byear{2024})
\doiurl{10.1038/s41567-023-02295-x}
\end{barticle}
\endbibitem

\bibitem[\protect\citeauthoryear{Doxzen et~al.}{2013}]{doxzen2013}
\begin{barticle}
\bauthor{\bsnm{Doxzen}, \binits{K.}},
\bauthor{\bsnm{Vedula}, \binits{S.R.K.}},
\bauthor{\bsnm{Leong}, \binits{M.C.}},
\bauthor{\bsnm{Hirata}, \binits{H.}},
\bauthor{\bsnm{Gov}, \binits{N.S.}},
\bauthor{\bsnm{Kabla}, \binits{A.J.}},
\bauthor{\bsnm{Ladoux}, \binits{B.}},
\bauthor{\bsnm{Lim}, \binits{C.T.}}:
\batitle{Guidance of collective cell migration by substrate geometry}.
\bjtitle{Integrative Biology}
\bvolume{5}(\bissue{8}),
\bfpage{1026}
(\byear{2013})
\doiurl{10.1039/c3ib40054a}
\end{barticle}
\endbibitem

\bibitem[\protect\citeauthoryear{Saraswathibhatla
  et~al.}{2020}]{saraswathibhatla2020}
\begin{barticle}
\bauthor{\bsnm{Saraswathibhatla}, \binits{A.}},
\bauthor{\bsnm{Galles}, \binits{E.E.}},
\bauthor{\bsnm{Notbohm}, \binits{J.}}:
\batitle{Spatiotemporal force and motion in collective cell migration}.
\bjtitle{Scientific Data}
\bvolume{7},
\bfpage{197}--\blpage{17}
(\byear{2020})
\doiurl{10.1038/s41597-020-0540-5}
\end{barticle}
\endbibitem

\bibitem[\protect\citeauthoryear{Kim et~al.}{2013}]{kim2013}
\begin{barticle}
\bauthor{\bsnm{Kim}, \binits{J.H.}},
\bauthor{\bsnm{{Serra-Picamal}}, \binits{X.}},
\bauthor{\bsnm{Tambe}, \binits{D.T.}},
\bauthor{\bsnm{Zhou}, \binits{E.H.}},
\bauthor{\bsnm{Park}, \binits{C.Y.}},
\bauthor{\bsnm{Sadati}, \binits{M.}},
\bauthor{\bsnm{Park}, \binits{J.-A.}},
\bauthor{\bsnm{Krishnan}, \binits{R.}},
\bauthor{\bsnm{Gweon}, \binits{B.}},
\bauthor{\bsnm{Millet}, \binits{E.}},
\bauthor{\bsnm{Butler}, \binits{J.P.}},
\bauthor{\bsnm{Trepat}, \binits{X.}},
\bauthor{\bsnm{Fredberg}, \binits{J.J.}}:
\batitle{Propulsion and navigation within the advancing monolayer sheet}.
\bjtitle{Nature Materials}
\bvolume{12}(\bissue{9}),
\bfpage{856}--\blpage{863}
(\byear{2013})
\doiurl{10.1038/nmat3689}
\end{barticle}
\endbibitem

\bibitem[\protect\citeauthoryear{Das et~al.}{2015}]{das2015}
\begin{barticle}
\bauthor{\bsnm{Das}, \binits{T.}},
\bauthor{\bsnm{Safferling}, \binits{K.}},
\bauthor{\bsnm{Rausch}, \binits{S.}},
\bauthor{\bsnm{Grabe}, \binits{N.}},
\bauthor{\bsnm{Boehm}, \binits{H.}},
\bauthor{\bsnm{Spatz}, \binits{J.P.}}:
\batitle{A molecular mechanotransduction pathway regulates collective migration
  of epithelial cells}.
\bjtitle{Nature Cell Biology}
\bvolume{17}(\bissue{3}),
\bfpage{276}--\blpage{287}
(\byear{2015})
\doiurl{10.1038/ncb3115}
{\href{https://arxiv.org/abs/25706233}{{25706233}}}
\end{barticle}
\endbibitem

\bibitem[\protect\citeauthoryear{{Cochet-Escartin}
  et~al.}{2014}]{cochet-Escartin2014}
\begin{barticle}
\bauthor{\bsnm{{Cochet-Escartin}}, \binits{O.}},
\bauthor{\bsnm{Ranft}, \binits{J.}},
\bauthor{\bsnm{Silberzan}, \binits{P.}},
\bauthor{\bsnm{Marcq}, \binits{P.}}:
\batitle{Border {{Forces}} and {{Friction Control Epithelial Closure
  Dynamics}}}.
\bjtitle{Biophysical Journal}
\bvolume{106}(\bissue{1}),
\bfpage{65}--\blpage{73}
(\byear{2014})
\doiurl{10.1016/j.bpj.2013.11.015}
\end{barticle}
\endbibitem

\bibitem[\protect\citeauthoryear{Rausch et~al.}{2013}]{rausch2013}
\begin{barticle}
\bauthor{\bsnm{Rausch}, \binits{S.}},
\bauthor{\bsnm{Das}, \binits{T.}},
\bauthor{\bsnm{Soin{\'e}}, \binits{J.R.}},
\bauthor{\bsnm{Hofmann}, \binits{T.W.}},
\bauthor{\bsnm{Boehm}, \binits{C.H.}},
\bauthor{\bsnm{Schwarz}, \binits{U.S.}},
\bauthor{\bsnm{Boehm}, \binits{H.}},
\bauthor{\bsnm{Spatz}, \binits{J.P.}}:
\batitle{Polarizing cytoskeletal tension to induce leader cell formation during
  collective cell migration}.
\bjtitle{Biointerphases}
\bvolume{8}(\bissue{1}),
\bfpage{32}
(\byear{2013})
\doiurl{10.1186/1559-4106-8-32}
\end{barticle}
\endbibitem

\bibitem[\protect\citeauthoryear{Chepizhko et~al.}{2018}]{chepizhko2018}
\begin{barticle}
\bauthor{\bsnm{Chepizhko}, \binits{O.}},
\bauthor{\bsnm{Lionetti}, \binits{M.C.}},
\bauthor{\bsnm{Malinverno}, \binits{C.}},
\bauthor{\bsnm{Giampietro}, \binits{C.}},
\bauthor{\bsnm{Scita}, \binits{G.}},
\bauthor{\bsnm{Zapperi}, \binits{S.}},
\bauthor{\bsnm{La~Porta}, \binits{C.A.M.}}:
\batitle{From jamming to collective cell migration through a boundary induced
  transition}.
\bjtitle{Soft Matter}
\bvolume{14}(\bissue{19}),
\bfpage{3774}--\blpage{3782}
(\byear{2018})
\doiurl{10.1039/C8SM00128F}
\end{barticle}
\endbibitem

\bibitem[\protect\citeauthoryear{Alert et~al.}{2019}]{alert2019}
\begin{barticle}
\bauthor{\bsnm{Alert}, \binits{R.}},
\bauthor{\bsnm{{Blanch-Mercader}}, \binits{C.}},
\bauthor{\bsnm{Casademunt}, \binits{J.}}:
\batitle{Active {{Fingering Instability}} in {{Tissue Spreading}}}.
\bjtitle{Physical Review Letters}
\bvolume{122}(\bissue{8}),
\bfpage{088104}
(\byear{2019})
\doiurl{10.1103/PhysRevLett.122.088104}
\end{barticle}
\endbibitem

\bibitem[\protect\citeauthoryear{Reffay et~al.}{2011}]{reffay11}
\begin{barticle}
\bauthor{\bsnm{Reffay}, \binits{M.}},
\bauthor{\bsnm{Petitjean}, \binits{L.}},
\bauthor{\bsnm{Coscoy}, \binits{S.}},
\bauthor{\bsnm{{Grasland-Mongrain}}, \binits{E.}},
\bauthor{\bsnm{Amblard}, \binits{F.}},
\bauthor{\bsnm{Buguin}, \binits{A.}},
\bauthor{\bsnm{Silberzan}, \binits{P.}}:
\batitle{Orientation and polarity in collectively migrating cell structures:
  Statics and dynamics.}
\bjtitle{Biophysical Journal}
\bvolume{100}(\bissue{11}),
\bfpage{2566}--\blpage{2575}
(\byear{2011})
\doiurl{10.1016/j.bpj.2011.04.047}
\end{barticle}
\endbibitem

\bibitem[\protect\citeauthoryear{Deng et~al.}{2021}]{deng2021}
\begin{barticle}
\bauthor{\bsnm{Deng}, \binits{Y.}},
\bauthor{\bsnm{Levine}, \binits{H.}},
\bauthor{\bsnm{Mao}, \binits{X.}},
\bauthor{\bsnm{Sander}, \binits{L.M.}}:
\batitle{Collective motility and mechanical waves in cell clusters}.
\bjtitle{European Physical Journal E}
\bvolume{44}(\bissue{11}),
\bfpage{1}--\blpage{15}
(\byear{2021})
\doiurl{10.1140/epje/s10189-021-00141-7}
\end{barticle}
\endbibitem

\bibitem[\protect\citeauthoryear{{P{\'e}rez-Gonz{\'a}lez}
  et~al.}{2019}]{perez-gonzalez2019}
\begin{barticle}
\bauthor{\bsnm{{P{\'e}rez-Gonz{\'a}lez}}, \binits{C.}},
\bauthor{\bsnm{Alert}, \binits{R.}},
\bauthor{\bsnm{{Blanch-Mercader}}, \binits{C.}},
\bauthor{\bsnm{{G{\'o}mez-Gonz{\'a}lez}}, \binits{M.}},
\bauthor{\bsnm{Kolodziej}, \binits{T.}},
\bauthor{\bsnm{Bazellieres}, \binits{E.}},
\bauthor{\bsnm{Casademunt}, \binits{J.}},
\bauthor{\bsnm{Trepat}, \binits{X.}}:
\batitle{Active wetting of epithelial tissues}.
\bjtitle{Nature Physics}
\bvolume{15}(\bissue{1}),
\bfpage{79}--\blpage{88}
(\byear{2019})
\doiurl{10.1038/s41567-018-0279-5}
\end{barticle}
\endbibitem

\bibitem[\protect\citeauthoryear{Mishra et~al.}{2019}]{mishra2019}
\begin{barticle}
\bauthor{\bsnm{Mishra}, \binits{A.K.}},
\bauthor{\bsnm{Campanale}, \binits{J.P.}},
\bauthor{\bsnm{Mondo}, \binits{J.A.}},
\bauthor{\bsnm{Montell}, \binits{D.J.}}:
\batitle{Cell interactions in collective cell migration}.
\bjtitle{Development}
\bvolume{146}(\bissue{23}),
\bfpage{172056}
(\byear{2019})
\doiurl{10.1242/dev.172056}
\end{barticle}
\endbibitem

\bibitem[\protect\citeauthoryear{Christiansen and
  Rajasekaran}{2006}]{christiansen2006}
\begin{barticle}
\bauthor{\bsnm{Christiansen}, \binits{J.J.}},
\bauthor{\bsnm{Rajasekaran}, \binits{A.K.}}:
\batitle{Reassessing {{Epithelial}} to {{Mesenchymal Transition}} as a
  {{Prerequisite}} for {{Carcinoma Invasion}} and {{Metastasis}}}.
\bjtitle{Cancer Research}
\bvolume{66}(\bissue{17}),
\bfpage{8319}--\blpage{8326}
(\byear{2006})
\doiurl{10.1158/0008-5472.CAN-06-0410}
\end{barticle}
\endbibitem

\bibitem[\protect\citeauthoryear{Friedl et~al.}{2012}]{Friedl2012}
\begin{barticle}
\bauthor{\bsnm{Friedl}, \binits{P.}},
\bauthor{\bsnm{Locker}, \binits{J.}},
\bauthor{\bsnm{Sahai}, \binits{E.}},
\bauthor{\bsnm{Segall}, \binits{J.E.}}:
\batitle{Classifying collective cancer cell invasion}.
\bjtitle{Nature Cell Biology}
\bvolume{14}(\bissue{8}),
\bfpage{777}--\blpage{783}
(\byear{2012})
\doiurl{10.1038/ncb2548}
\end{barticle}
\endbibitem

\bibitem[\protect\citeauthoryear{Tlili et~al.}{2018}]{tlili2018}
\begin{barticle}
\bauthor{\bsnm{Tlili}, \binits{S.}},
\bauthor{\bsnm{Gauquelin}, \binits{E.}},
\bauthor{\bsnm{Li}, \binits{B.}},
\bauthor{\bsnm{Cardoso}, \binits{O.}},
\bauthor{\bsnm{Ladoux}, \binits{B.}},
\bauthor{\bsnm{{Delano{\"e}-Ayari}}, \binits{H.}},
\bauthor{\bsnm{Graner}, \binits{F.}}:
\batitle{Collective cell migration without proliferation: Density determines
  cell velocity and wave velocity}.
\bjtitle{Royal Society Open Science}
\bvolume{5}(\bissue{5}),
\bfpage{172421}
(\byear{2018})
\doiurl{10.1098/rsos.172421}
\end{barticle}
\endbibitem

\bibitem[\protect\citeauthoryear{Hiraiwa}{2022}]{hiraiwa2022}
\begin{barticle}
\bauthor{\bsnm{Hiraiwa}, \binits{T.}}:
\batitle{Dynamic self-organization of migrating cells under constraints by
  spatial confinement and epithelial integrity}.
\bjtitle{European Physical Journal E}
\bvolume{45}(\bissue{2}),
\bfpage{1}--\blpage{16}
(\byear{2022})
\doiurl{10.1140/epje/s10189-022-00161-x}
\end{barticle}
\endbibitem

\bibitem[\protect\citeauthoryear{Khalilgharibi
  et~al.}{2016}]{khalilgharibi2016}
\begin{barticle}
\bauthor{\bsnm{Khalilgharibi}, \binits{N.}},
\bauthor{\bsnm{Fouchard}, \binits{J.}},
\bauthor{\bsnm{Recho}, \binits{P.}},
\bauthor{\bsnm{Charras}, \binits{G.}},
\bauthor{\bsnm{Kabla}, \binits{A.}}:
\batitle{The dynamic mechanical properties of cellularised aggregates}.
\bjtitle{Current Opinion in Cell Biology}
\bvolume{42},
\bfpage{113}--\blpage{120}
(\byear{2016})
\doiurl{10.1016/j.ceb.2016.06.003}
\end{barticle}
\endbibitem

\bibitem[\protect\citeauthoryear{Angelini et~al.}{2011}]{angelini2011}
\begin{barticle}
\bauthor{\bsnm{Angelini}, \binits{T.E.}},
\bauthor{\bsnm{Hannezo}, \binits{E.}},
\bauthor{\bsnm{Trepat}, \binits{X.}},
\bauthor{\bsnm{Marquez}, \binits{M.}},
\bauthor{\bsnm{Fredberg}, \binits{J.J.}},
\bauthor{\bsnm{Weitz}, \binits{D.A.}}:
\batitle{Glass-like dynamics of collective cell migration}.
\bjtitle{Proceedings of the National Academy of Sciences}
\bvolume{108}(\bissue{12}),
\bfpage{4714}--\blpage{4719}
(\byear{2011})
\doiurl{10.1073/pnas.1010059108}
\end{barticle}
\endbibitem

\bibitem[\protect\citeauthoryear{{Armengol-Collado}
  et~al.}{2024}]{armengol-collado2024}
\begin{barticle}
\bauthor{\bsnm{{Armengol-Collado}}, \binits{J.-M.}},
\bauthor{\bsnm{Carenza}, \binits{L.N.}},
\bauthor{\bsnm{Giomi}, \binits{L.}}:
\batitle{Hydrodynamics and multiscale order in confluent epithelia}.
\bjtitle{eLife}
\bvolume{13},
\bfpage{86400}
(\byear{2024})
\doiurl{10.7554/eLife.86400}
\end{barticle}
\endbibitem

\bibitem[\protect\citeauthoryear{Balasubramaniam
  et~al.}{2021}]{balasubramaniam2021}
\begin{barticle}
\bauthor{\bsnm{Balasubramaniam}, \binits{L.}},
\bauthor{\bsnm{Doostmohammadi}, \binits{A.}},
\bauthor{\bsnm{Saw}, \binits{T.B.}},
\bauthor{\bsnm{Narayana}, \binits{G.H.N.S.}},
\bauthor{\bsnm{Mueller}, \binits{R.}},
\bauthor{\bsnm{Dang}, \binits{T.}},
\bauthor{\bsnm{Thomas}, \binits{M.}},
\bauthor{\bsnm{Gupta}, \binits{S.}},
\bauthor{\bsnm{Sonam}, \binits{S.}},
\bauthor{\bsnm{Yap}, \binits{A.S.}},
\bauthor{\bsnm{Toyama}, \binits{Y.}},
\bauthor{\bsnm{M{\`e}ge}, \binits{R.-M.}},
\bauthor{\bsnm{Yeomans}, \binits{J.M.}},
\bauthor{\bsnm{Ladoux}, \binits{B.}}:
\batitle{Investigating the nature of active forces in tissues reveals how
  contractile cells can form extensile monolayers}.
\bjtitle{Nature Materials}
\bvolume{20}(\bissue{8}),
\bfpage{1156}--\blpage{1166}
(\byear{2021})
\doiurl{10.1038/s41563-021-00919-2}
\end{barticle}
\endbibitem

\bibitem[\protect\citeauthoryear{Tlili et~al.}{2015}]{tlili2015}
\begin{barticle}
\bauthor{\bsnm{Tlili}, \binits{S.}},
\bauthor{\bsnm{Gay}, \binits{C.}},
\bauthor{\bsnm{Graner}, \binits{F.}},
\bauthor{\bsnm{Marcq}, \binits{P.}},
\bauthor{\bsnm{Molino}, \binits{F.}},
\bauthor{\bsnm{Saramito}, \binits{P.}}:
\batitle{Colloquium: {{Mechanical}} formalisms for tissue dynamics}.
\bjtitle{European Physical Journal E}
\bvolume{38}(\bissue{5}),
\bfpage{1}--\blpage{31}
(\byear{2015})
\doiurl{10.1140/epje/i2015-15033-4}
\end{barticle}
\endbibitem

\bibitem[\protect\citeauthoryear{Tlili et~al.}{2020}]{tlili2020}
\begin{barticle}
\bauthor{\bsnm{Tlili}, \binits{S.}},
\bauthor{\bsnm{Durande}, \binits{M.}},
\bauthor{\bsnm{Gay}, \binits{C.}},
\bauthor{\bsnm{Ladoux}, \binits{B.}},
\bauthor{\bsnm{Graner}, \binits{F.}},
\bauthor{\bsnm{{Delano{\"e}-Ayari}}, \binits{H.}}:
\batitle{Migrating {{Epithelial Monolayer Flows Like}} a {{Maxwell Viscoelastic
  Liquid}}}.
\bjtitle{Physical Review Letters}
\bvolume{125}(\bissue{8}),
\bfpage{088102}
(\byear{2020})
\doiurl{10.1103/PhysRevLett.125.088102}
\end{barticle}
\endbibitem

\bibitem[\protect\citeauthoryear{Giuglaris}{2024}]{giuglaris2024}
\begin{botherref}
\oauthor{\bsnm{Giuglaris}, \binits{C.}}:
Controlling collective cell migration with multiscale cues ({{Contr{\^o}le}} de
  la migration cellulaire collective dans des environnements
  multi-{\'e}chelles), {{Paris}} {{Sciences}} et {{Lettres}}, {{Physique}} en
  {{Ile-de-France}}, under temporary embargo.
PhD thesis
(2024)
\end{botherref}
\endbibitem

\bibitem[\protect\citeauthoryear{Lucas and Kanade}{1981}]{lucas1981}
\begin{bchapter}
\bauthor{\bsnm{Lucas}, \binits{B.D.}},
\bauthor{\bsnm{Kanade}, \binits{T.}}:
\bctitle{An iterative image registration technique with an application to
  stereo vision}.
In: \bbtitle{Proceedings of the 7th Joint Conference on Artificial
  Intelligence, Vancouver, 24-28 August 1981},
pp. \bfpage{674}--\blpage{679}.
\bpublisher{Morgan Kaufmann Publishers},
\blocation{San Francisco}
(\byear{1981})
\end{bchapter}
\endbibitem

\bibitem[\protect\citeauthoryear{Bouguet}{1999}]{bouguet}
\begin{botherref}
\oauthor{\bsnm{Bouguet}, \binits{J.-Y.}}:
Pyramidal {{Implementation}} of the {{Lucas Kanade Feature Tracker
  Description}} of the Algorithm.
Intel Corporation Microprocessor Research Labs,
  http://robots.stanford.edu/cs223b04/algo\_tracking.pdf
(1999)
\end{botherref}
\endbibitem

\bibitem[\protect\citeauthoryear{Beatrici et~al.}{2023}]{beatrici2023}
\begin{barticle}
\bauthor{\bsnm{Beatrici}, \binits{C.}},
\bauthor{\bsnm{Kirch}, \binits{C.}},
\bauthor{\bsnm{Henkes}, \binits{S.}},
\bauthor{\bsnm{Graner}, \binits{F.}},
\bauthor{\bsnm{Brunnet}, \binits{L.}}:
\batitle{Comparing individual-based models of collective cell motion in a
  benchmark flow geometry}.
\bjtitle{Soft Matter}
\bvolume{19}(\bissue{29}),
\bfpage{5583}--\blpage{5601}
(\byear{2023})
\doiurl{10.1039/D3SM00187C}
\end{barticle}
\endbibitem

\bibitem[\protect\citeauthoryear{Cheddadi et~al.}{2011}]{Cheddadi2011}
\begin{barticle}
\bauthor{\bsnm{Cheddadi}, \binits{I.}},
\bauthor{\bsnm{Saramito}, \binits{P.}},
\bauthor{\bsnm{Dollet}, \binits{B.}},
\bauthor{\bsnm{Raufaste}, \binits{C.}},
\bauthor{\bsnm{Graner}, \binits{F.}}:
\batitle{Understanding and predicting viscous, elastic, plastic flows}.
\bjtitle{European Physical Journal E}
\bvolume{34},
\bfpage{1}--\blpage{15}
(\byear{2011})
\end{barticle}
\endbibitem

\bibitem[\protect\citeauthoryear{Hino et~al.}{2020}]{hino2020}
\begin{barticle}
\bauthor{\bsnm{Hino}, \binits{N.}},
\bauthor{\bsnm{Rossetti}, \binits{L.}},
\bauthor{\bsnm{{Mar{\'i}n-Llaurad{\'o}}}, \binits{A.}},
\bauthor{\bsnm{Aoki}, \binits{K.}},
\bauthor{\bsnm{Trepat}, \binits{X.}},
\bauthor{\bsnm{Matsuda}, \binits{M.}},
\bauthor{\bsnm{Hirashima}, \binits{T.}}:
\batitle{{{ERK-Mediated Mechanochemical Waves Direct Collective Cell
  Polarization}}}.
\bjtitle{Developmental Cell}
\bvolume{53}(\bissue{6}),
\bfpage{646}--\blpage{6608}
(\byear{2020})
\doiurl{10.1016/j.devcel.2020.05.011}
\end{barticle}
\endbibitem

\bibitem[\protect\citeauthoryear{Boocock
  et~al.}{2021}]{boocockTheoryMechanochemicalPatterning2021}
\begin{barticle}
\bauthor{\bsnm{Boocock}, \binits{D.}},
\bauthor{\bsnm{Hino}, \binits{N.}},
\bauthor{\bsnm{Ruzickova}, \binits{N.}},
\bauthor{\bsnm{Hirashima}, \binits{T.}},
\bauthor{\bsnm{Hannezo}, \binits{E.}}:
\batitle{Theory of mechano-chemical patterning and optimal migration in cell
  monolayers}.
\bjtitle{Nature Physics}
\bvolume{17},
\bfpage{267}--\blpage{274}
(\byear{2021})
\doiurl{10.1038/s41567-020-01037-7}
\end{barticle}
\endbibitem

\bibitem[\protect\citeauthoryear{Wu et~al.}{2022}]{wu2022}
\begin{barticle}
\bauthor{\bsnm{Wu}, \binits{H.}},
\bauthor{\bsnm{Shen}, \binits{Y.}},
\bauthor{\bsnm{Sivagurunathan}, \binits{S.}},
\bauthor{\bsnm{Weber}, \binits{M.S.}},
\bauthor{\bsnm{Adam}, \binits{S.A.}},
\bauthor{\bsnm{Shin}, \binits{J.H.}},
\bauthor{\bsnm{Fredberg}, \binits{J.J.}},
\bauthor{\bsnm{Medalia}, \binits{O.}},
\bauthor{\bsnm{Goldman}, \binits{R.}},
\bauthor{\bsnm{Weitz}, \binits{D.A.}}:
\batitle{Vimentin intermediate filaments and filamentous actin form unexpected
  interpenetrating networks that redefine the cell cortex}.
\bjtitle{Proceedings of the National Academy of Sciences}
\bvolume{119}(\bissue{10}),
\bfpage{2115217119}
(\byear{2022})
\doiurl{10.1073/pnas.2115217119}
\end{barticle}
\endbibitem

\bibitem[\protect\citeauthoryear{Nunes~Vicente et~al.}{2022}]{nunesvicente2022}
\begin{barticle}
\bauthor{\bsnm{Nunes~Vicente}, \binits{F.}},
\bauthor{\bsnm{Lelek}, \binits{M.}},
\bauthor{\bsnm{Tinevez}, \binits{J.-Y.}},
\bauthor{\bsnm{Tran}, \binits{Q.D.}},
\bauthor{\bsnm{{Pehau-Arnaudet}}, \binits{G.}},
\bauthor{\bsnm{Zimmer}, \binits{C.}},
\bauthor{\bsnm{{Etienne-Manneville}}, \binits{S.}},
\bauthor{\bsnm{Giannone}, \binits{G.}},
\bauthor{\bsnm{Leduc}, \binits{C.}}:
\batitle{Molecular organization and mechanics of single vimentin filaments
  revealed by super-resolution imaging}.
\bjtitle{Science Advances}
\bvolume{8}(\bissue{8}),
\bfpage{2696}
(\byear{2022})
\doiurl{10.1126/sciadv.abm2696}
\end{barticle}
\endbibitem

\bibitem[\protect\citeauthoryear{Balcioglu et~al.}{2020}]{balcioglu2020}
\begin{barticle}
\bauthor{\bsnm{Balcioglu}, \binits{H.E.}},
\bauthor{\bsnm{Balasubramaniam}, \binits{L.}},
\bauthor{\bsnm{Stirbat}, \binits{T.V.}},
\bauthor{\bsnm{Doss}, \binits{B.L.}},
\bauthor{\bsnm{Fardin}, \binits{M.-A.}},
\bauthor{\bsnm{M{\`e}ge}, \binits{R.-M.}},
\bauthor{\bsnm{Ladoux}, \binits{B.}}:
\batitle{A subtle relationship between substrate stiffness and collective
  migration of cell clusters}.
\bjtitle{Soft Matter}
\bvolume{16}(\bissue{7}),
\bfpage{1825}--\blpage{1839}
(\byear{2020})
\end{barticle}
\endbibitem

\bibitem[\protect\citeauthoryear{Latorre et~al.}{2018}]{latorre2018}
\begin{barticle}
\bauthor{\bsnm{Latorre}, \binits{E.}},
\bauthor{\bsnm{Kale}, \binits{S.}},
\bauthor{\bsnm{Casares}, \binits{L.}},
\bauthor{\bsnm{{G{\'o}mez-Gonz{\'a}lez}}, \binits{M.}},
\bauthor{\bsnm{Uroz}, \binits{M.}},
\bauthor{\bsnm{Valon}, \binits{L.}},
\bauthor{\bsnm{Nair}, \binits{R.V.}},
\bauthor{\bsnm{Garreta}, \binits{E.}},
\bauthor{\bsnm{Montserrat}, \binits{N.}},
\bauthor{\bsnm{{del Campo}}, \binits{A.}},
\bauthor{\bsnm{Ladoux}, \binits{B.}},
\bauthor{\bsnm{Arroyo}, \binits{M.}},
\bauthor{\bsnm{Trepat}, \binits{X.}}:
\batitle{Active superelasticity in three-dimensional epithelia of controlled
  shape}.
\bjtitle{Nature}
\bvolume{563}(\bissue{7730}),
\bfpage{203}--\blpage{208}
(\byear{2018})
\doiurl{10.1038/s41586-018-0671-4}
\end{barticle}
\endbibitem

\bibitem[\protect\citeauthoryear{Trogden et~al.}{2018}]{trogden2018}
\begin{barticle}
\bauthor{\bsnm{Trogden}, \binits{K.P.}},
\bauthor{\bsnm{Battaglia}, \binits{R.A.}},
\bauthor{\bsnm{Kabiraj}, \binits{P.}},
\bauthor{\bsnm{Madden}, \binits{V.J.}},
\bauthor{\bsnm{Herrmann}, \binits{H.}},
\bauthor{\bsnm{Snider}, \binits{N.T.}}:
\batitle{An image-based small-molecule screen identifies vimentin as a
  pharmacologically relevant target of simvastatin in cancer cells}.
\bjtitle{The FASEB Journal}
\bvolume{32}(\bissue{5}),
\bfpage{2841}
(\byear{2018})
\end{barticle}
\endbibitem

\bibitem[\protect\citeauthoryear{Green et~al.}{2020}]{green2020}
\begin{barticle}
\bauthor{\bsnm{Green}, \binits{Y.}},
\bauthor{\bsnm{Fredberg}, \binits{J.J.}},
\bauthor{\bsnm{Butler}, \binits{J.P.}}:
\batitle{Relationship between velocities, tractions, and intercellular stresses
  in the migrating epithelial monolayer}.
\bjtitle{Physical Review E}
\bvolume{101}(\bissue{6}),
\bfpage{062405}
(\byear{2020})
\doiurl{10.1103/PhysRevE.101.062405}
\end{barticle}
\endbibitem

\bibitem[\protect\citeauthoryear{Garcia et~al.}{2015}]{garcia2015}
\begin{barticle}
\bauthor{\bsnm{Garcia}, \binits{S.}},
\bauthor{\bsnm{Hannezo}, \binits{E.}},
\bauthor{\bsnm{Elgeti}, \binits{J.}},
\bauthor{\bsnm{Joanny}, \binits{J.-F.}},
\bauthor{\bsnm{Silberzan}, \binits{P.}},
\bauthor{\bsnm{Gov}, \binits{N.S.}}:
\batitle{Physics of active jamming during collective cellular motion in a
  monolayer}.
\bjtitle{Proceedings of the National Academy of Sciences}
\bvolume{112}(\bissue{50}),
\bfpage{15314}--\blpage{15319}
(\byear{2015})
\doiurl{10.1073/pnas.1510973112}
\end{barticle}
\endbibitem

\bibitem[\protect\citeauthoryear{Bischofs et~al.}{2004}]{bischofs2004}
\begin{barticle}
\bauthor{\bsnm{Bischofs}, \binits{B.}},
\bauthor{\bsnm{Safran}, \binits{S.A.}},
\bauthor{\bsnm{Schwarz}, \binits{U.S.}}:
\batitle{Elastic interactions of active cells with soft materials}.
\bjtitle{Physical Review E}
\bvolume{69}(\bissue{2}),
\bfpage{021911}
(\byear{2004})
\end{barticle}
\endbibitem

\bibitem[\protect\citeauthoryear{Chen et~al.}{2019}]{chen2019}
\begin{barticle}
\bauthor{\bsnm{Chen}, \binits{T.}},
\bauthor{\bsnm{{Callan-Jones}}, \binits{A.}},
\bauthor{\bsnm{Fedorov}, \binits{E.}},
\bauthor{\bsnm{Ravasio}, \binits{A.}},
\bauthor{\bsnm{Brugu{\'e}s}, \binits{A.}},
\bauthor{\bsnm{Ong}, \binits{H.T.}},
\bauthor{\bsnm{Toyama}, \binits{Y.}},
\bauthor{\bsnm{Low}, \binits{B.C.}},
\bauthor{\bsnm{Trepat}, \binits{X.}},
\bauthor{\bsnm{Shemesh}, \binits{T.}}:
\batitle{Large-scale curvature sensing by directional actin flow drives
  cellular migration mode switching}.
\bjtitle{Nature Physics}
\bvolume{15}(\bissue{4}),
\bfpage{393}--\blpage{402}
(\byear{2019})
\end{barticle}
\endbibitem

\bibitem[\protect\citeauthoryear{Chang et~al.}{2008}]{chang2008}
\begin{barticle}
\bauthor{\bsnm{Chang}, \binits{Y.-C.}},
\bauthor{\bsnm{Nalbant}, \binits{P.}},
\bauthor{\bsnm{Birkenfeld}, \binits{J.}},
\bauthor{\bsnm{Chang}, \binits{Z.-F.}},
\bauthor{\bsnm{Bokoch}, \binits{G.M.}}:
\batitle{{G}{E}{F}-{H}1 couples nocodazole-induced microtubule disassembly to
  cell contractility via {R}ho{A}.}
\bjtitle{Molecular Biology of the Cell}
\bvolume{19},
\bfpage{2147}--\blpage{2153}
(\byear{2008})
\end{barticle}
\endbibitem

\bibitem[\protect\citeauthoryear{Petrolli et~al.}{2019}]{petrolli2019}
\begin{barticle}
\bauthor{\bsnm{Petrolli}, \binits{V.}},
\bauthor{\bsnm{Le~Goff}, \binits{M.}},
\bauthor{\bsnm{Tadrous}, \binits{M.}},
\bauthor{\bsnm{Martens}, \binits{K.}},
\bauthor{\bsnm{Allier}, \binits{C.}},
\bauthor{\bsnm{Mandula}, \binits{O.}},
\bauthor{\bsnm{Herv{\'e}}, \binits{L.}},
\bauthor{\bsnm{Henkes}, \binits{S.}},
\bauthor{\bsnm{Sknepnek}, \binits{R.}},
\bauthor{\bsnm{Boudou}, \binits{T.}},
\bauthor{\bsnm{Cappello}, \binits{G.}},
\bauthor{\bsnm{Balland}, \binits{M.}}:
\batitle{Confinement-{{Induced Transition}} between {{Wavelike Collective Cell
  Migration Modes}}}.
\bjtitle{Physical Review Letters}
\bvolume{122}(\bissue{16}),
\bfpage{168101}
(\byear{2019})
\doiurl{10.1103/PhysRevLett.122.168101}
\end{barticle}
\endbibitem

\bibitem[\protect\citeauthoryear{Peyret et~al.}{2019}]{peyret2019}
\begin{barticle}
\bauthor{\bsnm{Peyret}, \binits{G.}},
\bauthor{\bsnm{Mueller}, \binits{R.}},
\bauthor{\bsnm{{d'Alessandro}}, \binits{J.}},
\bauthor{\bsnm{Begnaud}, \binits{S.}},
\bauthor{\bsnm{Marcq}, \binits{P.}},
\bauthor{\bsnm{M{\`e}ge}, \binits{R.-M.}},
\bauthor{\bsnm{Yeomans}, \binits{J.M.}},
\bauthor{\bsnm{Doostmohammadi}, \binits{A.}},
\bauthor{\bsnm{Ladoux}, \binits{B.}}:
\batitle{Sustained {{Oscillations}} of {{Epithelial Cell Sheets}}}.
\bjtitle{Biophysical Journal}
\bvolume{117}(\bissue{3}),
\bfpage{464}--\blpage{478}
(\byear{2019})
\doiurl{10.1016/j.bpj.2019.06.013}
\end{barticle}
\endbibitem

\bibitem[\protect\citeauthoryear{Yabunaka and Marcq}{2017}]{yabunaka2017}
\begin{barticle}
\bauthor{\bsnm{Yabunaka}, \binits{S.}},
\bauthor{\bsnm{Marcq}, \binits{P.}}:
\batitle{Emergence of epithelial cell density waves}.
\bjtitle{Soft Matter}
\bvolume{13},
\bfpage{7046}--\blpage{7052}
(\byear{2017})
\doiurl{10.1039/C7SM01172E}
\end{barticle}
\endbibitem

\bibitem[\protect\citeauthoryear{{Blanch-Mercader} and
  Casademunt}{2017}]{blanch-mercader2017}
\begin{barticle}
\bauthor{\bsnm{{Blanch-Mercader}}, \binits{C.}},
\bauthor{\bsnm{Casademunt}, \binits{J.}}:
\batitle{Hydrodynamic instabilities, waves and turbulence in spreading
  epithelia}.
\bjtitle{Soft Matter}
\bvolume{13}(\bissue{38}),
\bfpage{6913}--\blpage{6928}
(\byear{2017})
\doiurl{10.1039/C7SM01128H}
\end{barticle}
\endbibitem

\bibitem[\protect\citeauthoryear{Boocock et~al.}{2023}]{boocock2023}
\begin{barticle}
\bauthor{\bsnm{Boocock}, \binits{D.}},
\bauthor{\bsnm{Hirashima}, \binits{T.}},
\bauthor{\bsnm{Hannezo}, \binits{E.}}:
\batitle{Interplay between {{Mechanochemical Patterning}} and {{Glassy
  Dynamics}} in {{Cellular Monolayers}}}.
\bjtitle{PRX Life}
\bvolume{1}(\bissue{1}),
\bfpage{013001}
(\byear{2023})
\doiurl{10.1103/PRXLife.1.013001}
\end{barticle}
\endbibitem

\bibitem[\protect\citeauthoryear{Lin et~al.}{2021}]{lin2021}
\begin{barticle}
\bauthor{\bsnm{Lin}, \binits{S.-Z.}},
\bauthor{\bsnm{Zhang}, \binits{W.-Y.}},
\bauthor{\bsnm{Bi}, \binits{D.}},
\bauthor{\bsnm{Li}, \binits{B.}},
\bauthor{\bsnm{Feng}, \binits{X.-Q.}}:
\batitle{Energetics of mesoscale cell turbulence in two-dimensional
  monolayers}.
\bjtitle{Communications Physics}
\bvolume{4},
\bfpage{21}
(\byear{2021})
\doiurl{10.1038/s42005-021-00530-6}
\end{barticle}
\endbibitem

\bibitem[\protect\citeauthoryear{Giomi}{2015}]{giomi2015}
\begin{barticle}
\bauthor{\bsnm{Giomi}, \binits{L.}}:
\batitle{Geometry and topology of turbulence in active nematics}.
\bjtitle{Physical Review X}
\bvolume{5},
\bfpage{031003}
(\byear{2015})
\doiurl{10.1103/PhysRevX.5.031003}
\end{barticle}
\endbibitem

\bibitem[\protect\citeauthoryear{Cavagna et~al.}{2018}]{Cavagna2018}
\begin{barticle}
\bauthor{\bsnm{Cavagna}, \binits{A.}},
\bauthor{\bsnm{Giardina}, \binits{I.}},
\bauthor{\bsnm{Grigera}, \binits{T.S.}}:
\batitle{The physics of flocking: {{Correlation}} as a compass from experiments
  to theory}.
\bjtitle{Physics Reports}
\bvolume{728},
\bfpage{1}--\blpage{62}
(\byear{2018})
\doiurl{10.1016/j.physrep.2017.11.003}
\end{barticle}
\endbibitem

\bibitem[\protect\citeauthoryear{Vicsek and Zafeiris}{2012}]{vicsek2012}
\begin{barticle}
\bauthor{\bsnm{Vicsek}, \binits{T.}},
\bauthor{\bsnm{Zafeiris}, \binits{A.}}:
\batitle{Collective motion}.
\bjtitle{Physics Reports}
\bvolume{517}(\bissue{3}),
\bfpage{71}--\blpage{140}
(\byear{2012})
\doiurl{10.1016/j.physrep.2012.03.004}
\end{barticle}
\endbibitem

\bibitem[\protect\citeauthoryear{V{\'a}rkuti et~al.}{2016}]{varkuti2016}
\begin{barticle}
\bauthor{\bsnm{V{\'a}rkuti}, \binits{B.H.}},
\bauthor{\bsnm{K{\'e}pir{\'o}}, \binits{M.}},
\bauthor{\bsnm{Horv{\'a}th}, \binits{I.{\'A}.}},
\bauthor{\bsnm{V{\'e}gner}, \binits{L.}},
\bauthor{\bsnm{R{\'a}ti}, \binits{S.}},
\bauthor{\bsnm{Zsigmond}, \binits{{\'A}.}},
\bauthor{\bsnm{Hegyi}, \binits{G.}},
\bauthor{\bsnm{Lenkei}, \binits{Z.}},
\bauthor{\bsnm{Varga}, \binits{M.}},
\bauthor{\bsnm{{M{\'a}ln{\'a}si-Csizmadia}}, \binits{A.}}:
\batitle{A highly soluble, non-phototoxic, non-fluorescent blebbistatin
  derivative}.
\bjtitle{Scientific Reports}
\bvolume{6}(\bissue{1}),
\bfpage{26141}
(\byear{2016})
\end{barticle}
\endbibitem

\bibitem[\protect\citeauthoryear{Butler et~al.}{2002}]{butler2002}
\begin{barticle}
\bauthor{\bsnm{Butler}, \binits{J.P.}},
\bauthor{\bsnm{{Toli{\'c}-N{\o}rrelykke}}, \binits{I.M.}},
\bauthor{\bsnm{Fabry}, \binits{B.}},
\bauthor{\bsnm{Fredberg}, \binits{J.J.}}:
\batitle{Traction fields, moments, and strain energy that cells exert on their
  surroundings}.
\bjtitle{American Journal of Physiology-Cell Physiology}
\bvolume{282}(\bissue{3}),
\bfpage{595}--\blpage{605}
(\byear{2002})
\doiurl{10.1152/ajpcell.00270.2001}
\end{barticle}
\endbibitem

\bibitem[\protect\citeauthoryear{Hansen}{2010}]{hansen2010}
\begin{bbook}
\bauthor{\bsnm{Hansen}, \binits{P.C.}}:
\bbtitle{Discrete Inverse Problems: Insight and Algorithms}.
\bpublisher{Society for Industrial and Applied Mathematics},
\blocation{Philadelphia}
(\byear{2010}).
\doiurl{10.1137/1.9780898718836}
\end{bbook}
\endbibitem

\bibitem[\protect\citeauthoryear{Huber and Ronchetti}{2011}]{huber2011}
\begin{bbook}
\bauthor{\bsnm{Huber}, \binits{P.J.}},
\bauthor{\bsnm{Ronchetti}, \binits{E.M.}}:
\bbtitle{Robust {{Statistics}}}.
\bpublisher{John Wiley \& Sons},
\blocation{Hoboken}
(\byear{2011})
\end{bbook}
\endbibitem

\bibitem[\protect\citeauthoryear{}{}]{zotero-9062}
\begin{botherref}
{{HuberRegressor}}.
https://scikit-learn/stable/modules/generated/sklearn.linear\_model.HuberRegressor.html
\end{botherref}
\endbibitem

\end{thebibliography}

\newpage

\beginsupplement
 
\begin{center}
    \LARGE\textbf{Supplementary Materials}\\[1em]
  {\it \normalsize ``Geometric Confinement Reveals Scale-Free Velocity Correlations in Epithelial Cell Monolayer" }\\[1em]
    \large{Guillaume Duprez, M\'elina Durande, Fran\c{c}ois Graner, H\'el\`ene Delano\"e-Ayari}
\end{center}
%
%
%

\vspace{2em}


\section{Materials and Methods}
\label{sec:mat_meth}

\subsection{Cell monolayer experiments}

\subsubsection{Cell culture}
\label{sec:culture}

We use the same MDCK cell line as in Ref.~\cite{vedula2012}. Cells are seeded on the patterned substrates one to two days before starting the experiment, until they reach confluence. We then add mitomycin C (Sigma M4287) at a concentration of~ 
 8 $\mu$g/ml.

\subsubsection{Patterning on hard PDMS substrate}

We use the same protocol as described in detail in Ref.~\cite{tlili2018}.
Patterned PDMS stamps prepared from silanized wafers are incubated for 45~min at 37$^\circ$C or 60~min at room temperature with a 60~$\mu$L drop of 50~$\mu$g/mL fibronectin solution \modif{(Gibco Ref. 33010-018)}.

A thin layer of rigid PDMS (10\% of cross-linking agent) is spin-coated on a glass coverslip (\#1, diameter 30 mm). We use a velocity of 500~rpm for 10~s followed by 30~s at 1000~rpm. We aim to reach a thickness of around 200~$\mu$m. The glass coverslip is then glued (using Loctite Si 5398) to a custom-drilled Petri dish and cured overnight at 65$^\circ$C. Just before patterning, we then expose the sample to UV-ozone for approximately 20~min to activate the PDMS surface. After incubation, stamps are immersed extremely briefly in deionized water, then dried with pressurized air and pressed on the UV-activated PDMS surface to transfer fibronectin. A 2\% Pluronic F-127 (Sigma) solution is added to the Petri dish to chemically block the regions outside of the fibronectin pattern for 1~h at room temperature. The Pluronic solution is removed after 1~h and the Petri dish is rinsed three to six times with a PBS solution.

\modif{The dimensions of the racetrack widths and obstacle diameters 
are listed on Table~S1.}

\medskip
\medskip
\medskip

    \begin{center}
 \begin{tabular}{c|c|c|c}
$W$ & $D$ & {\bf V} & $N$  \\
\hline 
1000 & 300  & yes & 5 \\
1000 & 160 & yes & 4 \\
1000 & 300 & no & 7 \\
1000 & 160 & no & 7 \\
600 & 80 & yes & 7 \\
600 &80  & no & 5 \\
600 & 80 & no & 13 \\
\hline 
\end{tabular}
   \end{center}

\medskip

    \modif{ {\bf Table~S1: 
    List of 48 hard racetrack substrates.} $W$: Racetrack width ($\mu$m). $D$: obstacle diameter  ($\mu$m). {\bf V}: presence of chiral V-shaped obstacle design. $N$: number of racetracks. }




\subsection{Live Microscopy}

For long-term imaging, we employ phase contrast microscopy at 10X magnification. \modif{Pixel size is 0.645 $\mu$m.} Under the most optimistic assumptions, it is possible to acquire data from six samples in parallel, with each sample containing either four bands or four racetracks. This setup allows for up to 24 experiments to be conducted simultaneously during a single acquisition session. However, two main experimental limitations arise in practice. The first is that obtaining six perfectly synchronized and high-quality experiments is rare. The second limitation stems from the imaging time required to cover all samples within a time frame of less than 5~min. This constraint is critical to accurately capture the dynamics of cellular movement and to enable temporal averaging of the results.

To image a single band, three microscope fields of 2048×2048 pixels must be captured. In contrast, imaging a single racetrack requires between 10 and 20 fields. The positions of each field are manually marked, ensuring sufficient overlap between adjacent images. Subsequently, a custom Micro-Manager script, designed to enhance software stability, automatically moves through the fields at regular intervals. As a result, a sample contains between 12 fields (if composed exclusively of bands) and up to 80 fields (if containing racetracks).

\begin{figure}[!ht]
    \centering
    \includegraphics[width=0.6\textwidth]{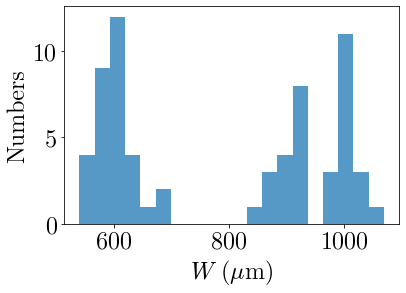}
    \caption{
    \vire{ Distribution of the racetrack widths. Total 47 patterns.}\modif{{\bf Racetrack measurements.}
    Distribution of 66 actual measurements on hard racetrack substrates (see Table~S1).
    }
    }
    \label{fig:distrib_width} 
\end{figure}

\begin{figure}[!ht]
	   \centering
        \includegraphics[width=0.6\textwidth]{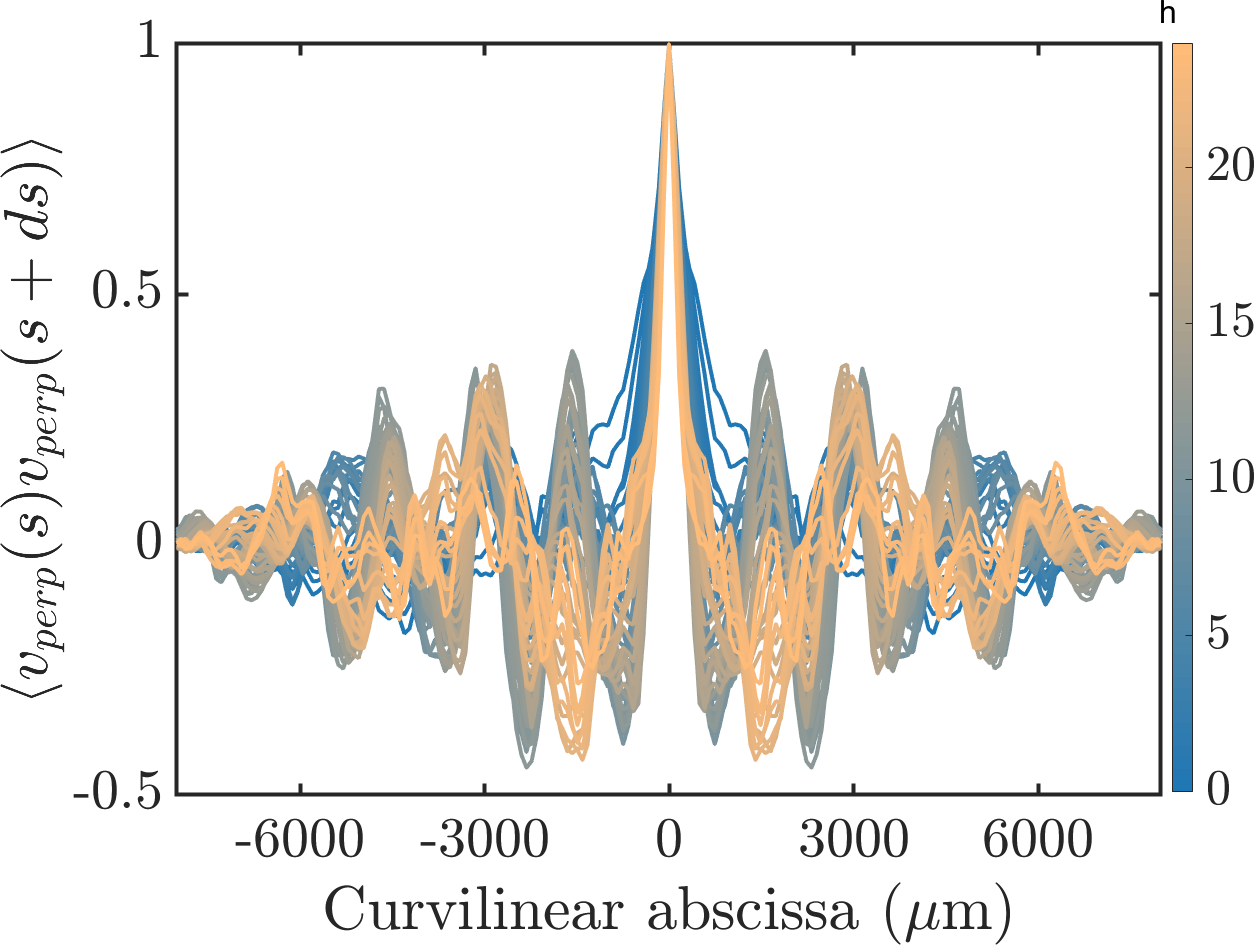}
	\caption{{\bf Perpendicular velocity autocorrelation.} Autocorrelation function along the racetrack midline (curvilinear abscissa $s$), of the  velocity component locally perpendicular to this midline, $v_{perp}(s)$. Color code: time (h). Same racetrack  as in  Fig.~\ref{fig:space_corr}a, $W=1000$~$\mu$m, midline perimeter 8200~$\mu$m, symmetrized $s \to -s$.}
        \label{fig:wave_corr_perp}
\end{figure}

\begin{figure}[t!]
    \centering
        \begin{subfigure}[t]{0.3\textwidth}
        \centering
        \includegraphics[width=\linewidth]{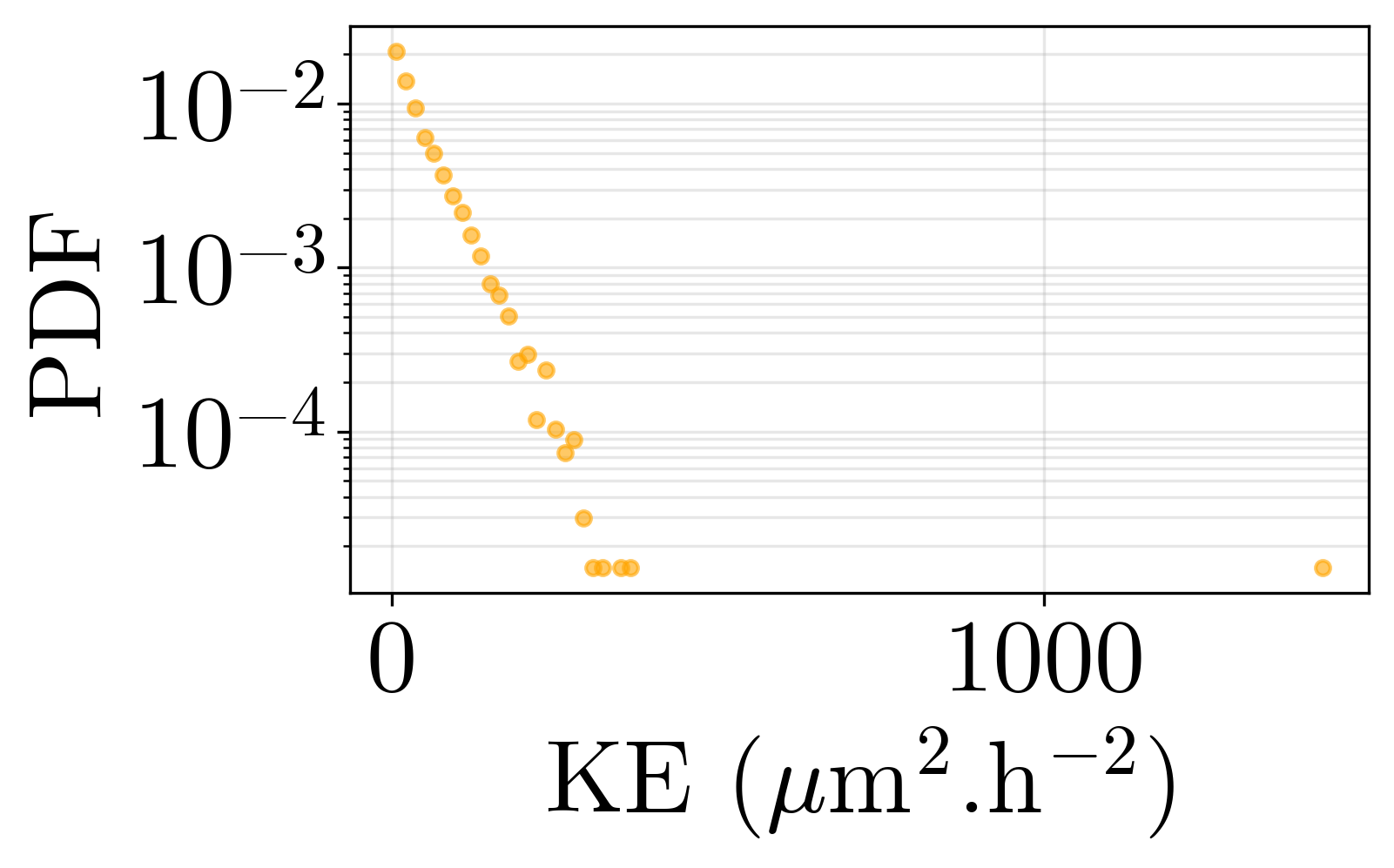}
        \caption{}
        \label{fig:pdf_nrj_H1}
    \end{subfigure}
    \hfill
     \begin{subfigure}[t]{0.3\textwidth}
        \centering
        \includegraphics[width=\linewidth]{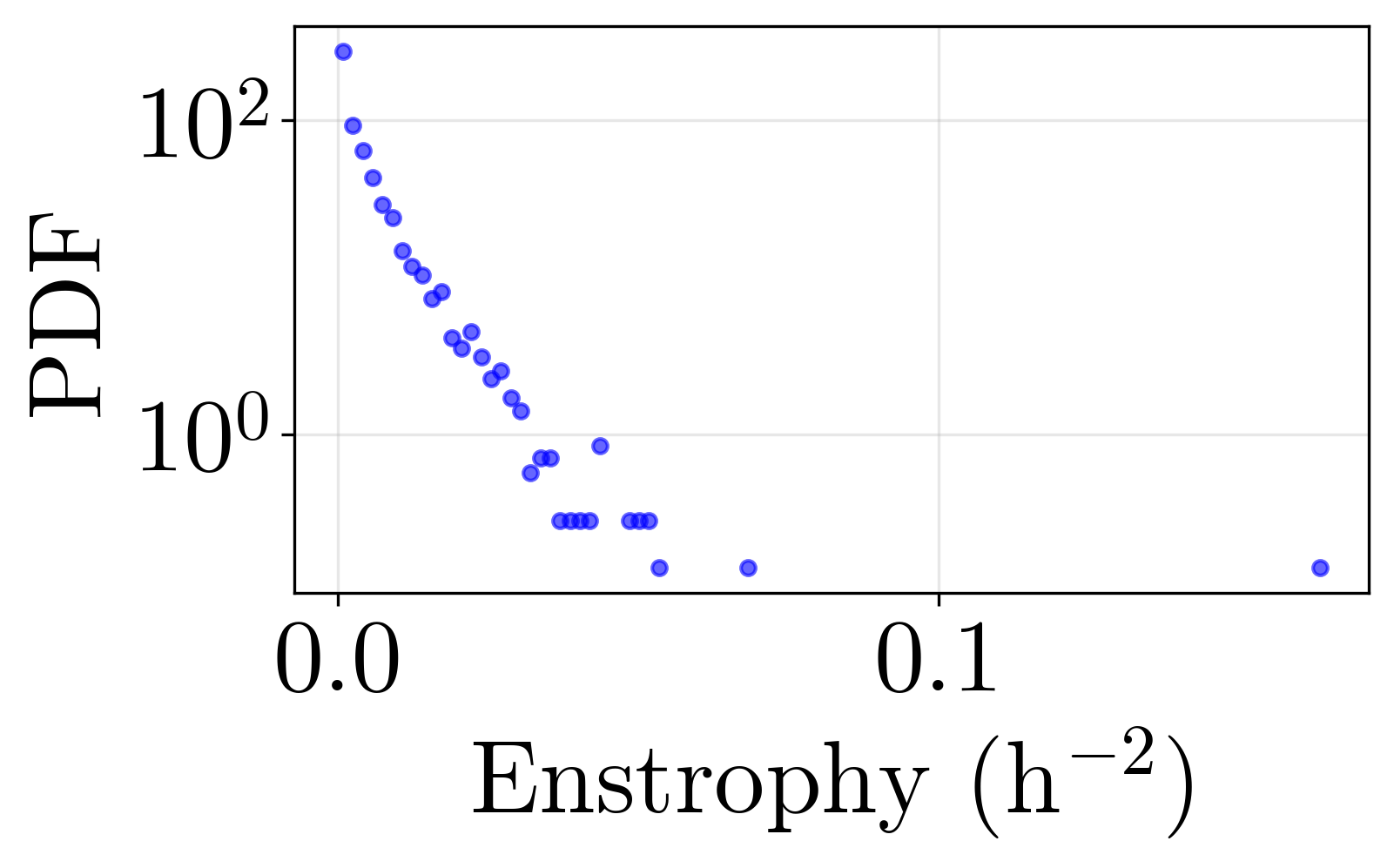}
        \caption{}
        \label{fig:pdf_ens_H1}
    \end{subfigure}
    \hfill
    \begin{subfigure}[t]{0.3\textwidth}
        \centering
        \includegraphics[width=\linewidth]{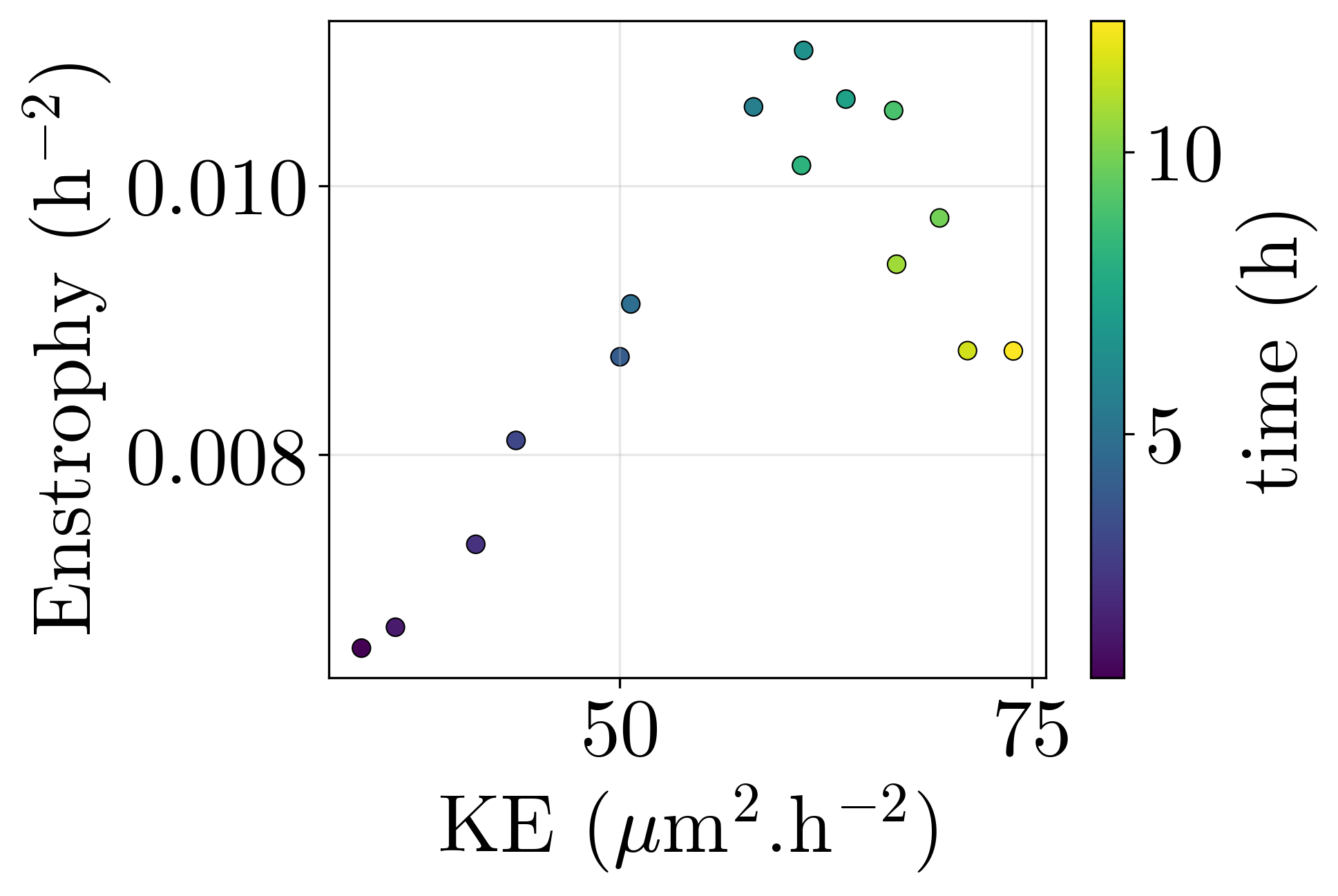}
        \caption{}
        \label{fig:ens_vs_nrj_H1}
    \end{subfigure}
    \begin{subfigure}[t]{0.3\textwidth}
        \centering
        \includegraphics[width=\linewidth]{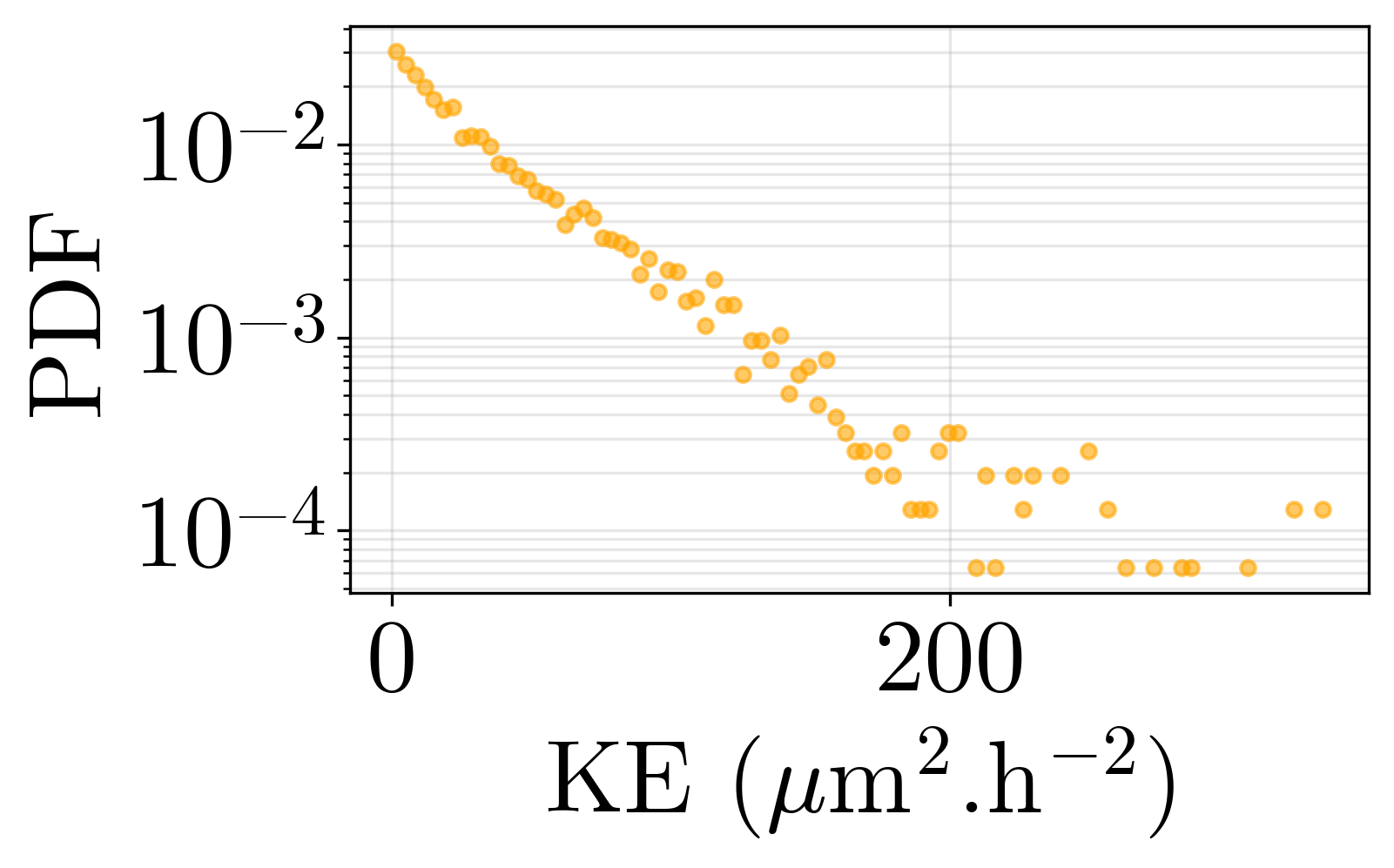}
        \caption{}
        \label{fig:pdf_nrj_H2}
    \end{subfigure}
    \hfill
     \begin{subfigure}[t]{0.3\textwidth}
        \centering
        \includegraphics[width=\linewidth]{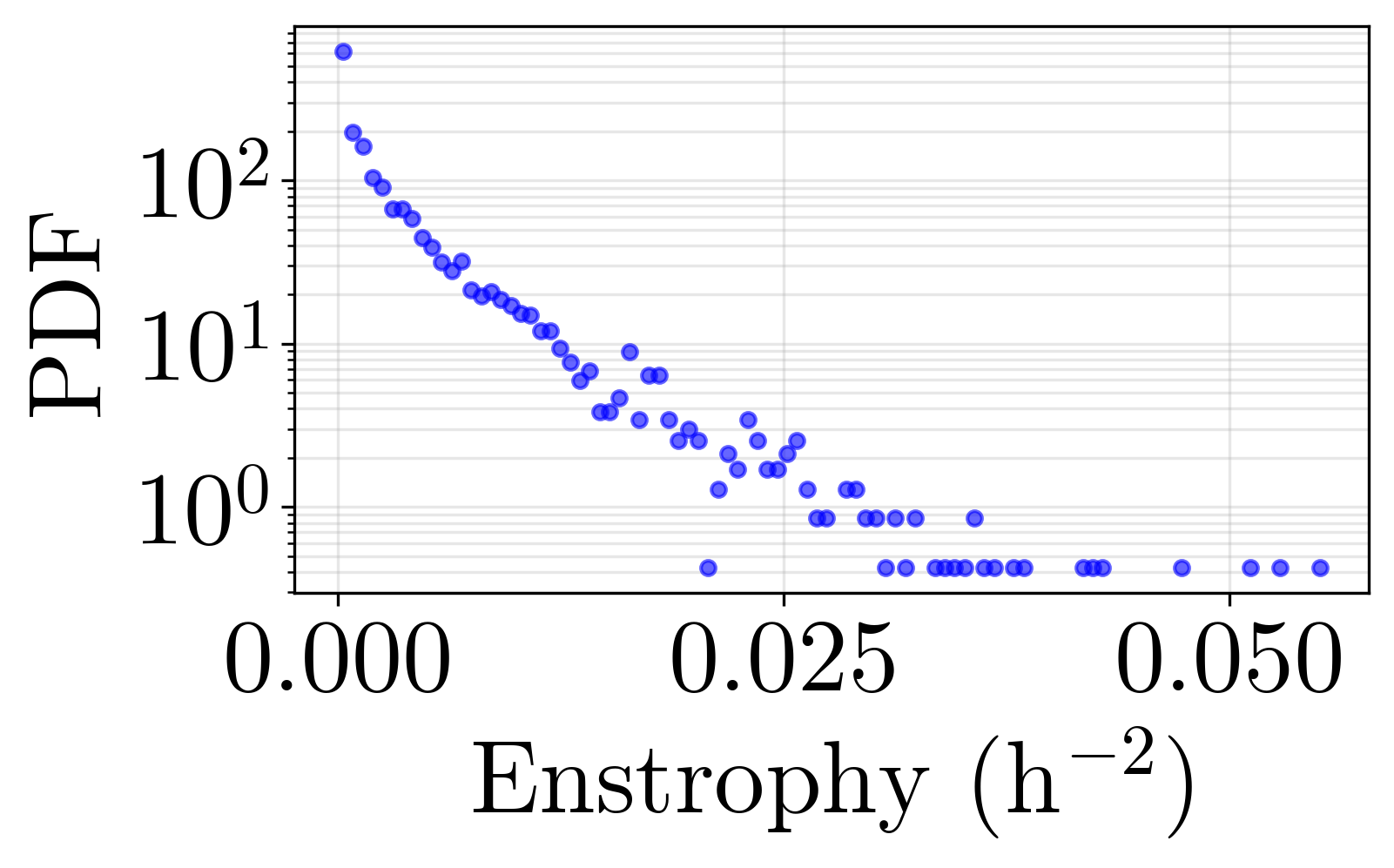}
        \caption{}
        \label{fig:pdf_ens_H2}
    \end{subfigure}
    \hfill
    \begin{subfigure}[t]{0.3\textwidth}
        \centering
        \includegraphics[width=\linewidth]{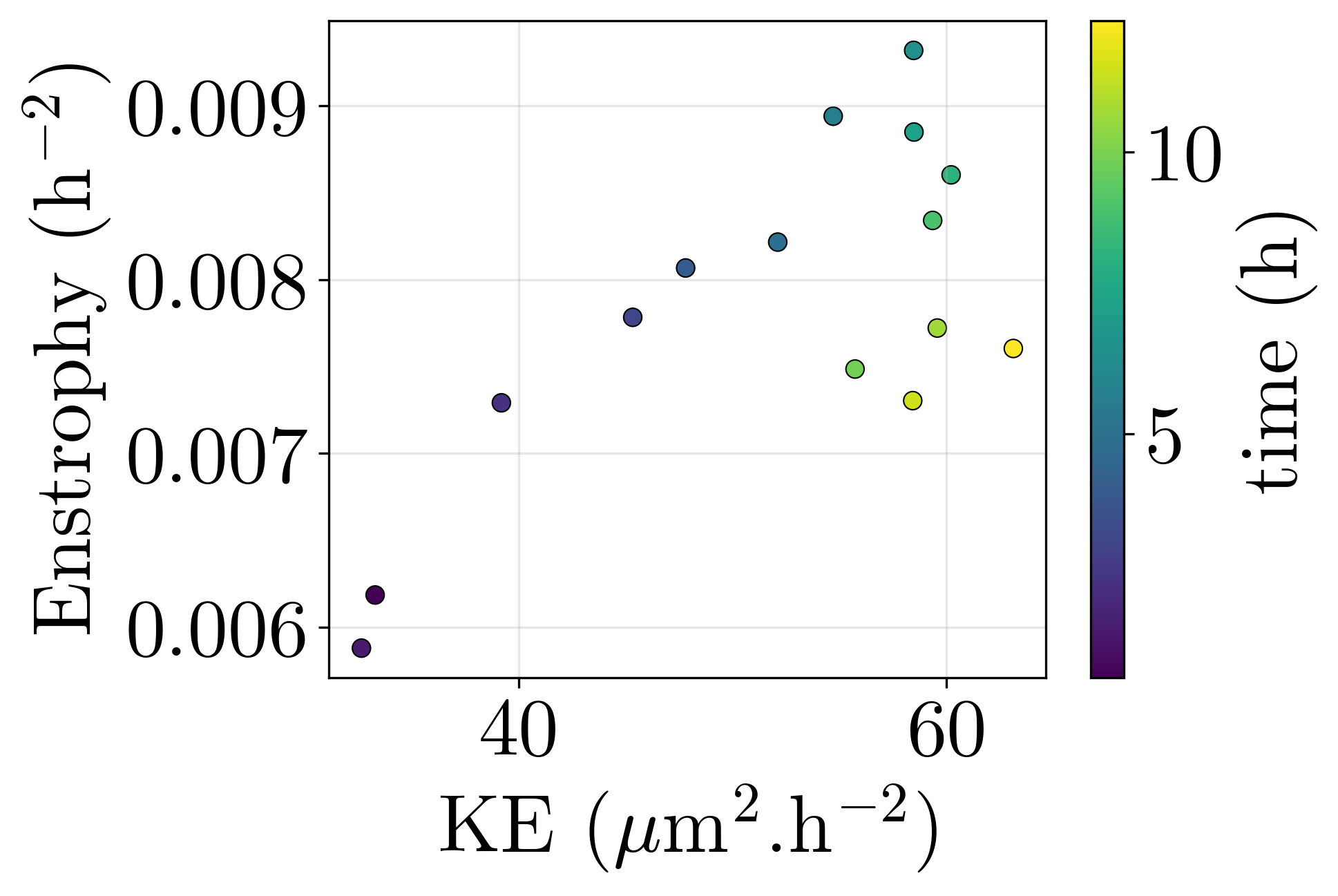}
        \caption{}
        \label{fig:ens_vs_nrj_H2}
    \end{subfigure}
    \begin{subfigure}[t]{0.3\textwidth}
        \centering
        \includegraphics[width=\linewidth]{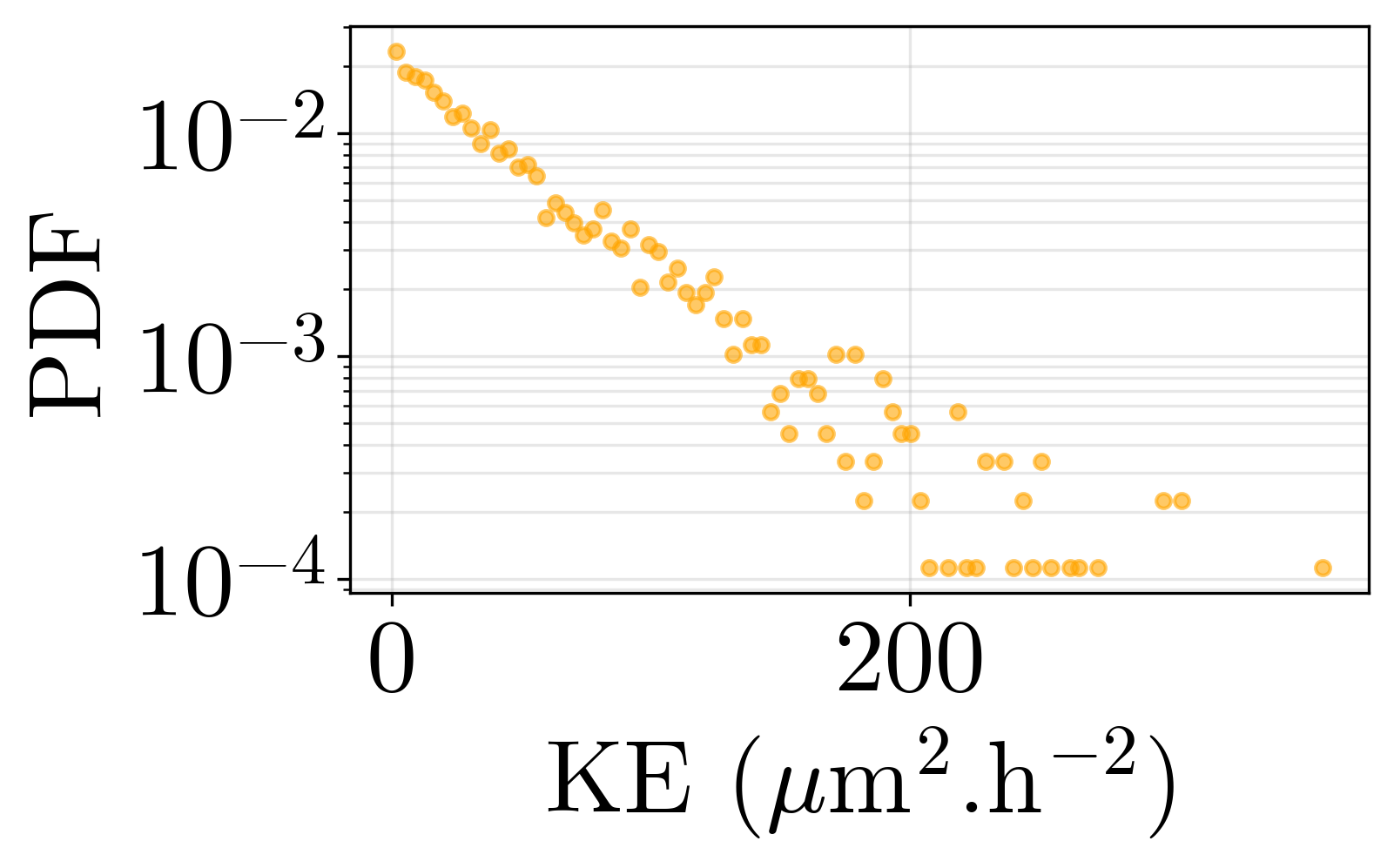}
        \caption{}
        \label{fig:pdf_nrj_H3}
    \end{subfigure}
        \hfill
    \begin{subfigure}[t]{0.3\textwidth}
        \centering
        \includegraphics[width=\linewidth]{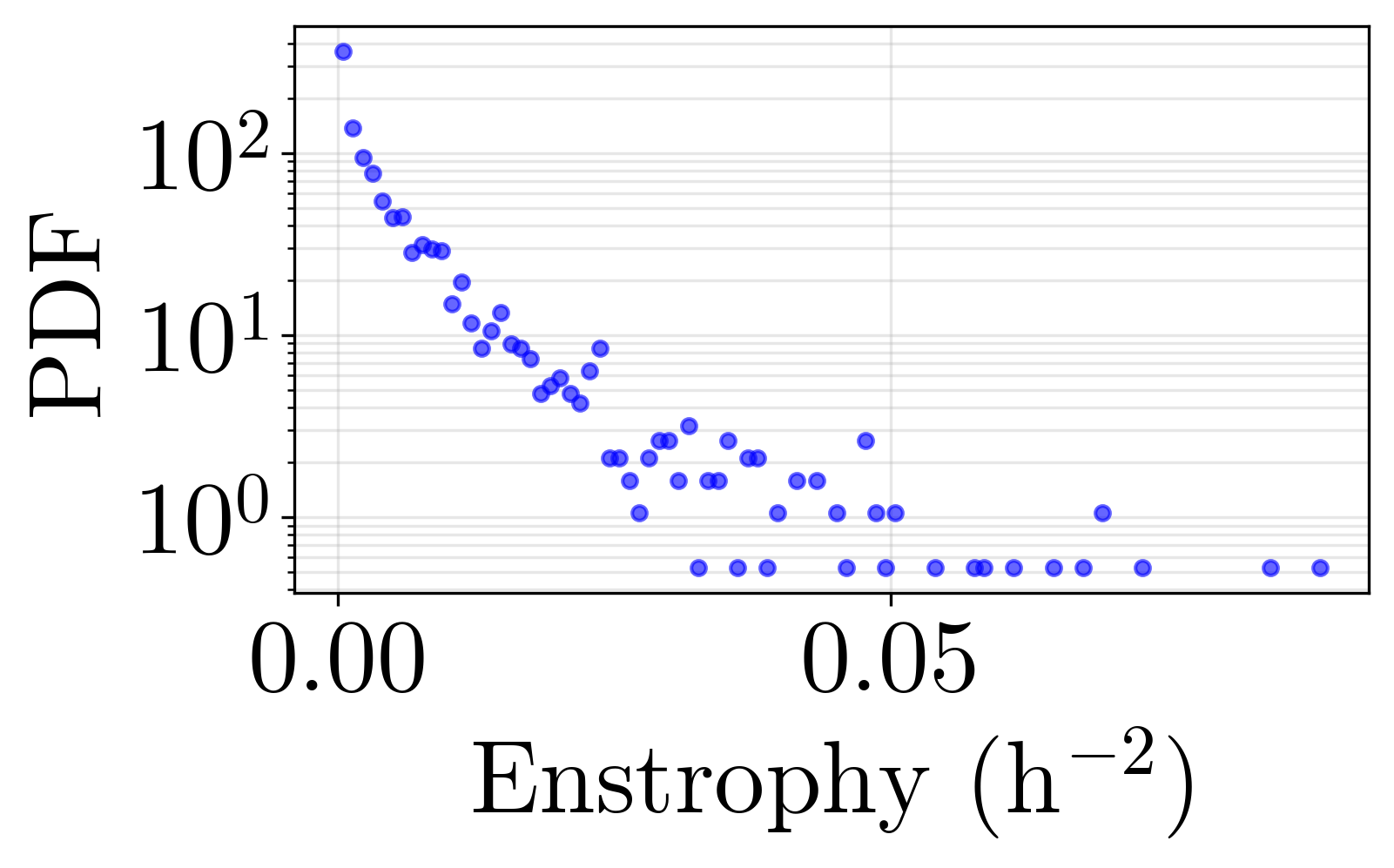}
        \caption{}
        \label{fig:pdf_ens_H3}
    \end{subfigure}
        \hfill
    \begin{subfigure}[t]{0.3\textwidth}
        \centering
        \includegraphics[width=\linewidth]{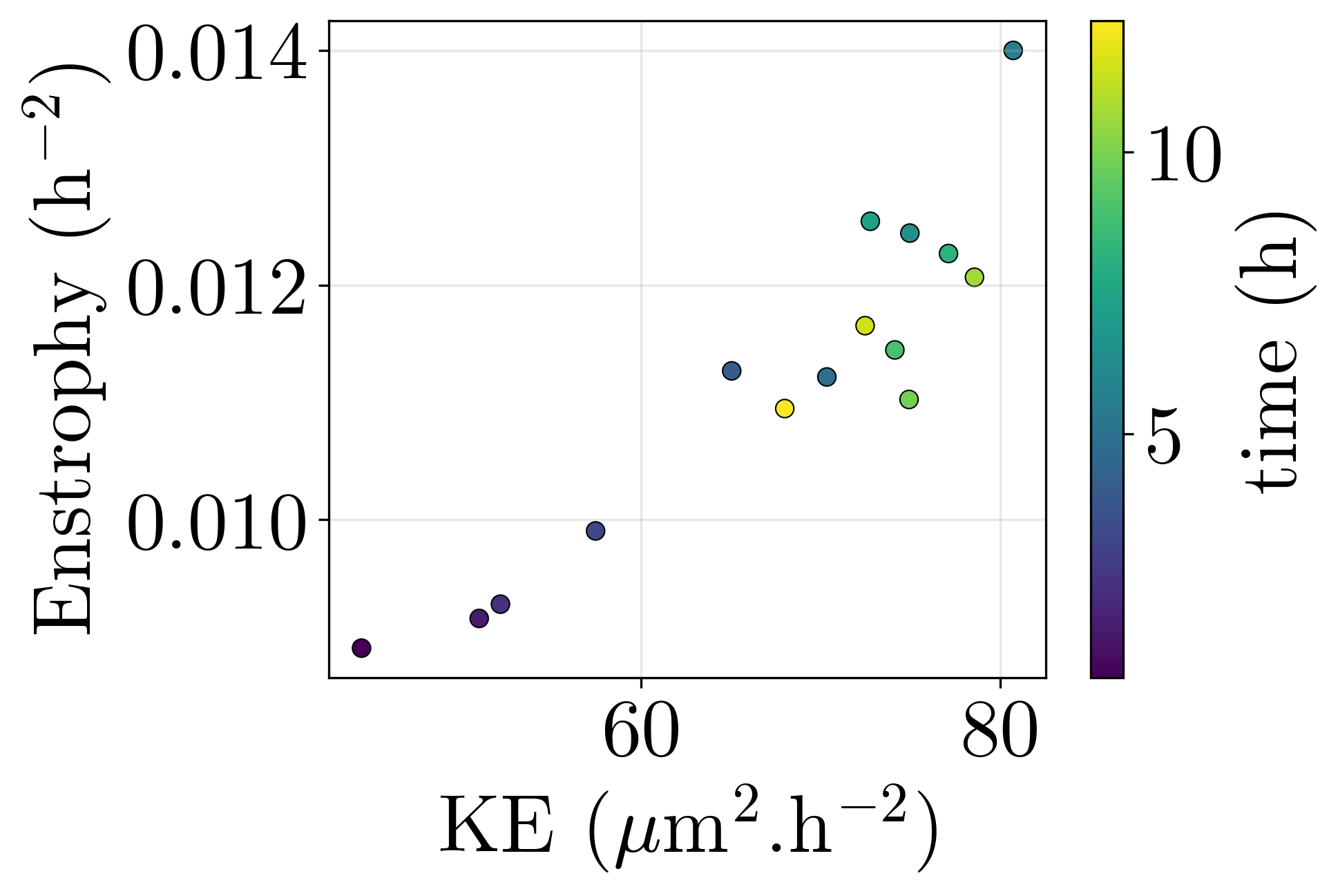}
        \caption{}
        \label{fig:ens_vs_nrj_H3}
    \end{subfigure}
        \begin{subfigure}[t]{0.3\textwidth}
        \centering
        \includegraphics[width=\linewidth]{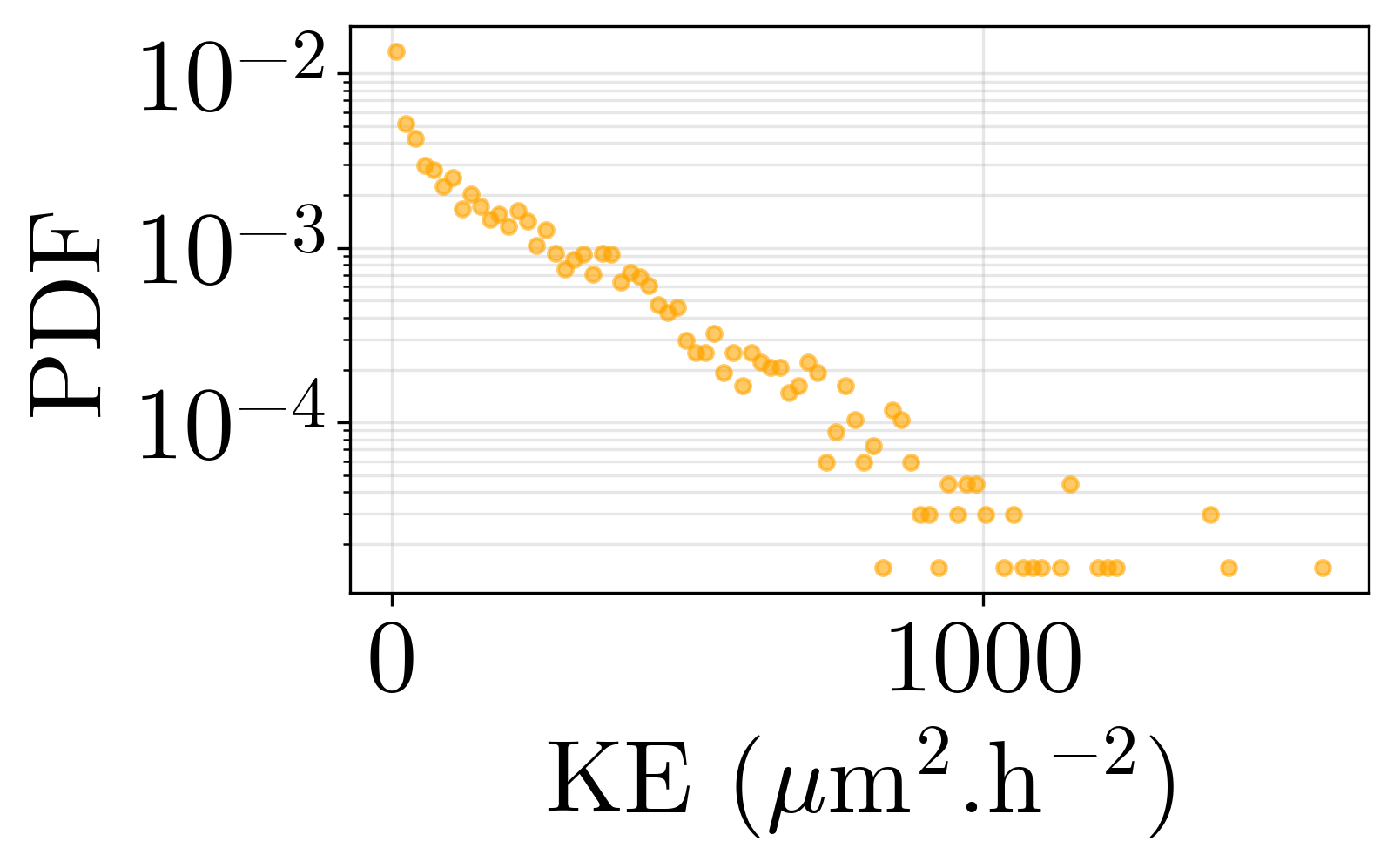}
        \caption{}
        \label{fig:pdf_nrj_H2V}
    \end{subfigure}
        \hfill
    \begin{subfigure}[t]{0.3\textwidth}
        \centering
        \includegraphics[width=\linewidth]{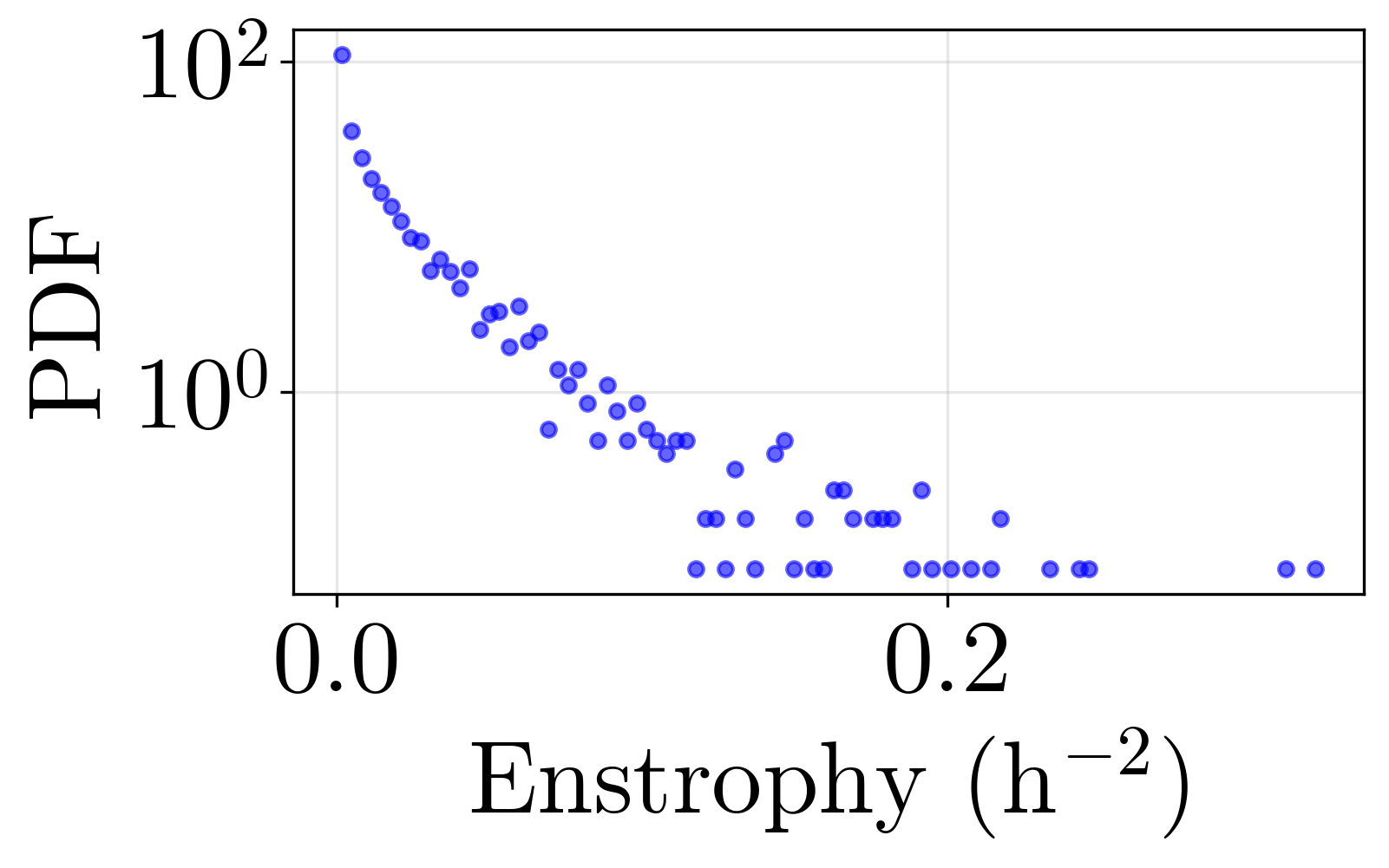}
        \caption{}
        \label{fig:pdf_ens_H2V}
    \end{subfigure}
        \hfill
    \begin{subfigure}[t]{0.3\textwidth}
        \centering
        \includegraphics[width=\linewidth]{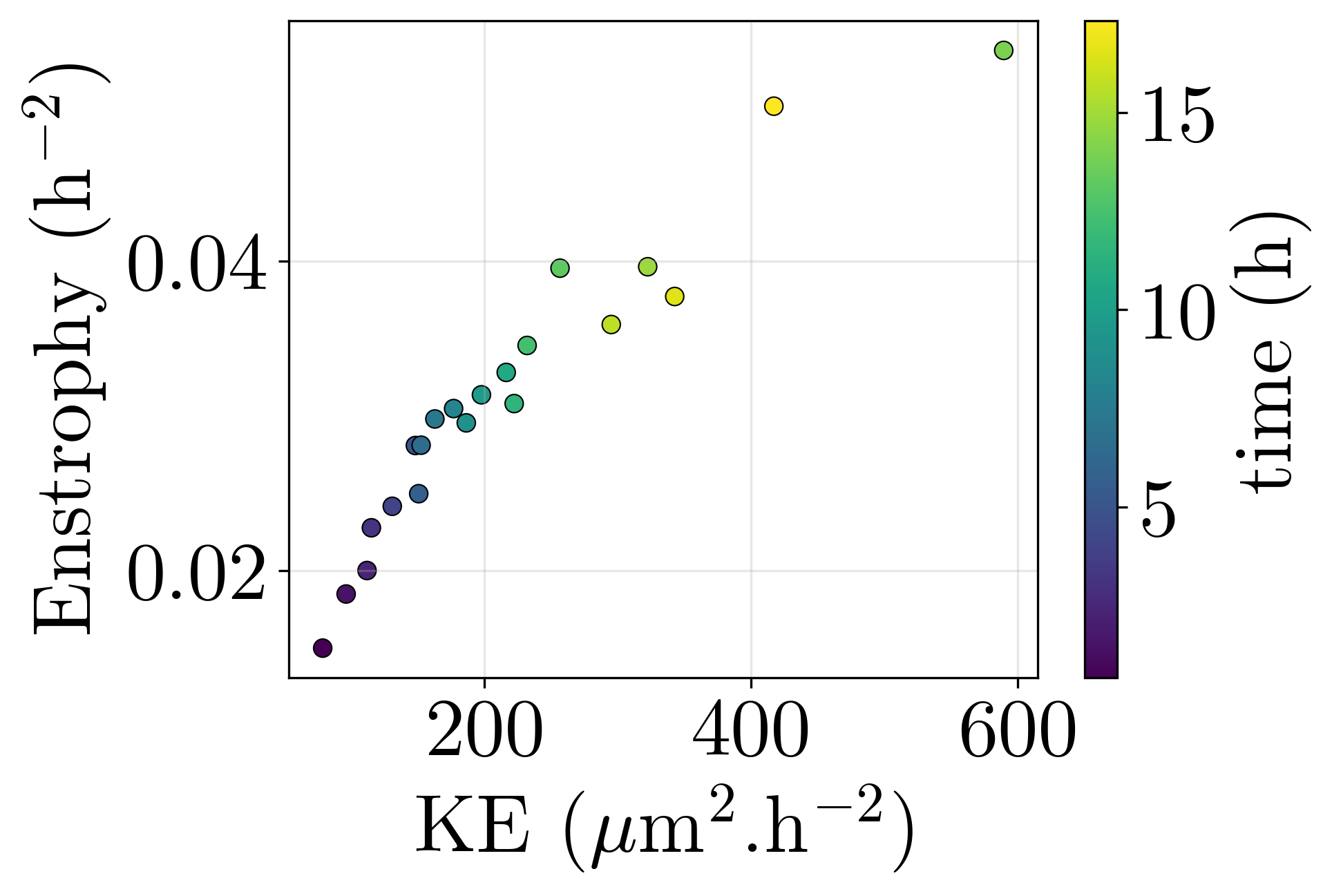}
        \caption{}
        \label{fig:ens_vs_nrj_H2V}
    \end{subfigure}
    \caption{\label{fig:energy_enstrophy_supp} 
    \modif{
        {\bf Kinetic energy and enstrophy for different racetracks geometries, not averaged over measurements.}
  Statistics averaged  in space over 64 pixels, in time over 50 frames. Same representation as Fig.~\ref{fig:energy_enstrophy}.
     {\bf (a-c)} Racetrack, width $W=1000$~$\mu$m, obstacle diameter $D=160$~$\mu$m. 
     (a) PDF of kinetic energy; (b) PDF of enstrophy; (c) Enstrophy versus kinetic energy for successive times (color-coded).  {\bf (d-f)} Same for  $W=1000$~$\mu$m, $D=300$~$\mu$m.
      {\bf (g-i)} Same for  $W=600$~$\mu$m, $D=160$~$\mu$m.
      {\bf (j-l)} Same for  $W=1000$~$\mu$m, $D=160$~$\mu$m,  and chiral $V$-shaped obstacle design.
     }}
\end{figure}

\modif{ The list of actual measurements on racetracks (Fig.~\ref{fig:distrib_width}) differs from that of racetrack substrates (Table~S1) for two reasons. First, there is some spreading in dimensions. Second, for each substrate we perform one or more measurements, selected as time sequences during which the velocity field is stable in time.}

\subsubsection{Soft PDMS substrate}

For traction force microscopy we used  Dow Corning Toray CY 52-276 A/ Dow Corning Toray CY 52-276 B, mixed at a ratio 5:5,  which exhibits a Young modulus of approximately 3~kPa (measured using a commercial rheometer).
Between 0.1 and 0.2~g of  this PDMS is deposited onto a commercial glass-bottom Petri dishes (Nunc 27~mm)  and allowed to spread naturally by viscosity over approximately one hour before curing at 65$^\circ$C for 17~min. This process results in a working thickness of the PDMS layer of about 100~$\mu$m.
 The surface is then silanized in liquid phase using Sigma reagent 440140 at 5\% in absolute ethanol to activate it for covalent bonding with the beads. After rinsing three times with 95\% ethanol to ensure homogeneous wetting without droplets, the substrate is dried in an oven at 60$^\circ$C for 30~min. The bead solution, diluted 1/250, is incubated on the surface at room temperature for 5~min, followed by three rinses with distilled water. To reduce the stickiness of the beads, they are passivated with a Tris solution for 20~min at room temperature, and finally dried at 60$^\circ$C for 15~min. 

\modif{This PDMS is very soft, fragile and sticky. Despite numerous attempts, we have not been able yet to find a method to place and remove a stamp without damaging either the cell monolayer or the PDMS substrate. This has prevented us from perform reproducible, unbiased open strip traction force experiments.}

\subsubsection{Immunofluorescence}

After approximately 24 hours of acquisition, the samples are fixed and stained to visualize different cellular components, as follows.
First, the samples are quickly rinsed with PBS. Next, they are fixed with 4\% PFA for 20~min. This is followed by three PBS rinses for 5~min each, during which the cells are permeabilized with 0.1\% Triton X-100. Finally, the samples are rinsed three more times with PBS for 5~min each.
 Vimentin  is stained with a primary antibody raised in rabbit (1/250 concentration) overnight, followed by a secondary goat anti-rabbit antibody (1/500 concentration) for 1 hour. The staining is performed at 4$^\circ$C when done overnight, and at room temperature otherwise. Finally, the samples are stored in a cool environment for several months, ensuring they do not dry out.

\subsubsection{Other drugs}\label{protodrogues}

Drugs other than mitomycin are added to the cell culture medium once migration has already significantly begun. Incubation is performed at 37$^\circ$C for 1 hour, followed by multiple rinses. 

To disrupt the microtubule network, Nocodazole (Sigma: M1404) at 0.5~$\mu$g/mL is used~\cite{chang2008}.
To affect the branched actin network, CK666 (Sigma: SML0006) at 100~$\mu$M is used~\cite{latorre2018,chen2019}. To disrupt the vimentin network, Simvastatin (Tocris: 1965/50) at 10~$\mu$M is used~\cite{trogden2018}. 
\modif{Finally, to affect myosin activity and cellular activity, para-amino-Blebbistatin (Axol ax494682) at 50 $\mu$M is used. It has the advantage of being much more stable and less phototoxic than traditional Blebbistatin~\cite{varkuti2016}.}

\subsection{Preparation of Products}\label{prepprod}

\subsubsection{Preparation of a Polyvinyl Alcohol (PVA) membrane}\label{PVA}

A solution is prepared by mixing 0.5 g of PVA in 20 mL of pure water, while stirring and heating (approximately 80$^\circ$C) until fully dissolved (typically within 5 to 6 hours). The solution is then filtered using a 0.22~$\mu$m filter and poured into a 10~cm diameter Petri dish. The membrane is dried under a fume hood with the lid slightly ajar to allow airflow while minimizing exposure. This method helps to prevent the formation of ripples on the surface of the membrane. The membrane is stored for one to two weeks, wrapped in parafilm to prevent excessive drying.

\subsubsection{Preparation of the bead solution}\label{solutionbille}

To prepare the bead solution, 3.8 mg/mL of sodium tetraborate and 5 mg/mL of boric acid are dissolved in distilled water with all necessary precautions.
This mixture is stirred until fully dissolved overnight under a fume hood. The undiluted bead solution is then sonicated for 10~min before being added to the medium at the desired concentration (1/250 or 1/500). This mixture remains stable for at least one month.

Immediately before an experiment, the bead solution is sonicated for 5~min to ensure proper homogenization and to detach any aggregated beads. Just before contact with the substrate, EDC (Sigma E7750)
is added at a concentration of 0.1 mg/mL. EDC degrades extremely quickly in a basic environment but acts as a potent catalyst when added to the medium immediately before use.

\subsubsection{Preparation of Pluronic solution}

A 2\% Pluronic solution is prepared by mixing 1 g of Pluronic in 50 mL of distilled water and stirring until fully dissolved. The solution is then filtered using a 0.22 $\mu$m filter.

\subsubsection{Indirect printing of protein patterns}\label{paragraph: impindirect}

For indirect printing on a soft substrate, the process begins by preparing the PVA membrane (see Section~\ref{PVA}). A PDMS stamp is prepared in the same way as for rigid PDMS surface stamping. Subsequently, the stamp is placed on the PVA membrane for approximately 5~min, allowing the adhesion proteins to adsorb onto the PVA surface. The membrane is then carefully cut around the stamps, and the stamps are peeled off the membrane. 

The next step involves placing the membrane face down onto the PDMS with great care to ensure that the proteins adsorb onto the PDMS rather than on the PVA. 
The PVA is dissolved while passivating the surface with a 2\% Pluronic solution for 2 hours. Finally, the surface is rinsed 6 to 7 times with PBS, taking care never to let the surface dry.

\subsubsection{Surface passivation}

At this stage, a glass surface with PDMS with the printed pattern is obtained. The final step before cell deposition involves passivating this surface to prevent cells from adhering to areas without adhesion proteins. For this purpose, Pluronic is used; this hydrophilic surfactant binds to the fibronectin-free areas and blocks cell access to these regions. The surface is incubated with a 2\% Pluronic solution in PBS for 2 hours before rinsing it 6 or 7 times with PBS to remove residual cytotoxicity.

\subsection{Data Analysis}

\subsubsection{Velocity measurements}
\label{sec:KLT}

The Kanade-Lucas-Tomasi (KLT) algorithm~\cite{lucas1981, bouguet} tracks object velocities by searching for pixels with high local intensity gradients. A fixed-size window of 64 pixels is defined, and the algorithm seeks the correspondence of points within this window in the subsequent image. The algorithm is pyramidal: recognition is performed on images of different successive resolutions with identical window sizes. Starting from the lowest resolution, large-scale movements are traced, and as the resolution increases, the displacement measurement is refined. This method allows for cell tracking, providing a Lagrangian approach to displacement. Eulerian analysis is possible by interpolating or averaging the results on a fixed grid.
In practice, the steps are as follows.

First, all microscopy fields at a given time are stitched together to create a large image. Points in these large images, for which the KLT is calculated, are defined by searching for local maxima in the last image. The KLT is used as a backward particle tracking method (from the end to the beginning). We select 50,000 points, which mainly correspond to cell edges, and exclude those that are too close to each other (less than 15 pixels apart).

Once all points are defined in the last image, the KLT calculation is performed by returning to the initial fields. 
Because of frame stitching, some overlap can occur; for points present in the overlap region, their velocities on both frames are averaged.

Finally, the points are moved according to the found velocity vector, and the KLT calculation steps are repeated. The local maxima are calculated only once, during the first pass of the algorithm.

\subsubsection{Traction Force Microscopy}
\label{sec:TFM}

The relationship between the traction field $\mathbf{t}(\mathbf{x})$ exerted by the cells and the substrate displacement field $\mathbf{u}(\mathbf{x})$ measured experimentally can be expressed as a convolution with the elastic Green tensor $G_{ij}$, where $i,j$ are cartesian coordinates:
\begin{equation}
u_i(\mathbf{x}) = \int G_{ij}(\mathbf{x} - \mathbf{x}')\, t_j(\mathbf{x}')\, d\mathbf{x}' ,
\end{equation}
 The coefficients of $\mathbf{G}$ are derived from the Boussinesq solution for an isotropic, linearly elastic half-space characterized by the Young modulus $E$ and the Poisson ratio $\nu$. We used Fast Fourier Transform, following Butler   \cite{butler2002} to compute cell-matrix stresses from the displacement field.

Direct inversion of $\mathbf{G}$ provides an unregularized solution that is highly sensitive to experimental noise, rendering the problem ill-posed and non-unique. To overcome this limitation, a truncated singular value decomposition (tSVD) regularization is applied \cite{hansen2010}. The singular values of $\mathbf{G}$ are ordered in decreasing magnitude, and the truncation point is identified from the maximum of the second derivative of their distribution, effectively filtering out the small eigenvalues associated with high-frequency components. This approach smooths the reconstructed traction field while preserving its essential spatial structure, ensuring a stable and robust inversion.

\subsubsection{Spatial correlation function}
\label{sec:measure_correlations}

Two parameters strongly affect the correlation function of the velocity field fluctuations: (i) the size $\mathrm{d}A$ of spatial averaging window, and (ii) the size of the box  on which  the correlation is calculated. 
Importantly, to avoid artifacts, in each box we normalize the fluctuations by the average velocity squared measured in this box.

The spatial correlation function is a function of the distance between two points. In the racetrack case, if we use  Cartesian coordinates the correlation will be performed without regard for the physical meaning: two points could be considered as correlated even if the direct path between them is the central zone void of cells.

\begin{figure}[!ht]
	\centering
            \begin{subfigure}[t]{0.47\textwidth}
        \centering
        \includegraphics[width=\linewidth]{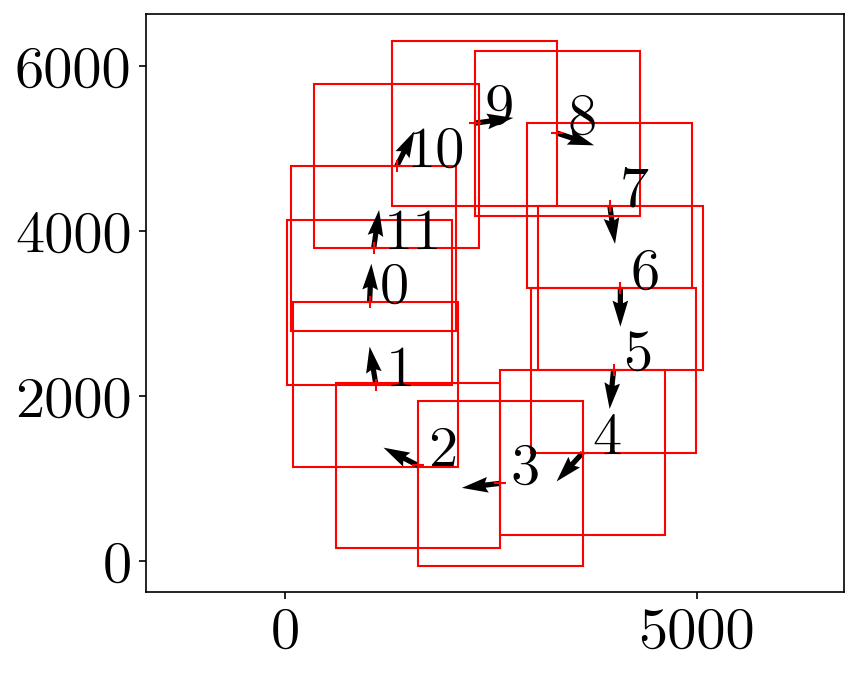}
        \caption{}
        \label{fig:hippogrid}  
    \end{subfigure}
     \begin{subfigure}[t]{0.52\textwidth}
        \centering
        \includegraphics[width=\linewidth]{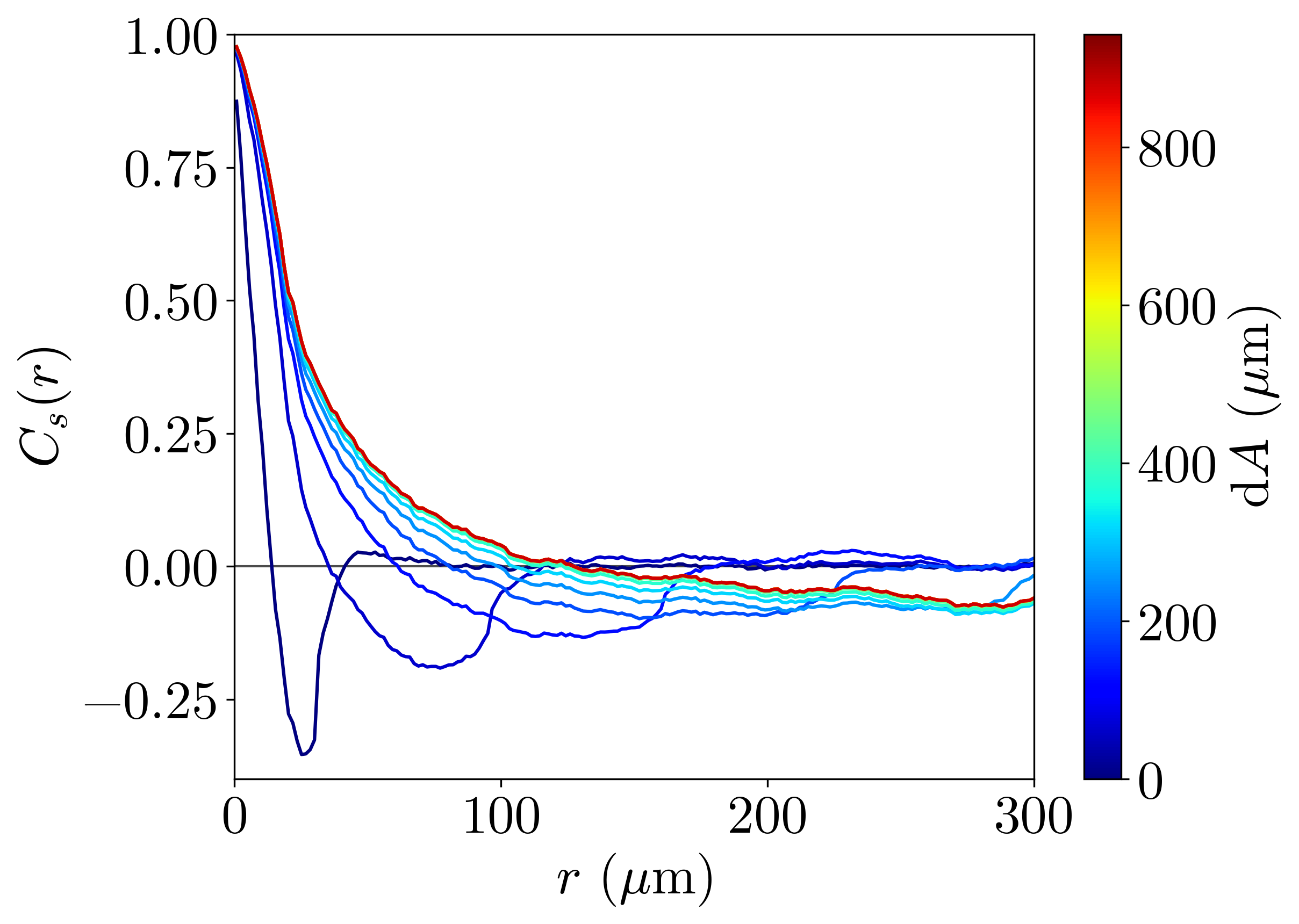}
        \caption{}
        \label{fig:corr_spatial_dA}
    \end{subfigure}
    
    \begin{subfigure}[t]{0.59\textwidth}
        \centering
        \includegraphics[width=\linewidth]{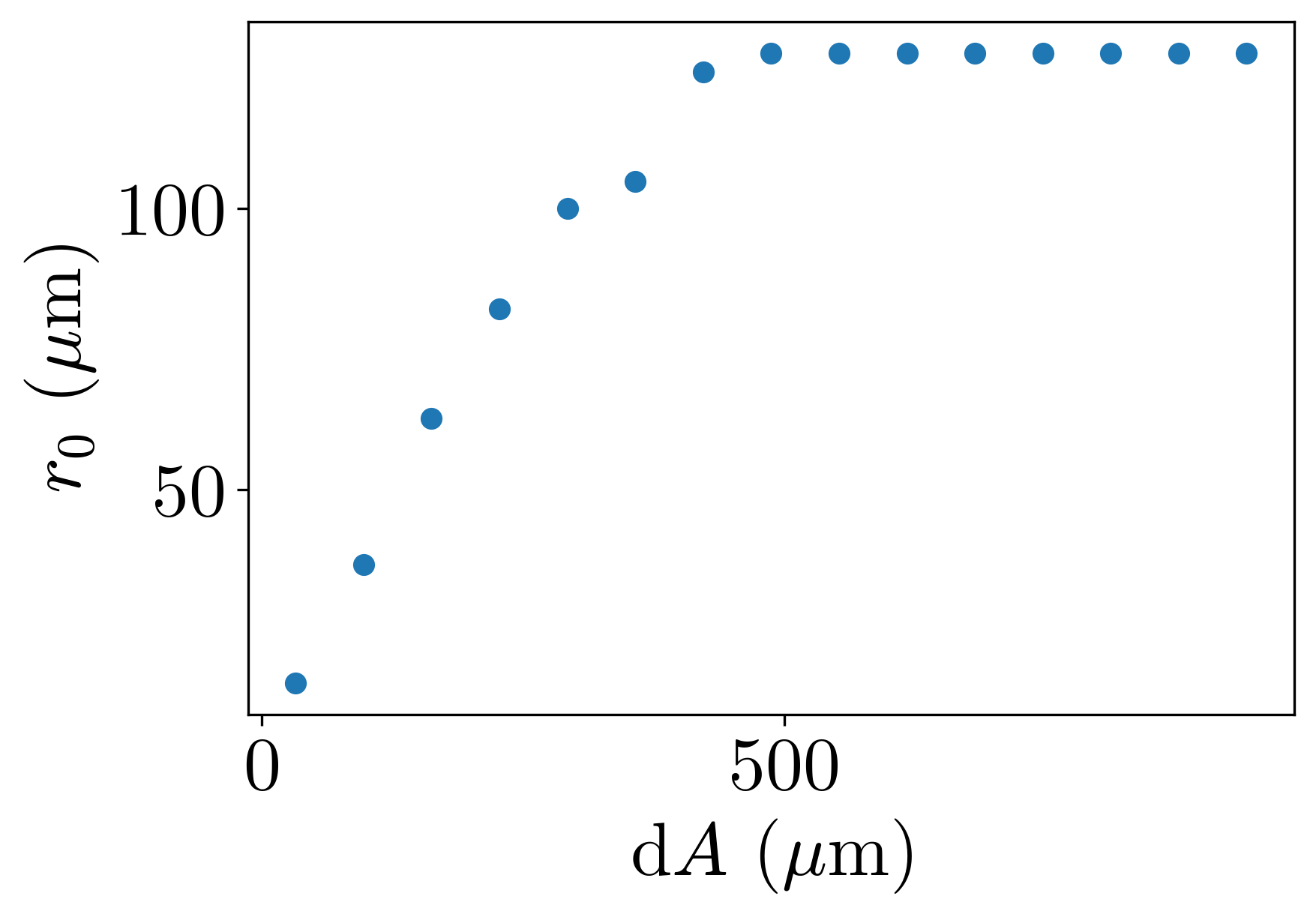}
        \caption{}
        \label{fig:r_0_dA}
    \end{subfigure}

	\caption{{\bf Averaging windows in a racetrack.} 
	{\bf (a)} Division of the racetracks into interrogation windows, here of size $\mathrm{d}A =1300$~$\mu$m. Graduations in $\mu$m. 
	{\bf (b)} Autocorrelation function for different values of averaging window size $\mathrm{d}A$, color coded ($\mu$m). 
	{\bf (c)} $r_0$ values obtained from each curve in (b) vs $\mathrm{d}A$.}
	\label{fig: dA_effect_classic}
	
\end{figure}

First, we use the classical method of calculation, by calculating the correlation function on squared interrogation windows  (Fig.~\ref{fig: dA_effect_classic}a). We first choose what we expect to be a sufficiently large one, i.e., 1300~$\mu$m  (Fig.~\ref{fig: dA_effect_classic}b). To increase the statistics, we repeat the same measurement all along the racetrack, and reach a plateau (Fig.~\ref{fig: dA_effect_classic}c). 

\begin{figure}[!ht]
    \centering
        \begin{subfigure}{0.99\textwidth}
        \includegraphics[width=1\textwidth]{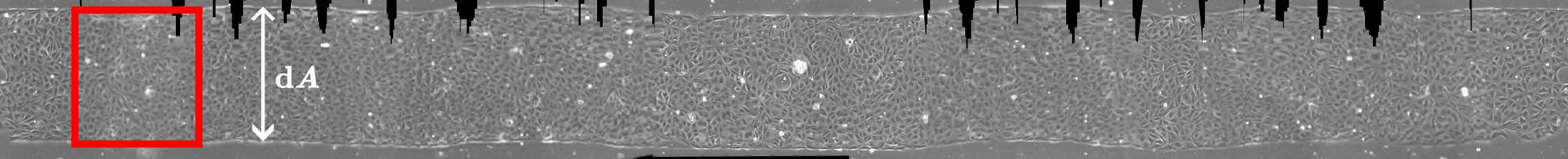}
        \caption{\label{fig:racetrack_flatten} } 
    \end{subfigure}

    \begin{subfigure}{0.6\textwidth}
        \includegraphics[width=1\textwidth]{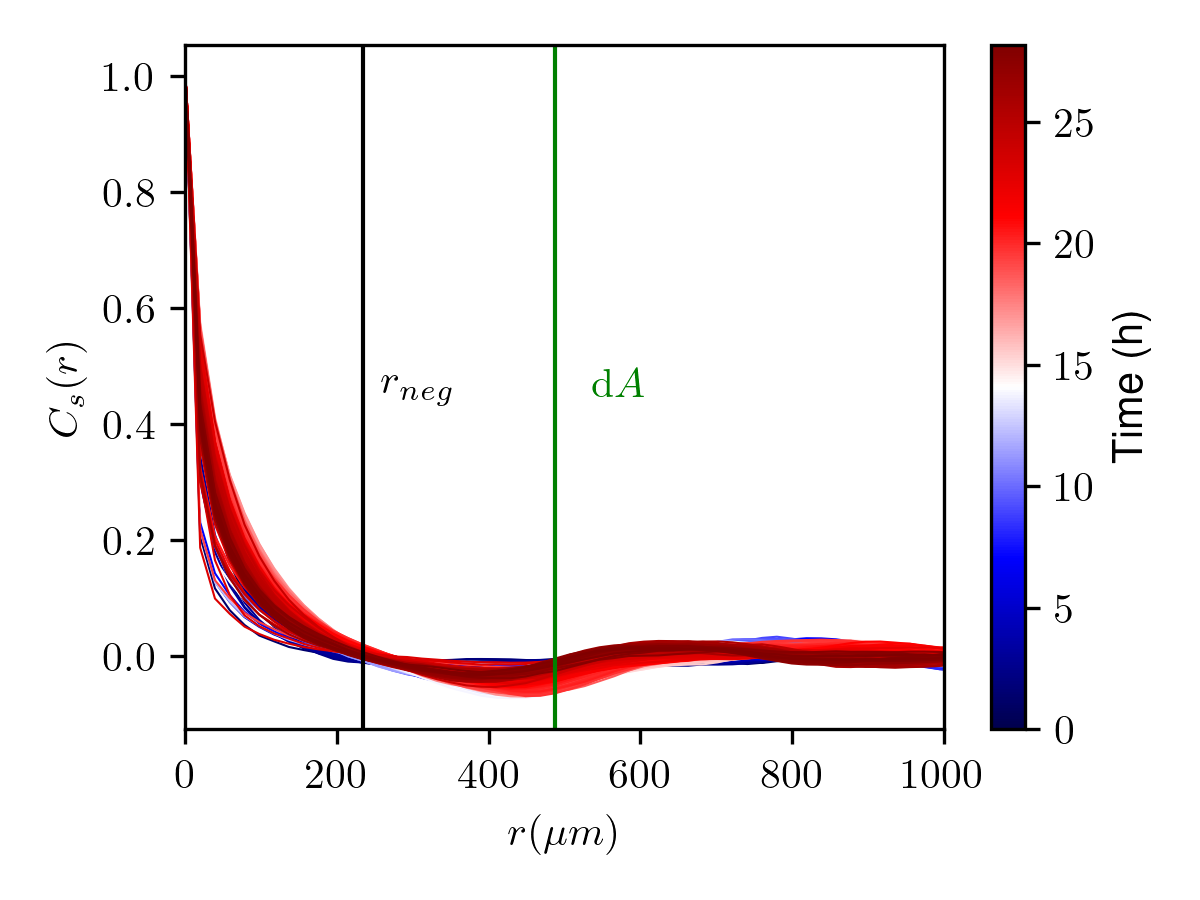}
        \caption{\label{fig:Cs_fs_curve_example}}  
    \end{subfigure}
    
    \begin{subfigure}{0.6\textwidth}
        \includegraphics[width=1\textwidth]{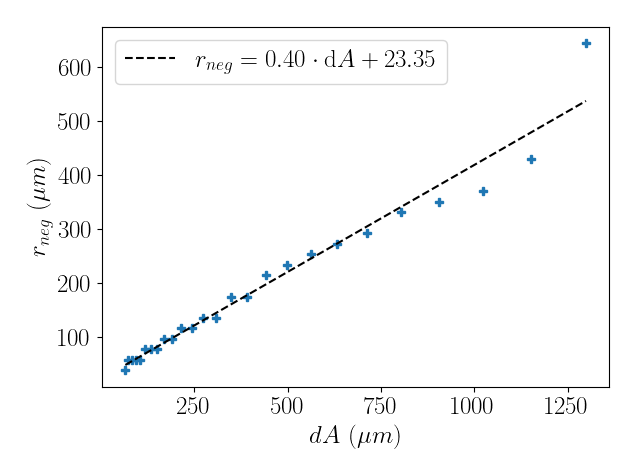}
        \caption{\label{fig:first_neg_value_Cs_fs} }  
    \end{subfigure}
    \caption{{\bf Effect of averaging window size in a racetrack.}
    {\bf (a)} Transformation of the racetrack images into the curvilinear system of coordinates, with periodic boundary conditions on the right and left sides. In this case, length $L=6800$~$\mu$m, width $W=600$~$\mu$m.
    {\bf (b)} Spatial correlation function of the velocity field fluctuations around the spatial average; $\mathrm{d}A$ is the average window size \modif{(vertical green line)}. Color code:  time (h). Black vertical line:  first negative value of the correlation function, $r_{\rm neg}$. 
    {\bf (c)} Increase of $r_{\rm neg}$ with the average window size $\mathrm{d}A$. The dashed line is a linear fit to the data.}
    \label{fig:effect_of_dA}
\end{figure}

\begin{figure}[t!]
    \centering
    \begin{subfigure}{0.49\textwidth}
        \centering
        \includegraphics[width=\linewidth]{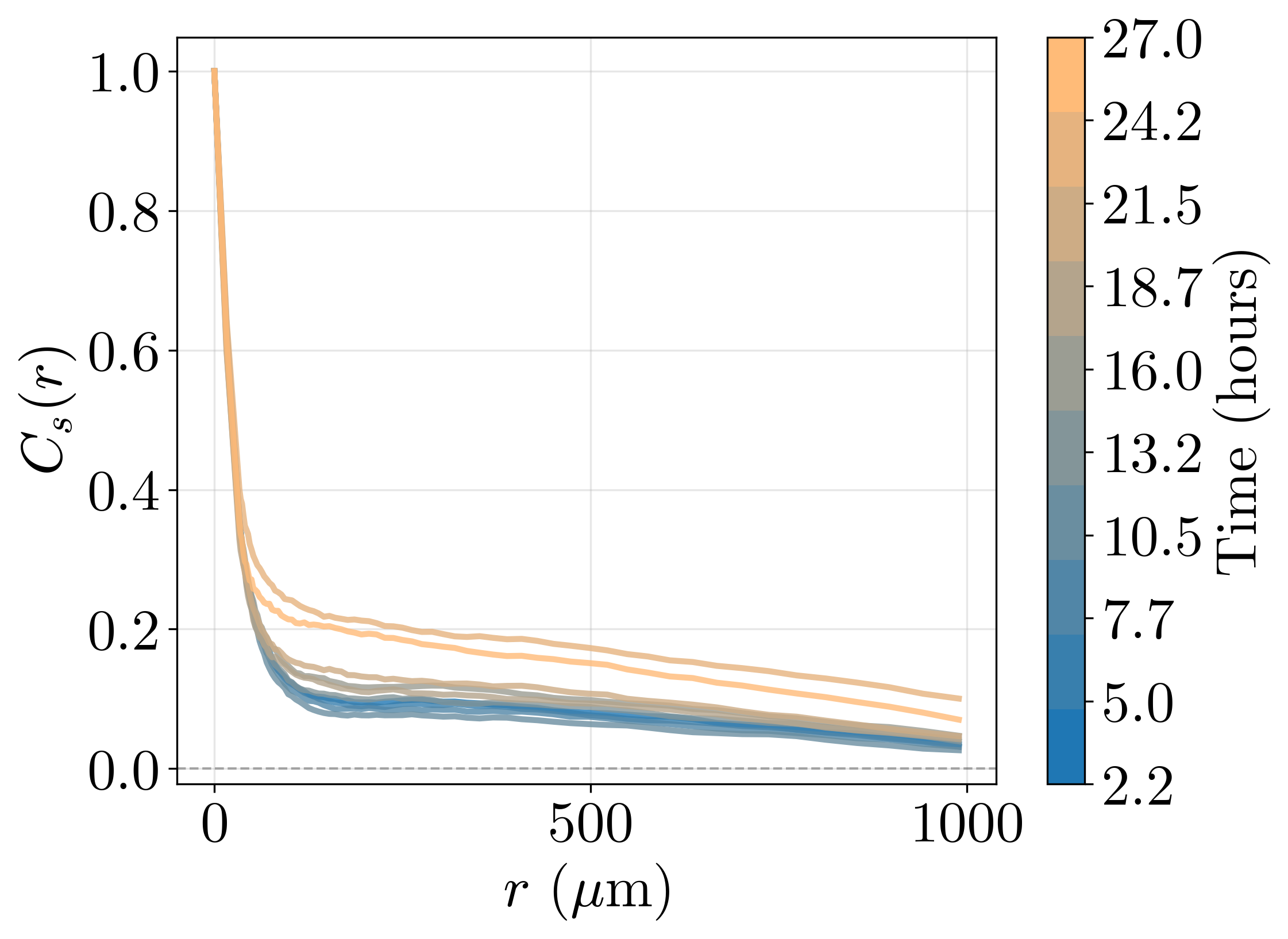}
        \caption{}
        \label{fig:corr_spat_lin_supp}
    \end{subfigure}
    \hfill
     \begin{subfigure}{0.49\textwidth}
        \centering
        \includegraphics[width=\linewidth]{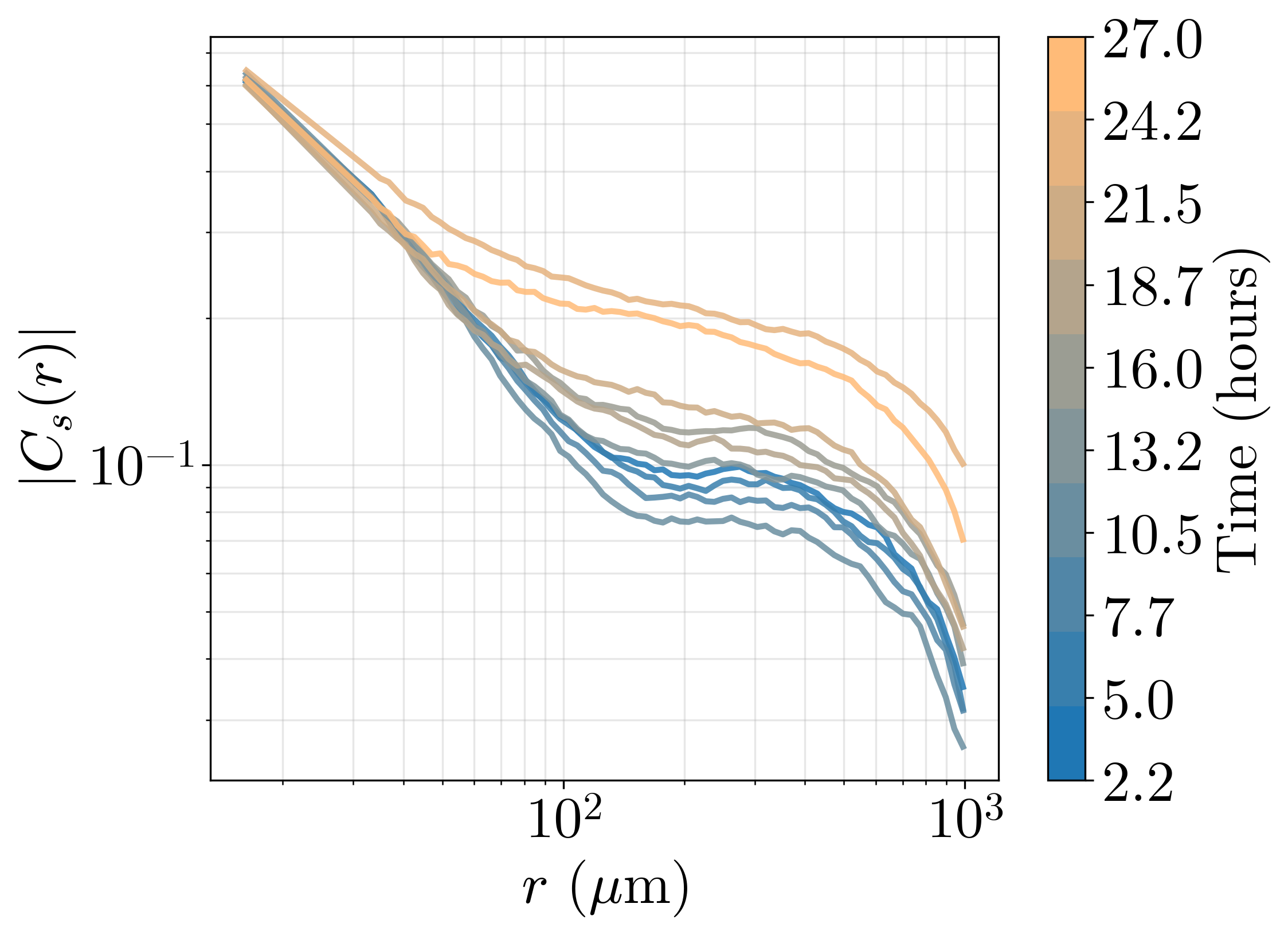}
        \caption{}
        \label{fig:corr_spat_log_supp}
    \end{subfigure}
    \begin{subfigure}[t]{0.49\textwidth}
        \centering
        \includegraphics[width=\linewidth]{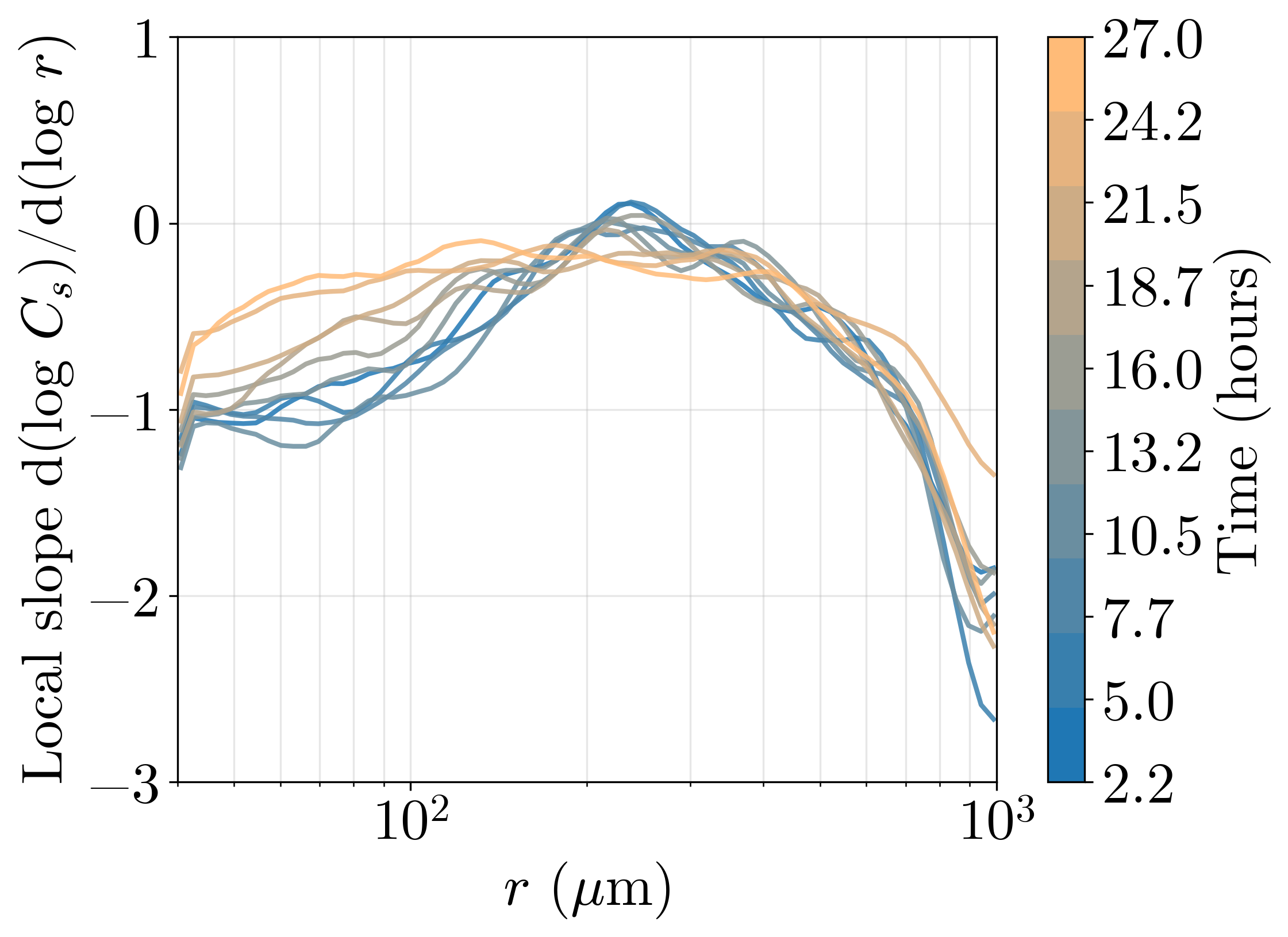}
        \caption{}
        \label{fig:corr_spat_slopes_supp}
    \end{subfigure}
    \hfill
    \begin{subfigure}[t]{0.49\textwidth}
        \centering
        \includegraphics[width=\linewidth]{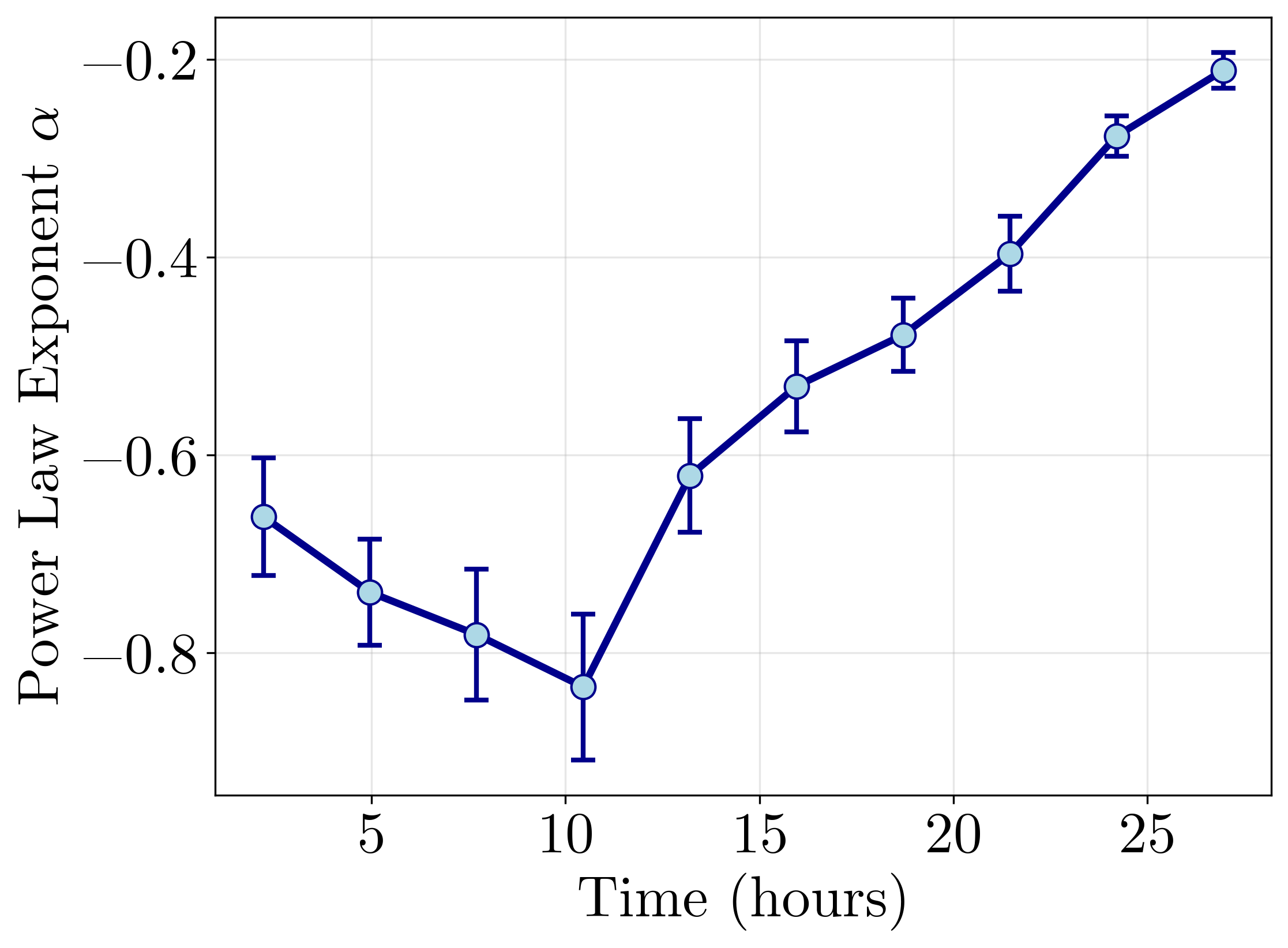}
        \caption{}
        \label{fig:corr_spat_alpha_supp}
    \end{subfigure}
    \caption{\label{fig:temporal_evo_supp} 
    \modif{
    {\bf Temporal evolution of the velocity correlation function}. Same as Fig.~\ref{fig:temporal_evo} on a racetrack with width $W=1000$~$\mu$m.
    {\bf (a)} Spatial correlation function of the velocity versus distance $r$, for successive times color-coded from 2.2 to 27~h.
    {\bf (b)} Same as (a) in log-log scale.
    {\bf (c)} Local exponent defined as the slope of (b), i.e. logarithmic derivative of (a).
    {\bf (d)} Time evolution of exponent $\alpha$, defined as the slope in (c), average ($\pm$~s.e.m.) between $r=30$ and 200~$\mu$m.
    }}    
\end{figure}

\begin{figure}[t!]
    \centering
    \begin{subfigure}{0.49\textwidth}
        \centering
        \includegraphics[width=\linewidth]{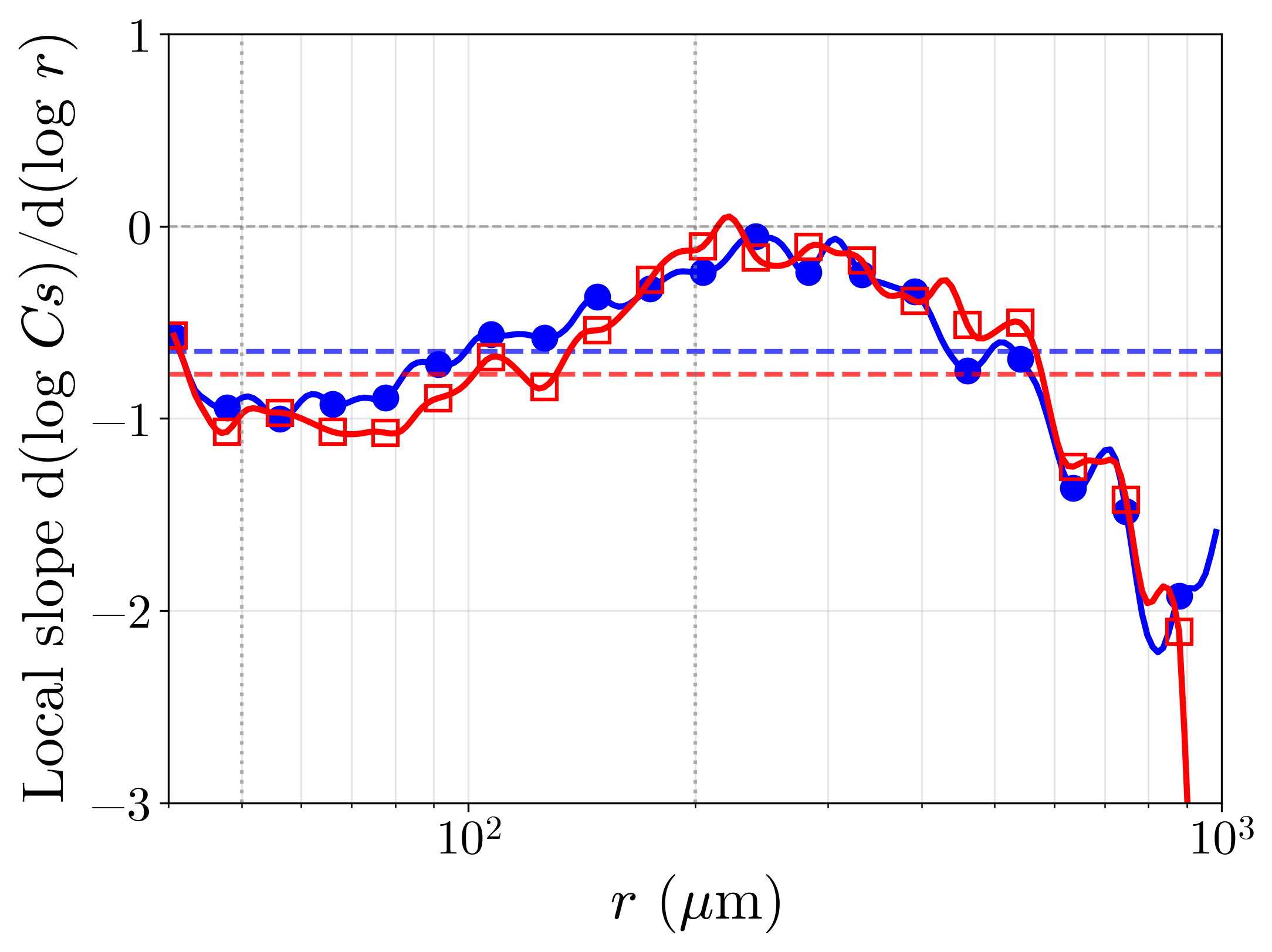}
        \caption{}
        \label{fig:drug_slope_control}
    \end{subfigure}
    \hfill
     \begin{subfigure}{0.49\textwidth}
        \centering
        \includegraphics[width=\linewidth]{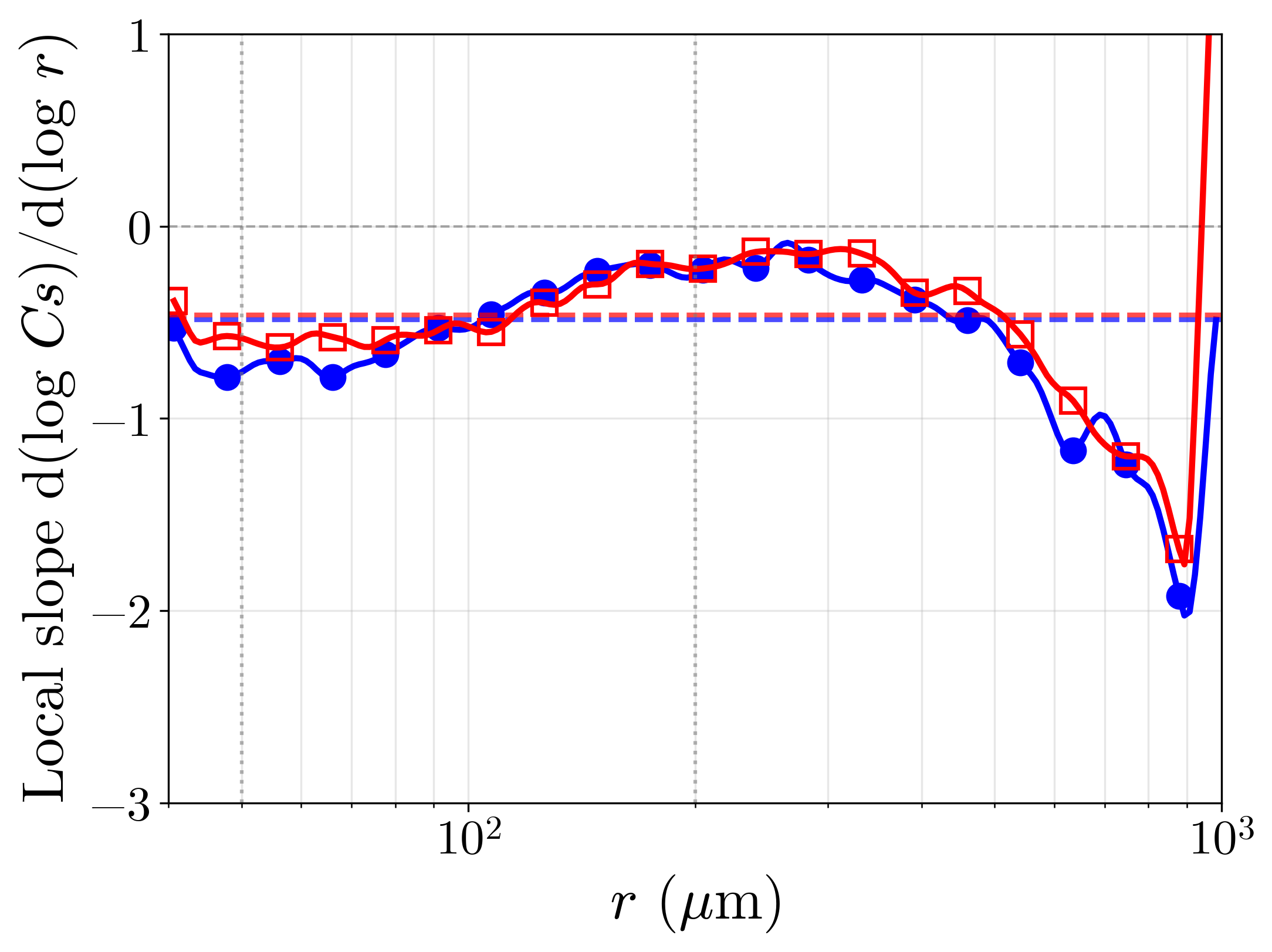}
        \caption{}
        \label{fig:drug_slope_CK666}
    \end{subfigure}
    \begin{subfigure}[t]{0.49\textwidth}
        \centering
        \includegraphics[width=\linewidth]{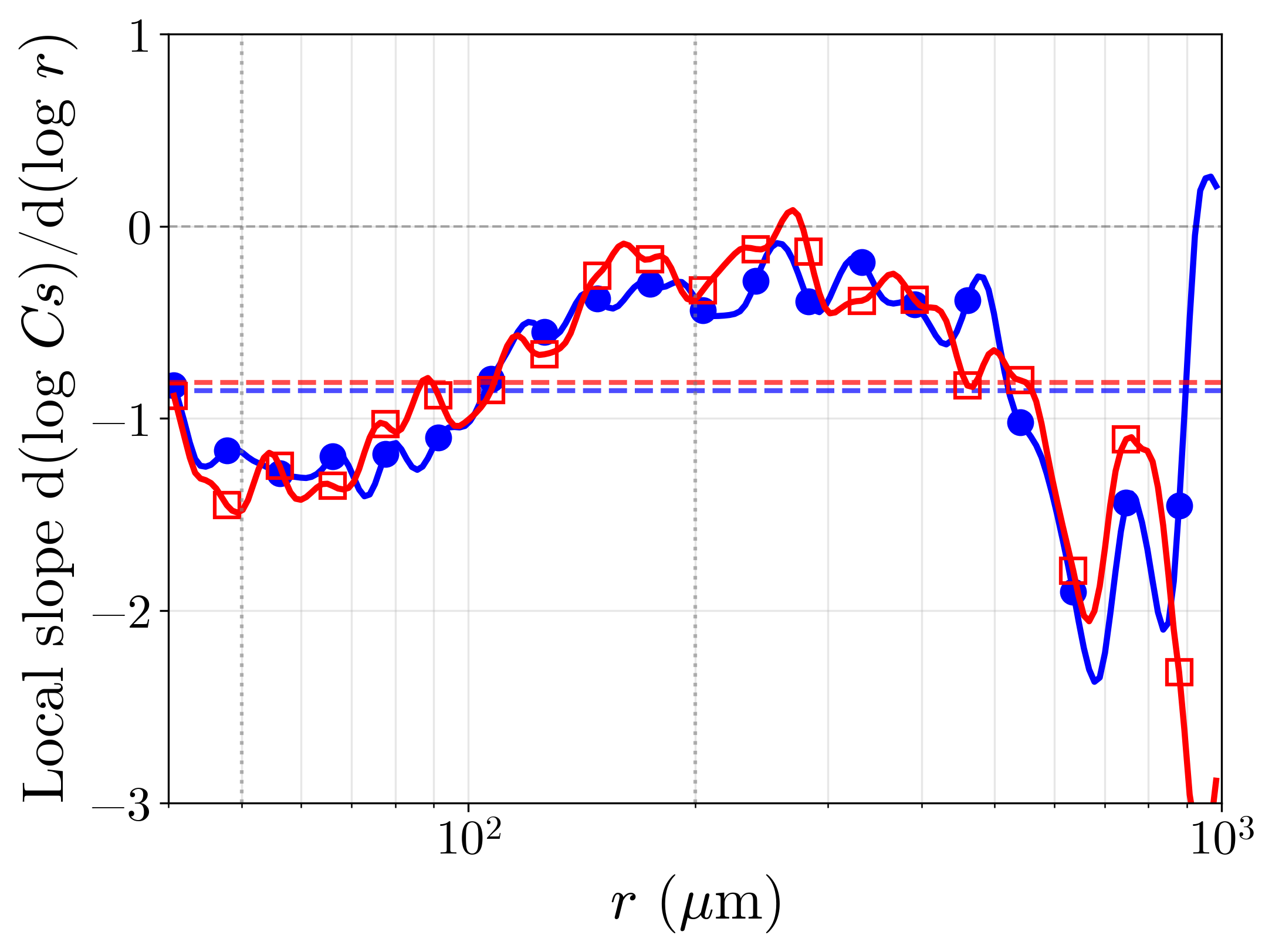}
        \caption{}
        \label{fig:drug_slope_noco}
    \end{subfigure}
    \hfill
    \begin{subfigure}[t]{0.49\textwidth}
        \centering
        \includegraphics[width=\linewidth]{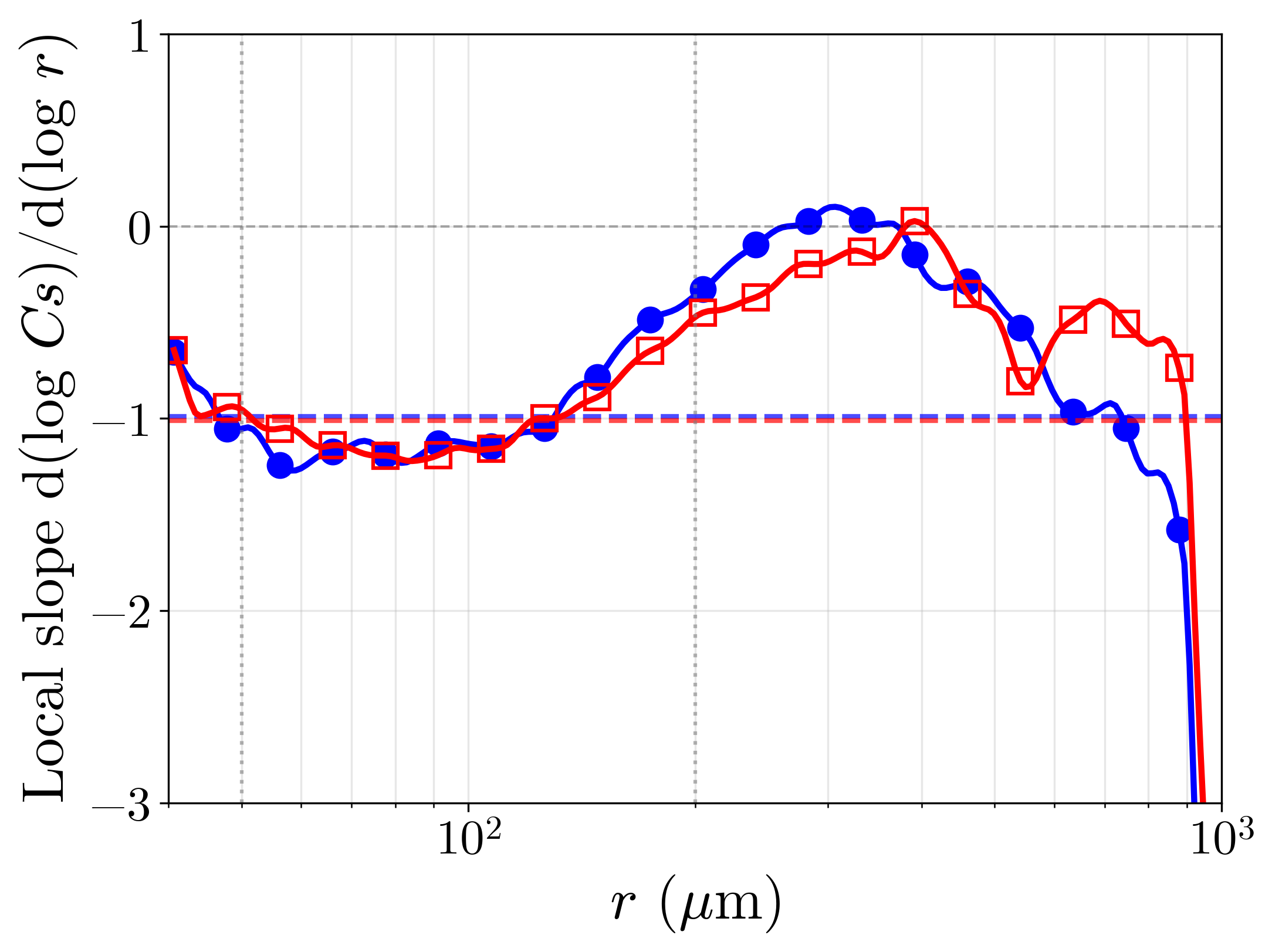}
        \caption{}
        \label{fig:drug_slope_vim}
    \end{subfigure}
    \caption{\label{fig:drugs_slopes} 
    \modif{
    {\bf Effects of drugs on the  spatial correlation exponents.} 
    {\bf (a)} Control, the medium of the well was renewed but without drug: average exponent $-0.65\pm 0.03$ before this intervention, $-0.77 \pm 0.03$ after it.
    {\bf (b)} CK666 was added to the medium:  $-0.49\pm0.02$ before, $-0.46\pm 0.02$ after.
    {\bf (c)} Nocodazole was added to the medium: $-0.86 \pm 0.05 $ before, $-0.81 \pm 0.05 $ after.
    {\bf (d)} Simvastatin was added to the medium: $-0.99\pm 0.03 $ before, $-1.01 \pm 0.02$ after.
    Blue circles before, red squares after.
    }}    
\end{figure}

Second,  to take advantage of the largest possible interrogation window, we unwrap and flatten the racetrack.
To do so, we define a local system of coordinates respectuous of the racetrack topology, using the curvilinear abscissa $s$ (i.e. the projected position onto the racetrack midline) and the distance to this line. 
We then obtain a rectangle with periodic boundary conditions (see Fig.~\ref{fig:effect_of_dA}a, and estimations of the error it entails). Performing again the correlation function calculation (Fig.~\ref{fig:effect_of_dA}b), this time we no longer reach
any plateau (Fig.~\ref{fig:effect_of_dA}c). 

Fitting a power law to  the correlation function  is performed using a robust regression technique: the Huber regressor~\cite{huber2011}, which is part of the \texttt{scikit-learn} library~\cite{zotero-9062}. It minimizes the least square residues, and modifies the loss function to down-weight outliers, providing a more reliable fit in the presence of noisy data.

\newpage

\section{Supplementary Figures}
\label{sec:supp_figures}

\newpage

\newpage

\newpage

\newpage

\newpage

\newpage

\clearpage

\section{Supplementary Movies}
\label{sec:supp_movies}


\medskip 
\medskip 
\medskip 
\medskip 

\begin{centering}
   \includegraphics[width=0.736\textwidth]{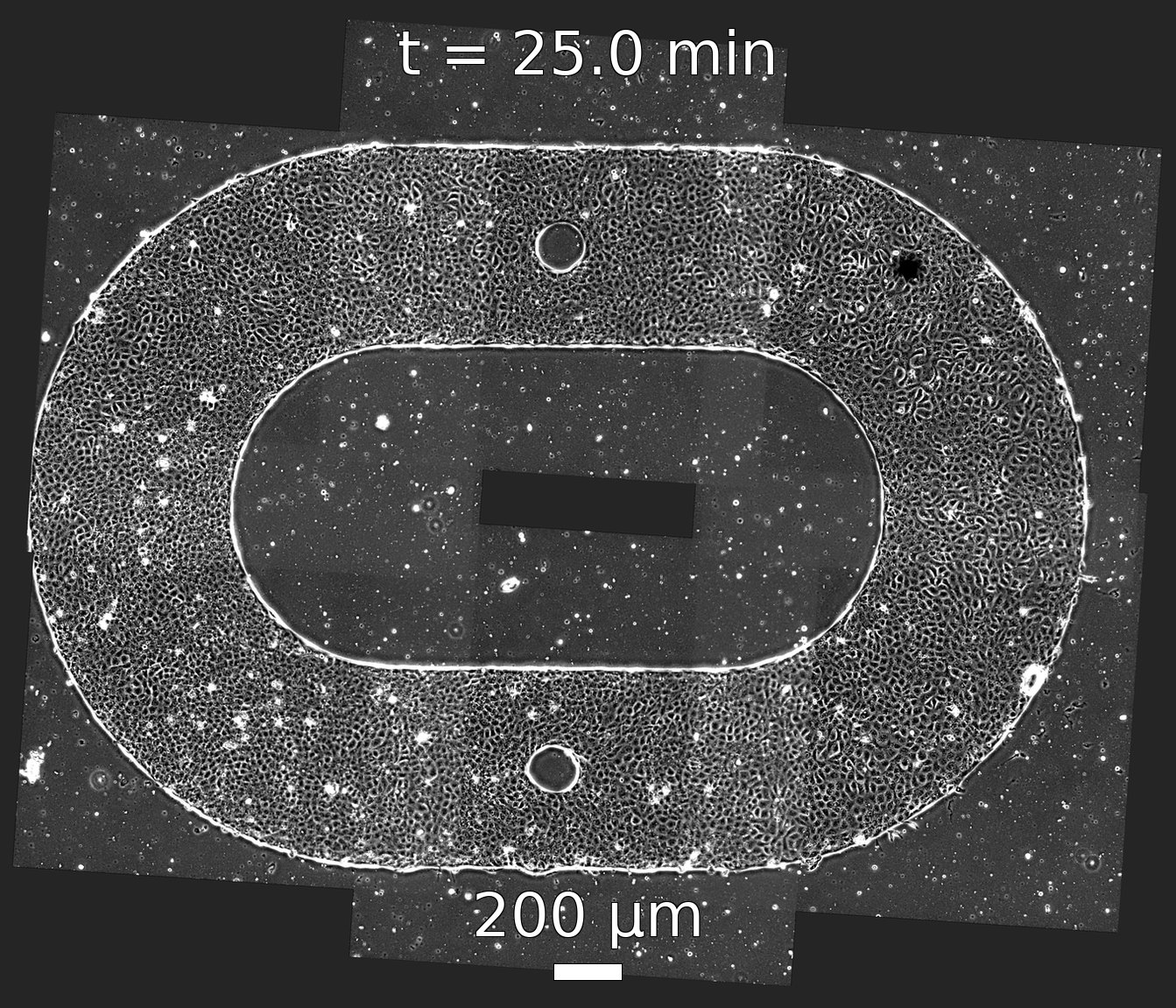}\\
     \modif{ {\bf Supplementary Movie 1. }Example of racetrack experiment. 
     Track width $W=600$~$\mu$m, obstacle diameter $D=160$~$\mu$m. 
     Interframe 5 min. At the end of the experiment, a drug is added: Simvastatin. }
\end{centering}

\newpage

\begin{centering}
   \includegraphics[width=0.917\textwidth]{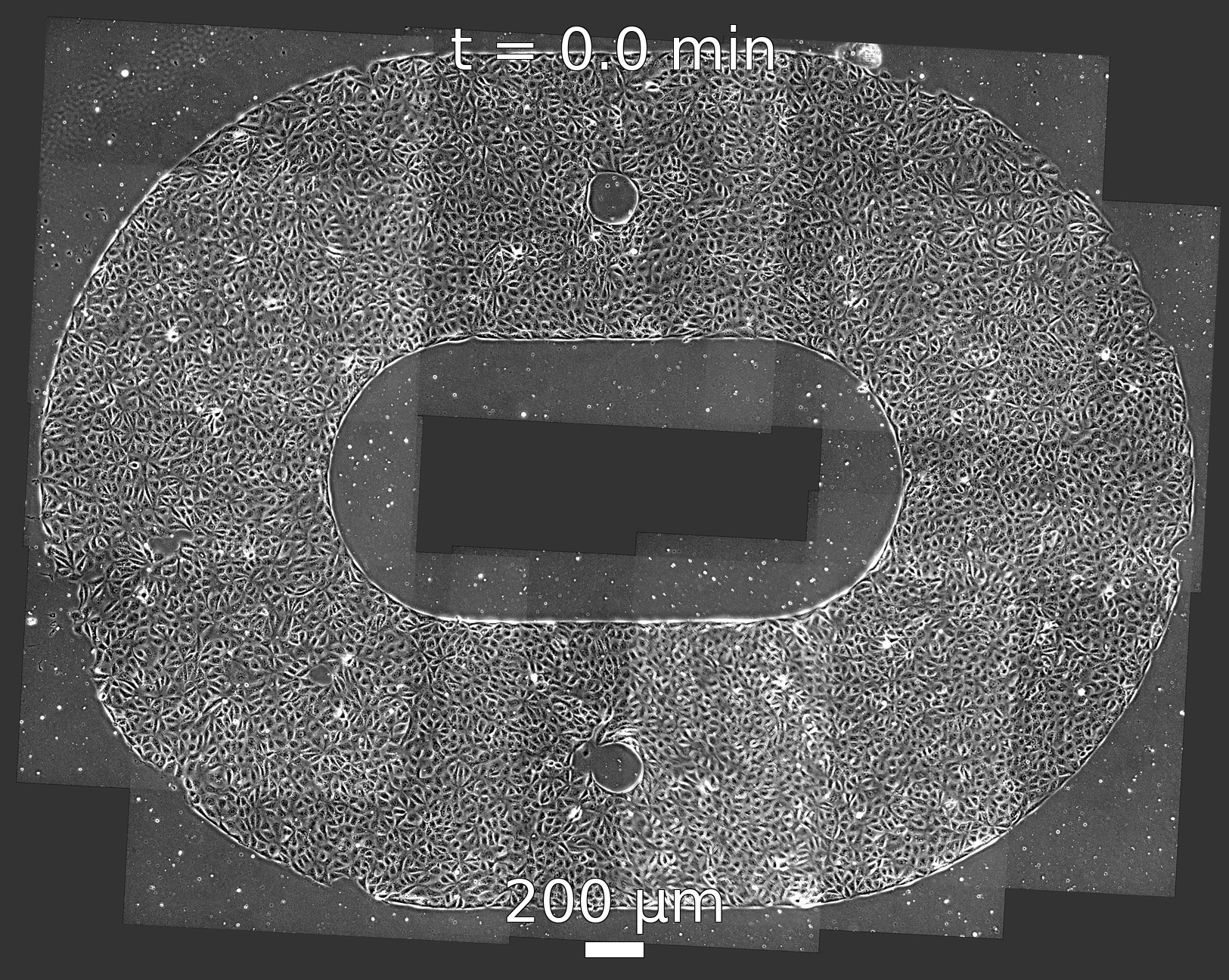}\\
\end{centering}
    \modif{ {\bf Supplementary Movie 2. } Other example of racetrack experiment, with different dimensions. 
     Track width $W=1000$~$\mu$m, obstacle diameter $D=160$~$\mu$m. 
     At the end of the experiment, CK666 is added. }

\newpage

\begin{centering}
   \includegraphics[width=0.9\textwidth]{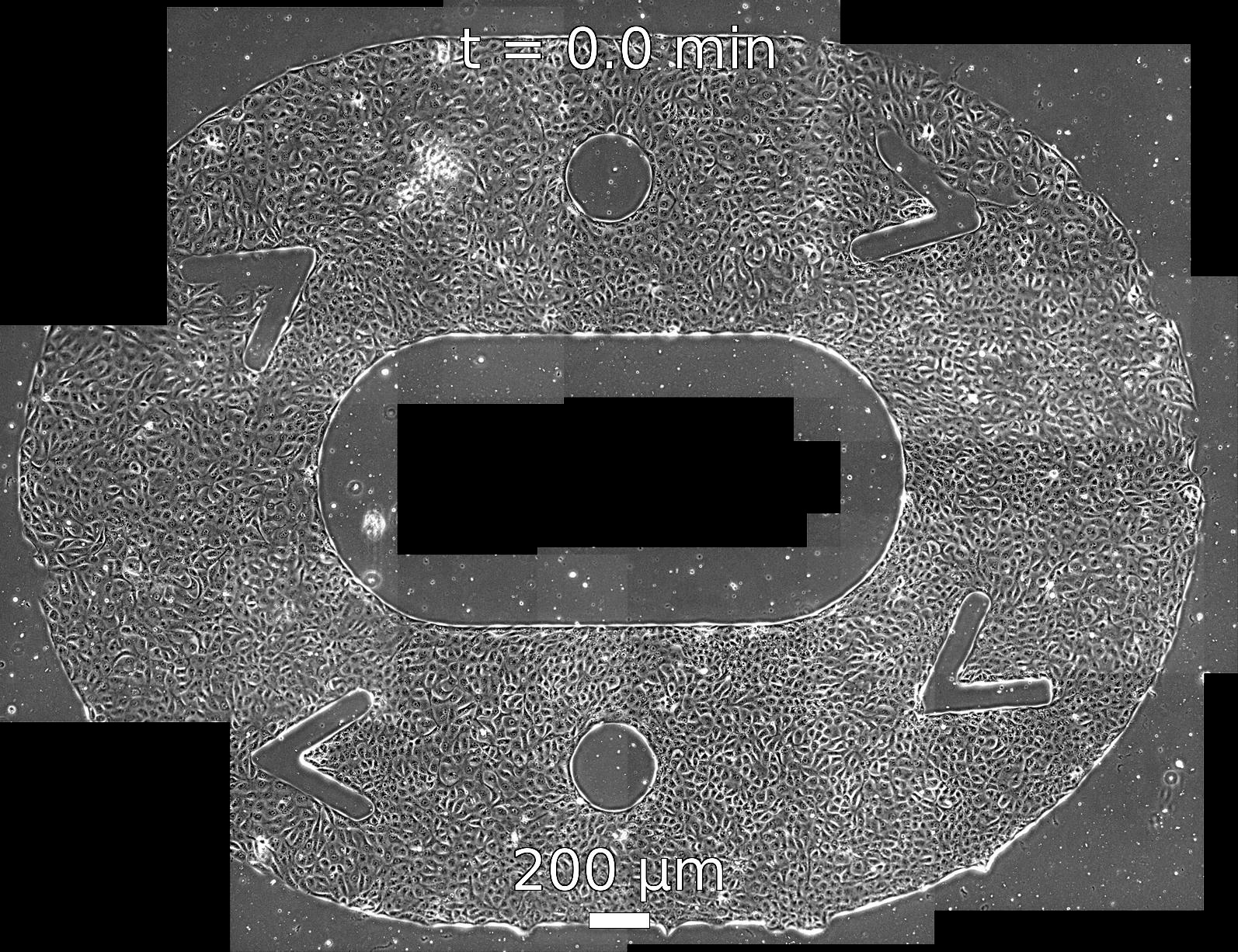}\\
\end{centering}
     \modif{ {\bf Supplementary Movie 3. } Other example of racetrack experiment, with chiral $V$-shaped obstacle design. 
        Track width $W=1000$~$\mu$m, obstacle diameter $D=300$~$\mu$m. 
     Image with a posteriori linear contrast adjustment. At the end of the experiment, Nocodazole is added. }

\end{document}